\documentclass{aastex631}

\usepackage{multirow}
\usepackage{booktabs}
\usepackage{amsmath}

\begin{document}
\title{Radio observations point to a moderately relativistic outflow in the fast X-ray transient EP241021a}

\author[0009-0004-9520-5822]{Muskan Yadav}
\affiliation{Dipartimento di Fisica, Universit\`a di Tor Vergata, Via della Ricerca Scientifica, 1, 00133 Rome, Italy} 

\author[0000-0002-1869-7817]{Eleonora Troja} 
\affiliation{Dipartimento di Fisica, Universit\`a di Tor Vergata, Via della Ricerca Scientifica, 1, 00133 Rome, Italy}

\author[0000-0003-4631-1528]{Roberto Ricci}
\affiliation{Dipartimento di Fisica, Universit\`a di Tor Vergata, Via della Ricerca Scientifica, 1, 00133 Rome, Italy}
\affiliation{INAF-Istituto di Radioastronomia, Via Gobetti, 101, 40129 Bologna, Italy}

\author[0000-0003-0691-6688]{Yu-Han Yang}
\affiliation{Dipartimento di Fisica, Universit\`a di Tor Vergata, Via della Ricerca Scientifica, 1, 00133 Rome, Italy}


\author[0000-0002-7721-8660]{Mark H. Wieringa}
\affiliation{CSIRO Space and Astronomy, P.O. Box 76, Epping, NSW 1710, Australia}

\author[0000-0002-9700-0036]{Brendan O'Connor}
\affiliation{McWilliams Center for Cosmology and Astrophysics, Department of Physics, Carnegie Mellon University, Pittsburgh, PA 15213, USA}

\author[0000-0001-7402-4927]{Yacheng Kang}
\affiliation{Dipartimento di Fisica, Universit\`a di Tor Vergata, Via della Ricerca Scientifica, 1, 00133 Rome, Italy}
\affiliation{Department of Astronomy, School of Physics, 
Peking University, Beĳing 100871, China}
\affiliation{Kavli Institute for Astronomy and Astrophysics, 
Peking University, Beĳing 100871, China}

\author[0000-0002-0216-3415]{Rosa L. Becerra}
\affiliation{Dipartimento di Fisica, Universit\`a di Tor Vergata, Via della Ricerca Scientifica, 1, 00133 Rome, Italy}

\author[0000-0001-9068-7157]{Geoffrey Ryan}
\affiliation{Perimeter Institute for Theoretical Physics, Waterloo, Ontario N2L 2Y5, Canada}

\author[0009-0001-0574-2332]{Malte Busmann}
\affiliation{University Observatory, Faculty of Physics, Ludwig-Maximilians-Universität München, Scheinerstr. 1, 81679 Munich, Germany}







\begin{abstract}

Fast X-ray transients (FXRTs) are short-lived X-ray outbursts with diverse progenitor scenarios, including compact object mergers, stellar core-collapses and tidal disruption events. The \textit{Einstein Probe} (EP) has enabled the rapid discovery and follow-up of dozens of FXRTs, revealing that while some of them overlap with traditional gamma-ray bursts (GRBs), a larger fraction of FXRTs have no associated gamma-ray counterpart down to deep limits. The origin of these gamma-ray dark FXRTs and their connection to the diverse landscape of stellar explosions remains an open question, which can be tackled through the study of their multi-wavelength counterparts and environment. 

In this paper, we present long-term radio observations of the gamma-ray dark EP241021a, which exhibits sustained radio emission for over 100~days, placing it among the longest-lived radio afterglows. We detect signature of interstellar scintillation in early epochs, allowing us to constrain the angular size and Lorentz factor of the emitting region. Our observations point to an outflow that is at least mildly relativistic with $\Gamma \gtrsim 4$. 
Afterglow modeling favors a moderately relativistic ($\Gamma \approx$40) top-hat jet ($\theta_c \approx$0.16 rad) interacting with a low-density interstellar medium. The derived beaming-corrected kinetic energy, $E_\mathrm{k}\sim 1.6 \times 10^{51}$~erg, and low radiative efficiency, $\eta_{\gamma} \lesssim$1\%, are consistent with a standard relativistic explosion which did not produce bright gamma-rays. 
Alternatively, a highly-relativistic ($\Gamma\gtrsim$100) structured jet remains consistent with our observations if seen substantially off-axis ($\theta_{\rm obs}>\theta_c$). In the latter case, the initial X-ray flare detected by EP would be caused by the lower-$\Gamma$ ejecta from the lateral wings intercepting our line of sight rather than by traditional prompt-emission mechanisms within the jet core. 

\end{abstract}


 \section{Introduction}
\label{sec:sec1}

Fast X-ray transients (FXRTs) are sudden and intense bursts of X-ray emission occurring on timescales ranging from a few seconds to several hours \citep{Arefiev2003,Bauer2017,QuirolaVasquez2022}. While FXRTs have been extensively investigated within the Galactic environment \citep{ConnorsAlanna1988,Smith2006,Sguera2016,Ferrigno2019}, their extragalactic counterparts remain poorly understood. 

Extragalactic FXRTs typically exhibit peak luminosities spanning $10^{42}$ to $10^{50}$~erg~s$^{-1}$, placing them among the most luminous transient phenomena in the X-ray regime. 
Space-based X-ray observatories such as \textit{Swift} \citep{Burrows2005}, \textit{Chandra} \citep{Weisskopf2002}, \textit{XMM-Newton} \citep{Jansen2001}, and \textit{eROSITA} \citep{Predehl2021} have played a critical role in their discovery and characterization (e.g., \citealp{Glennie2015,Xue2019Natur,Lin2020,Ai2021,Lin2022}). However, the long delay between the transient onset and its identification in archival searches often prevented placing meaningful constraints on their physical origin. A variety of progenitor scenarios, including compact object mergers, supernova shock break-outs, proto-magnetars, or tidal disruption events (TDEs) remain possible \citep{Xue2019Natur,Novara2020,Lin2022,Castaneda2024}. A notable case of a FXRT was reported by \citet{Soderberg2008Natur}, who discovered the X‐ray transient, XRF080109, in NGC~2770 at a distance of 27~Mpc. With a peak luminosity of $\approx\,6\times 10^{43}\,\mathrm{erg\,s^{-1}}$, XRF080109 featured a 600-s X-ray flash followed by the emergence of supernova SN2008D, providing the first robust link between FXRTs and the core-collapse of massive stars \citep{Soderberg2008Natur,Modjaz2008,Mazzali2008}. More recently, \citep{Bauer2017} reported the discovery of the FXRT CDF-S XT1 in the Chandra Deep Field-South, characterized by a rapid rise of $\sim\,100\,\mathrm{s}$, well described by a power-law decay with a slope of $\alpha\approx -1.53$, and a relatively hard spectrum with a photon index of 1.43. FXRT CDF-S XT1 was localized to a faint dwarf host galaxy at a redshift of $z=2.76$ \citep{QuirolaVasquez2024}, suggesting a much more energetic explosion than XRF080109.

The advent of the \textit{Einstein Probe} (EP; \citealt{Yuan2022,Yuan2025}) has significantly advanced our understanding of FXRTs. By enabling the real-time discovery and localization of these events, EP has facilitated the identification of dozens of new FXRTs. Some of them, such as EP240219A \citep{Yin2024}, EP240315A \citep{Liu2025NatAs, Ricci2025, Gillanders2024}, and EP240801a \citep{Jiang2025}, have been linked to classical gamma-ray bursts \citep[GRBs;][]{Kouveliotou1993,WoosleyBloom2006}.
However, most FXRTs do not feature a gamma-ray counterpart, i.e., gamma-ray-dark FXRTs \citep[e.g., EP240414a;][]{Sun2024}. 
A subset of these gamma-ray dark FXRTs may originate from extremely distant GRBs, whose spectral peaks are redshifted into the X-ray band \citep{Heise2003,Alessio2006}; in other cases, such as EP240414A \citep[$z=0.4$;][]{Sun2024,Srivastav2025,vanDalen2024}, the events occur at relatively low redshifts yet still lack detectable gamma-ray counterparts, further challenging our understanding of their physical origins. A leading class of models attributes such gamma-ray dark FXRTs to relativistic outflows powered by a central engine. For example, off-axis jets, dirty fireballs, and choked jets associated to either GRBs or TDEs can reproduce the observed timescales and luminosities of FXRTs 
\citep{2003ApJ...591.1097R,Paczynski1998, Dermer1999,MeszarosWaxman2001}. Alternatively, models involving supernova shock breakouts, classical TDEs, or fast blue optical transients (FBOTs) may produce X-ray transients without invoking relativistic jets \citep{Soderberg2008Natur,Holoien2016,2018ApJ...865L...3P}. 

In this context, radio observations serve as a powerful diagnostic for distinguishing among these scenarios. 
Unlike optical emission, the radio band probes fast-moving ejecta from early to very late ($\gtrsim1$~month) timescales, tracing long-lived synchrotron emission from relativistic electrons in the blast wave. These measurements provide valuable constraints on the energy budget and expansion dynamics of the outflow over extended periods \citep[see e.g.,][]{Ricci2025,Bright2025}. 
 As radio waves propagate through the interstellar medium (ISM), they are subject to various propagation effects, including interstellar scintillation (ISS; see e.g., \citealt{LeeLC1975,Armstrong1984}). 

This phenomenon has been extensively studied in compact radio sources such as pulsars and active galactic nuclei \citep[AGN;][]{Rickett1990, Goodman1997}, and has also been observed in several GRB afterglows, \citep{Frail1997Natur, Frail2000,Chandra2008,Granot2008}. 
ISS arises from small-scale electron density fluctuations in the ISM, which may induce rapid variations in signal intensity if the emitting source is sufficiently compact. Therefore, ISS provide important constraints on the angular size of the emitting region and, consequently, on the Lorentz factor of the outflow \citep{Heeschen1987, DennettThorpe2002Natur, Lovell2008, Rickett1995}, helping to determine whether the emission arises from a relativistic outflow. 

\begin{figure}
    \centering
    \includegraphics[width=0.5\linewidth]{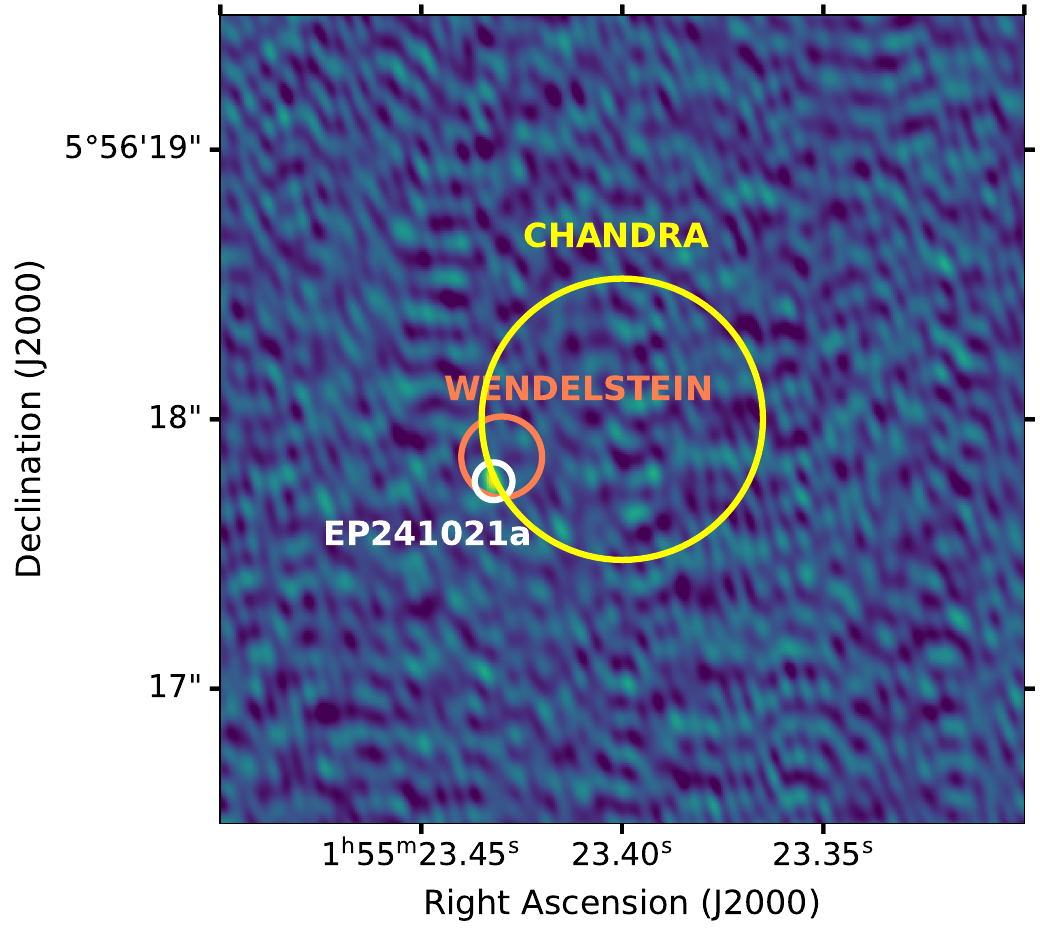}
    \caption{\textbf{e-MERLIN image of EP241021a.} The transient is co-located with the 90\% confidence-level from the Fraunhofer Telescope Wendelstein (orange circle; \citealt{Busmann2025}) at coordinates RA (J2000): 1$^{\mathrm{h}}$55$^{\mathrm{m}}$23.430$^{\mathrm{s}}$, Dec (J2000): +5$^\circ$56$^\prime$17.86$^{\prime\prime}$ with an uncertainty of 0.15$^{\prime\prime}$. It is also consistent with the 90\% C.L. Chandra X-ray localization (yellow circle; see Sec.~\ref{sec:chandra}).
    The e-MERLIN radio data (see Sec.~\ref{sec:e_MERLIN}) provide an even tighter positional constraint at 1$\sigma$ confidence, with milliarcsecond-level precision.}
    \label{fig:EP241021a}
\end{figure}

This paper presents a study of EP241021a, a FXRT discovered by the Wide-field X-ray Telescope (WXT) onboard the EP on October 10, 2024 at $T_0=\text{05:07:56 UT}$, with a duration of approximately $100\,\mathrm{s}$ \citep{Hu2024GCN}. Notably, no gamma-ray counterpart was detected \citep{Svinkin2024GCN} and the event was localized at a relatively low redshift of $z = 0.75$ \citep[see Fig.~\ref{fig:EP241021a};][]{Pugliese2024GCN,Zheng2024GCN,PerezFournon2024GCN}. We carried out extensive radio follow-up of the afterglow, which remained detectable for over 100 days post-burst in the observer's frame, placing it among the longest-lived radio afterglows to date \citep[$<$ 4\% of the events;][]{Chandra2008}.

 We identify evidence of ISS, offering a propagation-based explanation for the variability seen in the early-time radio light curve, which allows us to probe the physical properties of its relativistic outflow. Combined with constraints from afterglow modeling, these findings strongly support the presence of an outflow that is at least mildly relativistic with Lorentz factor of $\approx$4. 
 We first examine in Section~\ref{sec:sec2} the prompt emission properties. In Section~\ref{sec:sec3}, we provide a comprehensive overview of our radio monitoring campaign using ATCA and e-MERLIN. In Section~\ref{sec:sec4} we use the observed scintillation to constrain the source's angular size and the Lorentz factor. Section~\ref{sec:sec5} presents the results of our afterglow modeling and Section~\ref{sec:sec6} discusses their implications for the physical parameters of the event. Finally, in Section~\ref{sec:sec7}, we summarize our results.
Throughout this work, we adopt a flat $\Lambda$CDM cosmology model with the matter density parameter $\Omega_m = 0.315$, and the Hubble-Lemaître constant $H_0 = 67.4~\text{km\,s}^{-1}~\text{Mpc}^{-1}$ \citep{Planck2020}. 

\section{Prompt Emission}
\label{sec:sec2}

Simultaneous X-ray and gamma-ray observations of the prompt emission are crucial to assess whether an event resembles a typical GRB spectrum or is intrinsically softer, suggesting an X-ray flash (XRF) or another non-standard origin (e.g., a supernova shock breakout). This diagnostic is especially important in our analysis of EP241021a and similar gamma-ray dark FXRTs, as it helps clarify whether the absence of gamma-ray emission arises from instrumental sensitivity limits or from genuinely distinct spectral characteristics.

Our dataset comprises 84 EP events observed up to March 21, 2025. Of these, 55 events have complete prompt emission coverage from the \textit{Fermi} Gamma-ray Burst Monitor \citep[GBM;][]{Meegan2009} or were detections from Konus-Wind \citep{Aptekar1995}.

The remaining 29 events were either occulted by Earth or occurred while the detector was passing through the South Atlantic Anomaly (SAA), rendering prompt gamma-ray observations unavailable. In select cases, we extended the temporal search window by several minutes in an effort to recover partially observed events, thereby maximizing the number of FXRTs with usable prompt emission constraints.

\begin{figure}
    \centering
    \includegraphics[width=0.6\linewidth]{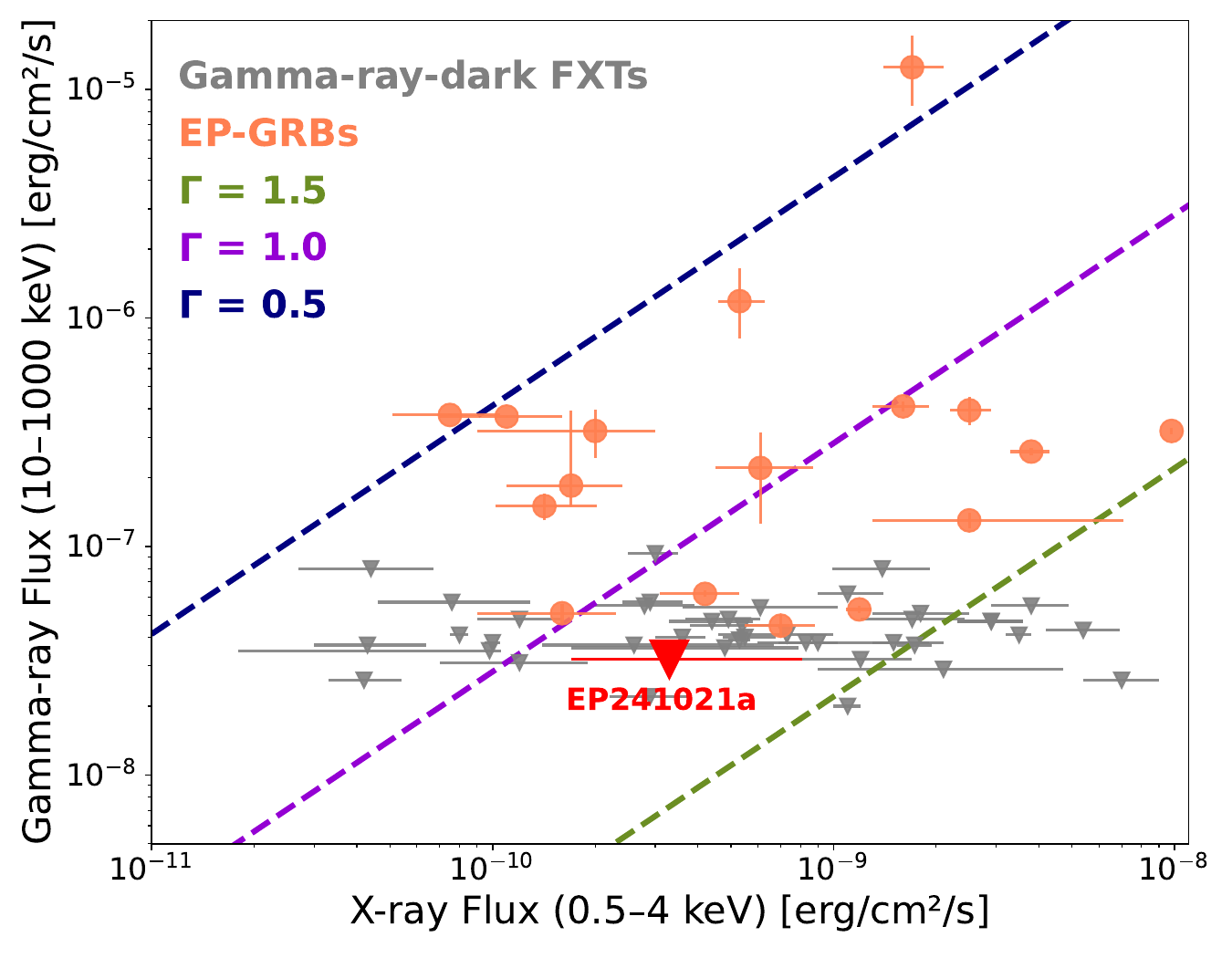}
    \caption{\textbf{Observed gamma-ray flux (10--1000~keV) versus X-ray flux (0.5--4~keV) for EP-discovered FXRTs.} EP sources with joint gamma-ray detections (EP-GRBs) are shown by orange circles, and gamma-ray dark FXRTs by upward gray triangles (which mark $3\sigma$ upper limits to the gamma-ray emission, based on 8.192~s time-averaged fluxes from \textit{Fermi}-GBM; (see Section~\ref{sec:sec2}). The X-ray fluxes represent the time-averaged, unabsorbed 0.5--4~keV fluxes derived from WXT observations; uncertainties are shown at the 90\% confidence level. EP241021a is indicated by the downward red triangle. Dashed lines represent power-law spectra with photon indices $\Gamma = 0.5$, $1.0$, and $1.5$.}
    \label{fig:prompt_flux}
\end{figure}

Fig.~\ref{fig:prompt_flux} compares the (observer frame) gamma-ray (10--1000~keV) and X-ray (0.5--4~keV) fluxes for a sample of FXRTs, including EP241021a. For this study, $3\sigma$ upper limits on gamma-ray fluxes were derived using the 
Gamma-ray Targeted Search (\textit{GTS})\footnote{\url{https://github.com/USRA-STI/gamma-ray-targeted-search}}, built around the Gamma-ray Data Tools \citep{GDT-Core}.
We adopt a Band function\footnote{The Band function is a phenomenological model of GRB spectra \citep{Band1993}. See Equation~\ref{eq:band}. defined as:
\begin{equation}
\label{eq:band}
N(E) = 
\begin{cases}
A \left( \dfrac{E}{100\,\mathrm{keV}} \right)^{\alpha} \exp\left( -\dfrac{E}{E_0} \right), & E < (\alpha - \beta)E_0 \\
A \left( \dfrac{(\alpha - \beta)E_0}{100\,\mathrm{keV}} \right)^{\alpha - \beta} \exp(\beta - \alpha) \left( \dfrac{E}{100\,\mathrm{keV}} \right)^{\beta}, & E \geq (\alpha - \beta)E_0
\end{cases}
\end{equation}
}
with the peak energy given by $E_{\rm peak} = (2 + \alpha) E_0$, and we adopt typical GRB parameters $\alpha = -1.0$, $\beta = -2.3$, and $E_{\rm peak} = 230~\mathrm{keV}$ (\textit{Normal} template) to calculate the upper limits of energy fluxes in the case of non-detections. For events with detections, the time-averaged gamma-ray fluxes were taken directly from the values reported through GCN circulars (see Appendix~\ref{sec:appA}). The dashed lines in Fig.~\ref{fig:prompt_flux} indicate constant photon indices ($\Gamma = 0.5$, $1.0$, and $1.5$), assuming a simple power-law spectrum. We observe that most gamma-ray bright FXRTs lie at $\Gamma \sim 1.0$, consistent with the region typically occupied by standard GRBs \citep{Poolakkil2021}.

For EP241021a, we used the X-ray flux (0.5--4~keV) estimated by \citet{Hu2024GCN} of $(3.3^{+1.5}_{-1.7}) \times 10^{-10}~\mathrm{erg~cm^{-2}~s^{-1}}$. Using the targeted search, we place a $3\sigma$ upper limit on the gamma-ray flux of EP241021a as \( <3.2 \times 10^{-8}~\mathrm{erg~cm^{-2}~s^{-1}} \) for a timescale of 8.192 s. At a redshift of $z=0.75$, this corresponds to a limit on the isotropic-equivalent gamma-ray energy of $E_{\gamma, \rm iso} \lesssim 7.2\times10^{50}$ erg, consistent with the value inferred by \citet{Busmann2025} and at the lower end of the distribution for cosmological GRBs, although still consistent with it.

A small subset of EP events fall below the $\Gamma = 1.5$ line, which lies at the higher end of the X-ray flux distribution. These events are particularly intriguing, as they are bright in X-rays yet lack detectable gamma-ray emission.
This suggests that their gamma-ray faintness is not due to instrumental biases and may instead reflect unusually softer spectra. These are the best candidates to search for very high-redshift bursts, 
off-axis and choked jets, as well as exotic types of explosions. 

The two classes of gamma-ray bright and gamma-ray dark transients mostly overlap in the region of $\Gamma = 1.0-1.5$, where EP241021a also lies. Therefore, based solely on its high-energy properties, it is not possible to robustly identify EP241021a as either a GRB-like explosion or a peculiar X-ray rich transient. 

In this context, the X-ray to gamma-ray flux ratio of EP241021a remains consistent with the broader EP-GRB population (Fig.~\ref{fig:prompt_flux}). This indicates that prompt emission properties alone may often be insufficient to unambiguously determine the nature of FXRTs. 

\section{Afterglow Observations}
\label{sec:sec3}

\subsection{Australia Telescope Compact Array}
\textit{Australia Telescope Compact Array} (ATCA) observations were conducted over 11 epochs between October 28, 2024, and January 25, 2025, covering frequencies from 2.1~GHz to 16.7~GHz. The bandwidth in all the observations was 2~GHz in the Compact Array Broadband Backend 1~MHz per frequency channel mode. The primary and bandpass calibrator was 1934-638 for all observations at 2.1, 5.5 and 9~GHz. At 16.7~GHz, 1921–293 served as the bandpass calibrator observations, while 1934-638 remained the primary calibrator. The phase calibrator 0146+056 was used to correct complex gains. The 21.2 GHz data were taken at the same time of the 16.7 GHz ones but the 21.2 GHz flux densities are not presented in this work because high system temperature values and large flux density scale corrections from the primary calibrator resulted in unreliable flux density values for the phase calibrator and the target source. Data reduction was performed using the Miriad \citep{SaultTeuben1995} software package, which was used to flag Radio Frequency Interferences (RFIs), calibrate the complex gains, bootstrap the flux density scale, and image the fringe visibilities. A null robustness parameter value was used as a good trade-off between uniform and natural weighting in the imaging strategy. Final cleaned radio maps were inspected by Karma kvis \citep{KVIS} and ds9 \citep{SAO2000} visualisation softwares. When a detection was found the peak flux was extracted, otherwise, a 3-$\sigma$ upper limit was derived from the Root Mean Square (RMS) noise in the restored maps. The uncertainties on the flux densities were estimated by summing in quadrature the map RMS noise and 5\% of the peak flux density, accounting for residual complex gain calibration errors.


\begin{table}
    \centering
    \caption{\textbf{Summary of EP241021a: ATCA Flux Density Measurements at 2.1, 5.5, 9, and 17~GHz.} The columns are as follows: Epoch, Start Time (UT), $T - T_0$~(days) representing the time elapsed since the burst detection, and $S_{\nu}$ representing the peak flux density at each observing frequency.}
    \renewcommand{\arraystretch}{1.5}
    \setlength{\tabcolsep}{0.18cm}{\begin{tabular}{ccccc}
        \hline
        Epoch & Start Time (UT) & $T - T_0$ (d) & $\nu$ (GHz) & $S_{\nu}$ ($\mu$Jy) \\
        \hline
        1 & 2024-10-29T10:43:25 & 8.3  & 5.5 & 356 $\pm$ 18 \\
         &  &  & 9   & 413 $\pm$ 21 \\
        \hline
        2 & 2024-11-08T8:52:55 & 18.3 & 5.5 & 434 $\pm$ 22 \\
         &  &  & 9   & 543 $\pm$ 27 \\
         &  &  & 17  & 366 $\pm$ 35 \\
        \hline
        3 & 2024-11-11T8:34:05 & 21.3 & 2.1 & 339 $\pm$ 31 \\
        \hline
        4 & 2024-11-15T7:36:25 & 25.3 & 5.5 & 802 $\pm$ 42 \\
         &  &  & 9   & 596 $\pm$ 31 \\
         &  &  & 17  & 347 $\pm$ 41 \\
        \hline
        5 & 2024-11-21T7:30:05 & 32.3 & 2.1 & $< 117$ \\
         &  &  & 5.5 & 937 $\pm$ 52 \\
         &  &  & 9   & 987 $\pm$ 54 \\
        \hline
        6 & 2024-11-24T7:01:00 & 34.1 & 17 & 450 $\pm$ 46 \\
        \hline
        7 & 2024-12-07T8:29:45 & 47.3 & 5.5 & 620 $\pm$ 45 \\
         &  &  & 9   & 565 $\pm$ 43 \\
         &  &  & 17  & 361 $\pm$ 50 \\
        \hline
        8 & 2024-12-08T8:06:35 & 48.3 & 2.1 & 761 $\pm$ 228 \\
        \hline
        9 & 2024-12-23T7:01:25 & 63.2 & 2.1 & 403 $\pm$ 55 \\
         &  &  & 5.5 & 875 $\pm$ 49 \\
         &  &  & 9   & 870 $\pm$ 43 \\
        \hline
        10 & 2024-12-30T5:36:55 & 70.1 & 2.1 & 395 $\pm$ 118 \\
         &  &  & 5.5 & 492 $\pm$ 34 \\
         &  &  & 9   & 366 $\pm$ 25 \\
         &  &  & 17  & 180 $\pm$ 54 \\
        \hline
        11 & 2025-01-25T5:14:35 & 96.0 & 2.1   & 250 $\pm$ 60 \\
         &  &  & 5.5   & 428 $\pm$ 30 \\
          &  &  & 9 & 335 $\pm$ 24 \\
        \hline
    \end{tabular}}
    \label{tab:flux_density_ATCA}
\end{table}

\begin{table}
    \centering
    \caption{\textbf{Summary of EP241021a: e-MERLIN Flux Density Measurements at 5~GHz.} The columns are as follows: $T - T_0$ representing the time elapsed since the burst detection, and $S_{\nu}$ representing the peak flux density at the corresponding frequency.}
    \renewcommand{\arraystretch}{1.5}
    \setlength{\tabcolsep}{0.55cm}{\begin{tabular}{cccc}
        \hline
        Epoch & $T - T_0$ (d) & $\nu$ (GHz)  & $S_{\nu}$ ($\mu$Jy) \\
        \hline
        1 & 48.8 & 5 & 355 $\pm$ 34 \\
        2 & 74.6 & 5 & 454 $\pm$ 41 \\
        3 & 110.7 & 5 & 206 $\pm$ 35 \\
        4 & 136.5 & 5 & 118 $\pm$ 27 \\
        \hline
    \end{tabular}}
    \label{tab:flux_density_grouped}
\end{table}

\subsection{e-MERLIN}
\label{sec:e_MERLIN}

EP241021a was observed with \textit{enhanced Multi Element Remotely Linked Interferometer Network} (e-MERLIN) in four epochs at the centre frequency of 5~GHz with a total bandwidth of 512~MHz divided in 4 sub-bands of 128~MHz each. The data were correlated by the e-MERLIN software correlator. The resulting fringe visibilities were calibrated in the {\sc CASA} software v.5.8 using the e-MERLIN CASA pipeline \citep{emerlin_casa_pipeline,Moldon2021} and imaged in CASA v.5.5.0 using a Briggs weighting scheme and a robustness value of 0.5. The target was observed together with 1331+3030 (alias 3C286), 1407+2827, 0154+0823 as the flux calibrator , the pointing calibrator and the bandpass calibrator, respectively. The target was fit with a 2D Gaussian in {\sc CASA} and the peak flux was quoted in the flux density measurement, being it point-like in the $100 \times 40$ mas final cleaned maps. The e-MERLIN position obtained by 2D Gaussian fitting to the cleaned map of EP241021a is RA = 01$^{\mathrm{h}}$ 55$^{\mathrm{m}}$ 23.4325$^{\mathrm{s}}$, Dec = +05$^\circ$ 56$^\prime$ 17.774$^{\prime\prime}$ (J2000) with a positional $1\sigma$ uncertainity of (3, 9) mas in RA and DEC, respectively. Flux uncertainties were estimated as detailed in Table~\ref{tab:flux_density_ATCA}. Detections and the earlier flux upper limit obtained by this work are shown in Table~\ref{tab:flux_density_grouped}.

\subsection{Very Large Array}
\label{sec:vla}
Very Large Array (VLA) observations were performed on 2025 June 6th (232 days after the EP trigger) at 6 GHz central frequency and a 4-GHz bandwidth with a duration of 12 min on-source under the project SN078192 (PI: Troja) in C array configuration. The primary calibrator was 3C48 and the phase calibrator was J0149+0555. The raw data was downloaded from the NRAO archive and calibrated using the VLA CASA pipeline built in CASA v.6.6.1 using standard settings. The calibrated dataset was imaged in CASA v.5.5.0 using the task {\it tclean} with the Briggs parameter set to 0.5 and 2000 clean iterations. The cleaned image of the target was inspected using the CASA viewer. A 2D Gaussian fit was performed with the built-in CASA viewer fitter and the rms noise in the cleaned map was extracted using the task {\it imstat} in a region of the map away from any bright source. As the source was point-like we used the peak flux to estimate the source flux density. A flux of 107 $\pm$ 11 $\mu$Jy  was measured, where the flux uncertainty was the result of the sum in quadrature of the statistical noise and a multiplicative term (5\% of the source flux density) associated with residual gain calibration error.     

\subsection{X-ray Observations}
\label{sec:chandra}
Observations of EP241021a were conducted with the Follow-up X-ray Telescope (FXT) onboard \textit{Einstein Probe}, starting $\approx$36.5 hours after the EP-WXT trigger \citep{Wang2024GCN}. Follow-up observations from the \textit{Swift} X-ray Telescope (XRT) were reported by \citet{Busmann2025}, revealing a long-lasting X-ray counterpart. 
To determine whether the long-lived nature of the X-ray emission was possibly due to contamination from a nearby source, we analyzed public data from the \textit{Chandra X-ray Observatory} contained in the Chandra Data Collection~\dataset[DOI: 10.25574/cdc.415]{https://doi.org/10.25574/cdc.415}, which observed EP241021a approximately 14 days after trigger (ObsID: 30598; PI: Jonker) for a total exposure of 9.9 ks.
A single X-ray source is detected within the XRT localization at coordinates RA = 01$^{\mathrm{h}}$ 55$^{\mathrm{m}}$ 23.3984$^{\mathrm{s}}$, Dec = +05$^\circ$ 56$^\prime$ 17.498$^{\prime\prime}$ (RA = 28.847$^\circ$, Dec = +5.938$^\circ$) with $90\%$ positional uncertainty of $0.52^{\prime\prime}$.
Source counts were extracted using the \textit{CIAO} software package (v4.16)\footnote{\url{https://cxc.harvard.edu/ciao/}}, employing a circular aperture with a radius of 1.1~arcseconds, corresponding to approximately 90\% encircled energy of the Chandra point spread function (PSF). Aperture photometry was performed in the 0.5--8.0 keV band. The resulting spectrum was modeled using an absorbed power-law model (\texttt{tbabs*powerlaw}) in \textsc{XSPEC}, with the hydrogen column density ($N_\mathrm{H}$) fixed to the Galactic value reported in \citet{Hu2024GCN}. We find a best-fit photon index of $\Gamma = 1.3 \pm 0.7$, and an unabsorbed model flux of $(6.1^{+1.4}_{-1.2}) \times 10^{-14}$~erg~cm$^{-2}$~s$^{-1}$ in the 0.3--10 keV band. 
This value is consistent with the nearly simultaneous XRT measurements \citep{Busmann2025} and confirms that the late-time flattening is an intrinsic property of the source.


\begin{figure*}
    \centering
    \includegraphics[width=\textwidth]{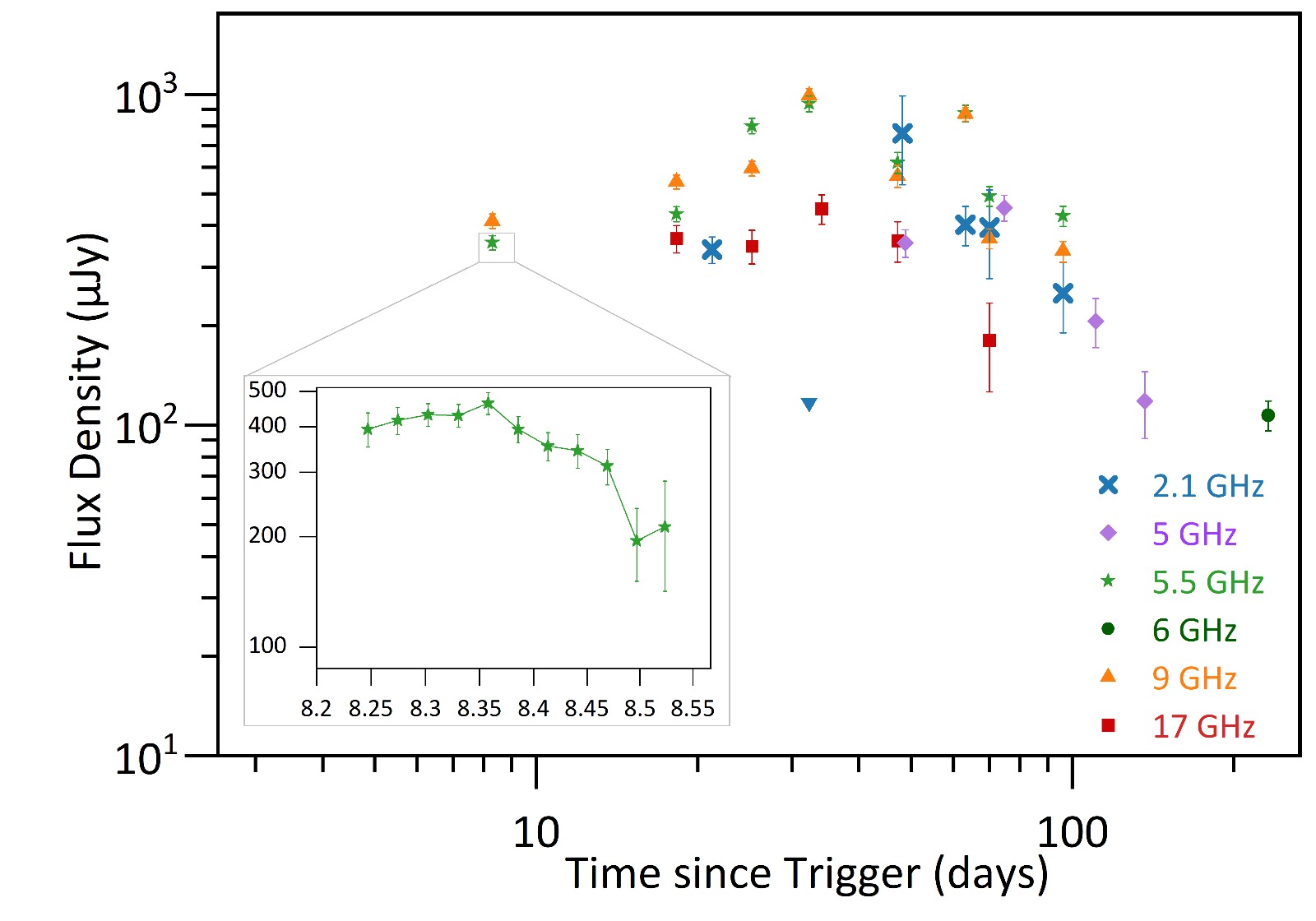} 
    \caption{\textbf{ATCA and e-MERLIN radio observations of EP241021a.} Flux densities are listed in Table~\ref{tab:flux_density_ATCA} and~\ref{tab:flux_density_grouped}. The inset panel displays the intra-observation light curve from the first epoch at 5.5~GHz, segmented into 40-min intervals to highlight short-term variability.} 
    \label{fig:variability}
\end{figure*}

\section{Constraints from radio observations}
\label{sec:sec4}

As mentioned in the Section~\ref{sec:sec1}, afterglows from compact radio sources are subject to ISS, which can produce significant flux modulations due to distorted wavefronts \citep{Walker1998}. Specifically, such flux variations can arise from diffractive and refractive scintillation, each sensitive to different angular scales and frequency bandwidths \citep{Narayan1992,Goodman1997}. By identifying the epochs when diffractive scintillation ceases and refractive scintillation becomes suppressed, we can place lower and upper bounds, respectively, on the angular size of the source. These size constraints, when combined with the angular diameter distance to the source, allow us to infer limits on the Lorentz factor of the outflow, as demonstrated in previous studies of GRBs and other relativistic transients \citep[e.g.,][]{Frail2000,Cenko2008,Zhang2023}. Motivated by this, in Section~\ref{sec:sec4.1}, we present evidence for ISS during the radio monitoring of EP241021a. In Section~\ref{sec:sec4.2}, we use the ISS behavior to constrain the source's angular size and Lorentz factor. We calculate the brightness temperature in Section~\ref{sec:sec4.3} using early-time radio data.

\subsection{Evidence for ISS}
\label{sec:sec4.1}

The strength and timescale of ISS effects depend on both the observing frequency and the angular size of the source \citep[e.g.,][]{LiuYulan2022}. ISS manifests on two distinct timescales: diffractive ISS (DISS), which induces large-amplitude and rapid fluctuations over minutes due to small-scale interference patterns; and refractive ISS (RISS), which causes slower variability over weeks to months as the signal traverses larger-scale density structures in the scattering screen \citep{Walker1998, Lewandowski2011, Granot2014, Liu2022}. These scintillation effects can produce substantial variations in both flux and spectral shape, especially during the early stages of an afterglow. Initially, when the GRB jet is highly-collimated, DISS dominates the variability; as the emitting region expands, DISS becomes suppressed, and RISS emerges as the dominant contributor \citep{Frail1997Natur}. Consequently, different radio facilities, operating at various frequencies and timescales, might detect variability on distinct timescales. 

To search for ISS signatures, we analyzed intra-epoch variability within individual ATCA observations by dividing them into 40~minutes segments. In the first epoch, we detect significant short-timescale variability ($\sim$5$\sigma$) down to 5.5~GHz, indicative of DISS. No such rapid variability is observed in subsequent epochs, which is consistent with the source expanding beyond the angular scale necessary for DISS to occur, roughly 18 days after the burst. Moreover, the absence of significant variability beyond epoch 7 suggests that RISS has also been suppressed as the emitting region continues to expand (See Fig.\ref{fig:variability}). Since both DISS and RISS depend sensitively on the angular size of the source, the temporal evolution of these scintillation effects constrain the outflow velocity \citep{Frail1997Natur, Frail2000, Chandra2008, Granot2014}.

Assuming that scattering occurs in a thin screen along the line of sight, the screen is located at a distance $ D_{\text{scr}} \sim (h_z/2) \sin^{-1} |b| \approx 0.6$~kpc from the observer \citep{Reynolds1989ApJ, Galama2003}, where $h_z$ is the scale height of the ionised gas layer ($\sim 1$~kpc) and $b=-53.47$ is the Galactic latitude in degrees.
We use the NE2001 model of the Galactic distribution of free electrons \citep{Cordes2002, Walker1998} to derive the  scattering measure (SM) of $1.3 \times 10^{-4} \, \text{kpc} \, \text{m}^{-20/3}$ and a transition frequency of $\nu_0 \sim 6$~GHz. 
Below this frequency, density irregularities in the ISM cause strong phase perturbations leading to significant scattering, whereas above this frequency, scattering effects become weaker.

The decorrelation bandwidth (\(\Delta\nu_{\rm dc}\)) for diffractive scintillation represents the frequency range over which the intensity variations caused by diffraction remain correlated \citep{Goodman1997}:
\begin{equation}
\Delta\nu_{\text{dc}} = 2.2 \left( \frac{\nu}{5.5 \text{ GHz}} \right)^{22/5} \left( \frac{D_{\text{scr}}}{\text{kpc}} \right)^{-1} \left( \frac{\rm SM}{10^{-4} \text{ m}^{-20/3} \text{ kpc}} \right)^{-6/5} \text{ GHz}\,.
\label{eq:deltanu_dc}
\end{equation}

Since \( \Delta\nu_{\text{obs}} \sim \Delta\nu_{\text{dc}} \) = 2.5~GHz, diffractive scintillation is expected to have a significant influence at $\approx$5.5 GHz and be progressively suppressed at lower frequencies. Moreover, 
the 2.1~GHz frequency data have a lower signal-to-noise ratio which precludes a time-variability analysis. 


\subsection{Angular size measurements and Lorentz factor}
\label{sec:sec4.2}
The angular scales can be expressed in terms of the angular size of the first Fresnel zone evaluated at the transition frequency $\nu_0$. The characteristic angular size is given by:
\begin{equation}
\theta_{F_0} = 2.1 \times 10^4 \, {\rm SM}^{0.6} \, \nu_0^{-2.2} \, \mu\text{as}\,,
\end{equation}
where $\theta_{F_0}$ denotes the angular size of the first Fresnel zone at the scattering screen, yielding a value of $\theta_{F_0} = 2.0$~$\mu$as \citep{Granot2014} for EP241021a.
After the second epoch, we observe no significant short-term variation in the measured flux density, indicating that DISS has ceased by $\approx$18 days post-burst. 
Thus, the source angular size $\theta_s$ must be larger than a critical threshold, $\theta_d = \theta_{F_0} \times (\nu / \nu_0)^{6/5}$~$\mu$as = 1.8~$\mu$as  at 5.5 GHz \citep{Granot2014}.
This lower limit on the angular size can then be converted into constraints on the Lorentz factor  as $\Gamma \gtrsim \theta_d D_A/c \Delta t_{rf}\approx 1.5$ at 10 d (rest-frame), 
where the angular diameter distance $D_A = D_L / (1+z)^2 = 4.8 \times 10^{27}$~cm, $D_L$ is the luminosity distance, 
\(c\) the speed of light, and $\Delta t_{rf}$ is the time since the burst in the rest-frame.
Considering that the Lorentz factor of an expanding fireball
decreases in time as $\Gamma \propto t^{-3/8}$ in an uniform density medium \citep{Piran1999}, we derive a lower limit on the initial Lorentz factor of $\Gamma \gtrsim 3.8$ at 1.5 d (observer's frame), since at this time the optical and X-ray afterglow are seen to fade in time \citep{Busmann2025}, thus indicating that the blastwave is already in its deceleration phase. 
Our limit is rather conservative and the resulting conclusion of a mildly relativistic outflow does not strongly depend on the specific choice of input parameters. For instance, we verified that using the default settings of the NE2001 model, our limit would increase by a factor of $\approx$2. 

In the case of refractive scintillation, the scattering disk is much larger than the first Fresnel zone, and this results in ${\theta_r = \theta_{\mathrm{F0}} (\nu / \nu_0)^{-11/5} = 2.3 \,\mu\mathrm{as}}$ at 5.5 GHz \citep{Granot2014}. 
Large amplitude fluctuations are expected at 5.5 GHz, 
with modulation index $m_r = (\nu / \nu_0)^{17/30} \sim 0.95$ \citep{Granot2014}, and at 2.1 GHz with $m_r \sim 0.6$. This effect is observed up to  $\approx25$ d
at 5.5 GHz and up to $\approx32$ d at 2.1 GHz. It explains the large discrepancy between the measured fluxes at different frequencies, but is likely quenched afterwards.

By imposing $\theta_s \lesssim \theta_r$ at 25 d (5.5 GHz), we derive a rather tight upper limit on the physical size of the source $D_s \lesssim \theta_r D_A \approx 6 \times 10^{16}$ cm. Following the steps outlined above,  we derive an upper limit on the Lorentz factor of $\Gamma\lesssim$4.3 at 1.5 d (observer's frame). 

Therefore, the observed variability of the radio counterpart consistently points to an outflow that is at least mildly relativistic, with Lorentz factor $\Gamma \approx$4. 
Unfortunately, without a clear identification of the deceleration time \citep[e.g.,][]{Molinari2007}, we cannot exclude the presence of an ultrarelativistic outflow at earlier times. 

\begin{figure*}
    \centering
    \includegraphics[width=0.6\textwidth]{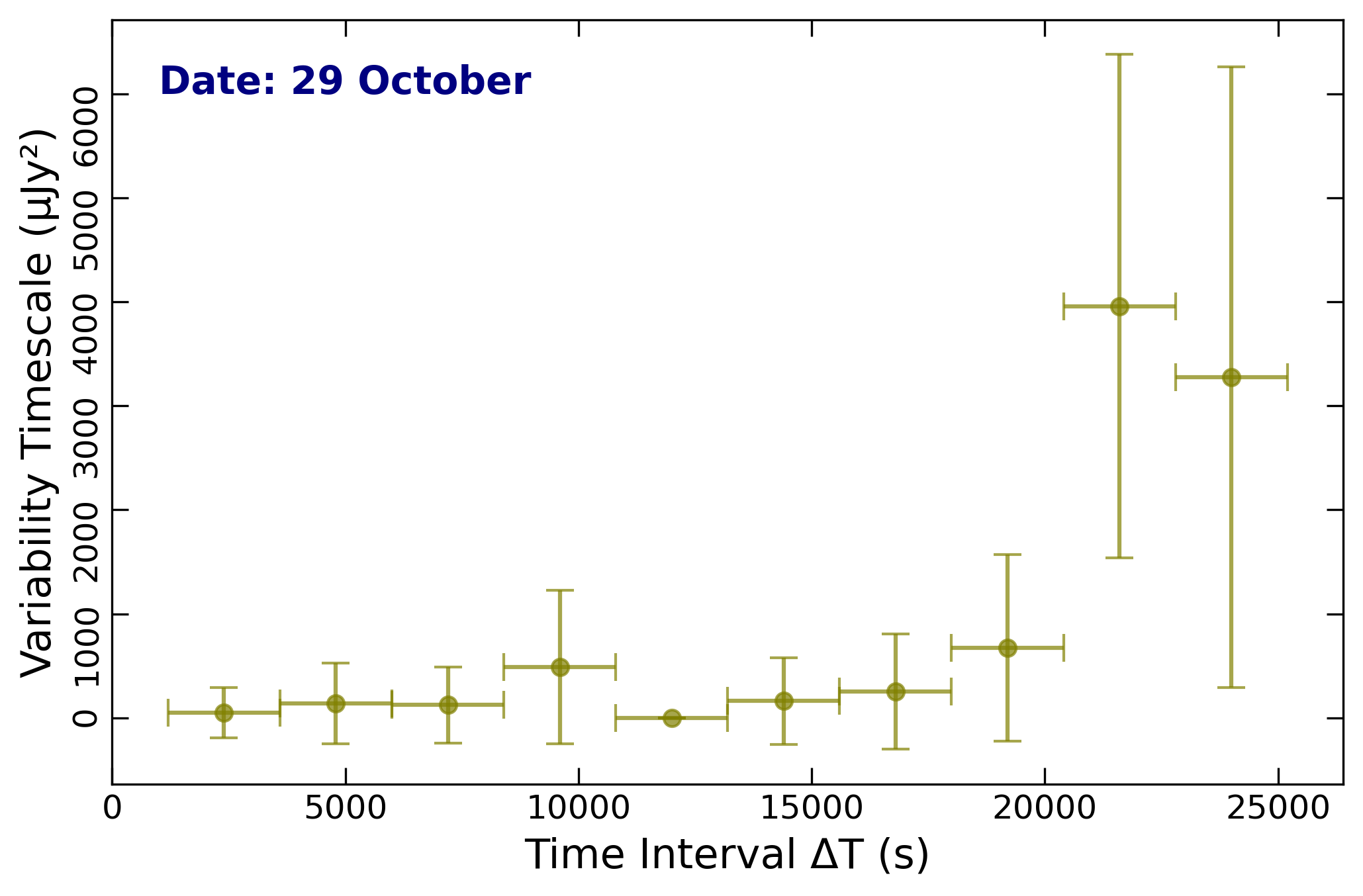}
    \caption{\textbf{Variability timescale} on October 29, 2024 with data sampled at 40-minute intervals.}
    \label{fig:variabilityfunction}
\end{figure*}

\subsubsection{Variability function}

The above estimates depend on the adopted ISM model, in particular the scattering measure. 
An alternative way to estimate this quantity directly from the data is by examining the variability timescale using intensity structure functions \citep{Simonetti1985,Hughes1992,Ciaramella2004,Chandra2008},
\begin{equation}
V = \frac{1}{N-1} \sum_{i=1}^{N-1} \left[ F(t_i+\Delta t) - F(t_i) \right]^2\,,
\end{equation}
where \( F(t) \) is the flux density at time \( t \), \( \Delta t \) is the fixed time interval between consecutive measurements, and \( N \) is the total number of observations. This variability metric quantifies the mean squared deviation of the flux measurements from the reference value, serving as an indicator of the temporal variability induced by the scintillation. 
We tentatively identify a break with 20--30\% accuracy at $\Delta T \approx 2 \times 10^4$~s (see Fig.~\ref{fig:variabilityfunction}), which places a lower limit to the true timescale of diffractive scintillation $t_{\rm diff}$. 
For a relative transverse velocity \( v_{\text{rel}} \sim 30 \)\,km\,s$^{-1}$, we derive a scattering measure:

\begin{multline}
{\rm SM} \lesssim 0.4 \left( \frac{\nu}{5.5 \text{ GHz}} \right)^{2} \left( \frac{t_{\text{diff}}}{20,000 \text{ s}} \right)^{-5/3} \left( \frac{v _{\text{\rm rel}}}{30 \text{ km s}^{-1}} \right)^{-5/3}\\ \times10^{-4} \, \text{m}^{-20/3} \text{ kpc\,,}
\end{multline}
which is smaller than the NE2001 value.
As the critical threshold depends on the SM as $\theta_d \propto SM^{3/5} \approx$0.9 $\mu$as \citep{Cordes2002}, the resulting constraint on the Lorentz factor would only slightly weaken to $\Gamma\,\approx$2 at 1.5 d, which is still mildly relativistic.


\subsection{Brightness Temperature}
\label{sec:sec4.3}
The brightness temperature $T_b$ provides an estimate of the intensity of radiation emitted by a source, assuming it behaves like a blackbody. Its value can be derived from early-time radio observations to aid constraining the relativistic nature of the outflow \citep{Anderson2014}. 

For EP241021a, we use the first ATCA detection at $\sim$8 days for calculating $T_b$ as \citep{Longair2011}:
\begin{equation}
T_b = 1.153 \times 10^{-8} \, D_L^2 S_{\nu} \nu^{-2} t^{-2} (1+z)^{-1} \approx 3.3 \times 10^{13} K, 
\end{equation}
where 
\( S_{\nu} \) is the flux in Jy, \( \nu \) is the frequency in Hz, \( t \) is the time since the trigger in seconds, \( z \) is the redshift. Since the maximum possible emitting region size is constrained by the speed of light (\( ct \)), comparing the observed $T_b$ with the inverse-Compton limit (\( T_B \approx 10^{12} \)\,K) allows us to estimate the minimum bulk Lorentz factor (\( \Gamma \)) of the emitting material. This relationship is given by \citep{Galama1999Natur}:
\begin{equation}
\Gamma \gtrsim \left( \frac{T_b}{T_B} \right)^{1/3}\,\approx 3.
\end{equation}
Extrapolating this value back in time, using the relation $\Gamma \propto t^{-3/8}$ as discussed in the previous section, we derive a lower limit on the Lorentz factor $\Gamma \gtrsim$5.5 at 1.5 d. This is slightly larger than the constraints from ISS, however, considering the uncertainties in the models and underlying assumptions, both these results appear in agreement and consistently point to an outflow that is at least mildly relativistic. 

\section{Multi-wavelength Afterglow Modeling}
\label{sec:sec5}

In the previous sections, we demonstrated that radio observations point to an expanding outflow with at least mildly relativistic velocity. The interaction of this fireball with the surrounding medium is known to produce  broadband synchrotron radiation \citep{Sari1998,Wijers1999,Granot2002}, known as afterglow, whose brightness and temporal evolution depend, among other factors, on the outflow structure, its geometry, and the observer's viewing angle \citep{Rossi1993,Sari1999, Troja2019,Ryan2020,Ryan2024}. 
This afterglow emission is generally described with a power-law form: $F_\nu \propto \nu^{-\beta} t^{-\alpha}$, where $\alpha$ and $\beta$ obey to standard closure relations \citep{Sari1999,ZhangMeszarosREVIEW}.
We use this empirical model to derive basic physical constraints from the afterglow observations. 

The evolution of the radio spectral index, shown in Fig.~\ref{fig:spectral-index}, reveals the following:

\begin{figure}
    \centering
    \includegraphics[width=0.6\textwidth]{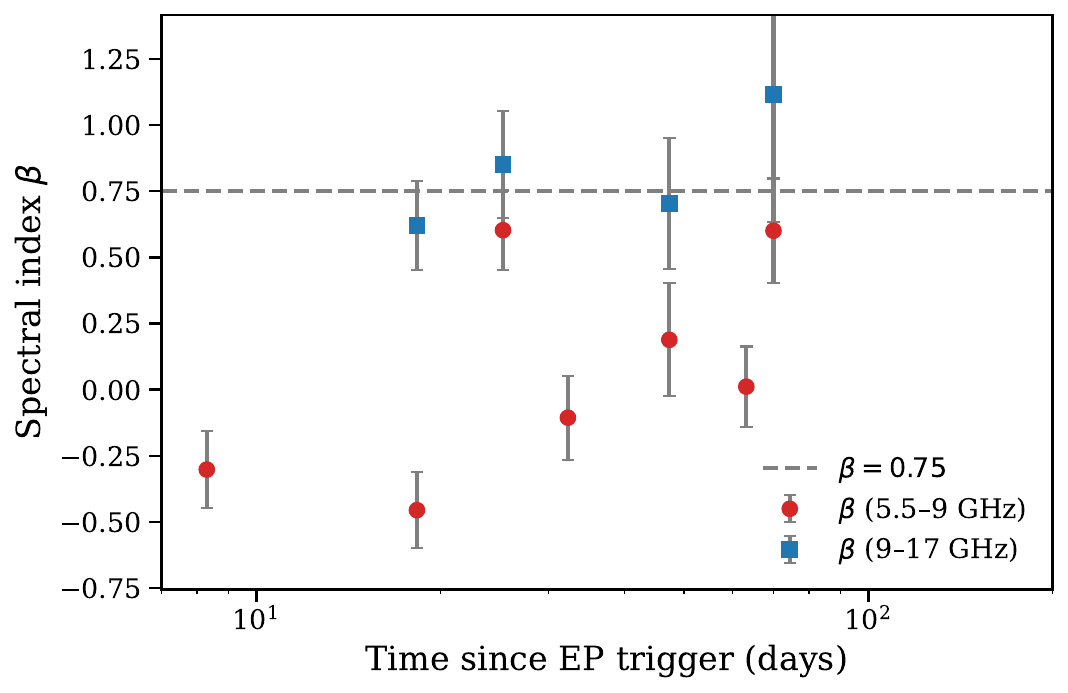}
    \caption{
\textbf{Spectral index evolution} between 5.5–9 GHz (red circles) and 9–17 GHz (blue squares) over time since the EP trigger. The dashed line indicates $\beta = 0.75$.}
    \label{fig:spectral-index}
\end{figure}

\begin{itemize}
\item the spectral index between 9 and 17 GHz remains constant around the value $\beta_2 \sim 0.75$. 
This is consistent with the synchrotron regime $\nu_m < \nu_r < \nu_c$ and implies a characteristic synchrotron frequency $\nu_m \lesssim 9$ GHz after $T_0 + 18$ days, and a spectral index $p = 2\beta_2 + 1 \approx 2.5$ of the shock-accelerated electrons.

\item the spectral index between 5.5 and 9 GHz is seen to vary. At early times, its value might be affected by ISS, which would explain its rapid variations.  However, it displays a consistent trend of increasing from   $\beta_1 \sim -1/3$ to  $\beta_2 \sim 0.75$. 
This is consistent with the characteristic synchrotron frequency $\nu_m $ crossing this frequency range and progressively moving below 5.5 GHz after $T_0 + 70$ days, as expected for an external forward shock expanding into a uniform medium. 
\end{itemize}

The observed behavior is broadly consistent with expectations of an external forward shock expanding into a uniform medium, for which 
$\nu_m \propto t^{-3/2}$ and $F_m \approx$ constant, and rules out any major contribution from reverse shock. 

We use the best-sampled radio light curve at 5.5 GHz, including our 6 GHz observations, to characterize the afterglow temporal evolution. A smoothly broken power-law fit \citep{Ricci2025} yields a rising slope of $\alpha_1 = 0.7 \pm 0.08$ (68\% confidence level), peaking at $\sim$40 d (rest frame), followed by a decay with slope $\alpha_2 = 1.9 \pm 0.1$ (68\% c.l.).

The early rise, although likely affected by ISS, is consistent with the expected behavior for $\nu_r \lesssim \nu_m$, 
as also suggested by the spectral index $\beta_1\approx$-1/3. 
The steep late-time decay helps further constraining models as it disfavors quasi-spherical outflows, such as mildly relativistic cocoon or choked jet models \citep[e.g.,][]{Troja2019}, and points to a collimated outflow.

Our empirical analysis shows that the radio emission in the range 2-17 GHz is broadly consistent with the evolution of optically thin synchrotron radiation from a population of shock-accelerated electrons. This allows us to place an upper limit on the self-absorption frequency $\nu_{sa} \lesssim$5 GHz, which in turn provides an independent constraint on the bulk Lorentz factor, $\Gamma \approx$4 \citep{BarniolDuran2013}.
In contrast, the optical and near-infrared counterpart of EP241021a display a much more complex behavior, with a prominent  bump at $\approx$7 d followed by a shallow long-lasting decay \citep{Busmann2025}. 
Non-standard evolution is also observed at X-ray energies
(Sec. \ref{sec:chandra}). Therefore, we exclude the late-time ($\gtrsim$7 d) optical and X-ray data from our modeling and test whether a basic afterglow model can consistently describe the radio dataset and the early optical and X-ray observations. 

\begin{figure*}[t]
    \centering
    \begin{minipage}[t]{0.45\textwidth}
        \centering
        \includegraphics[width=\textwidth]{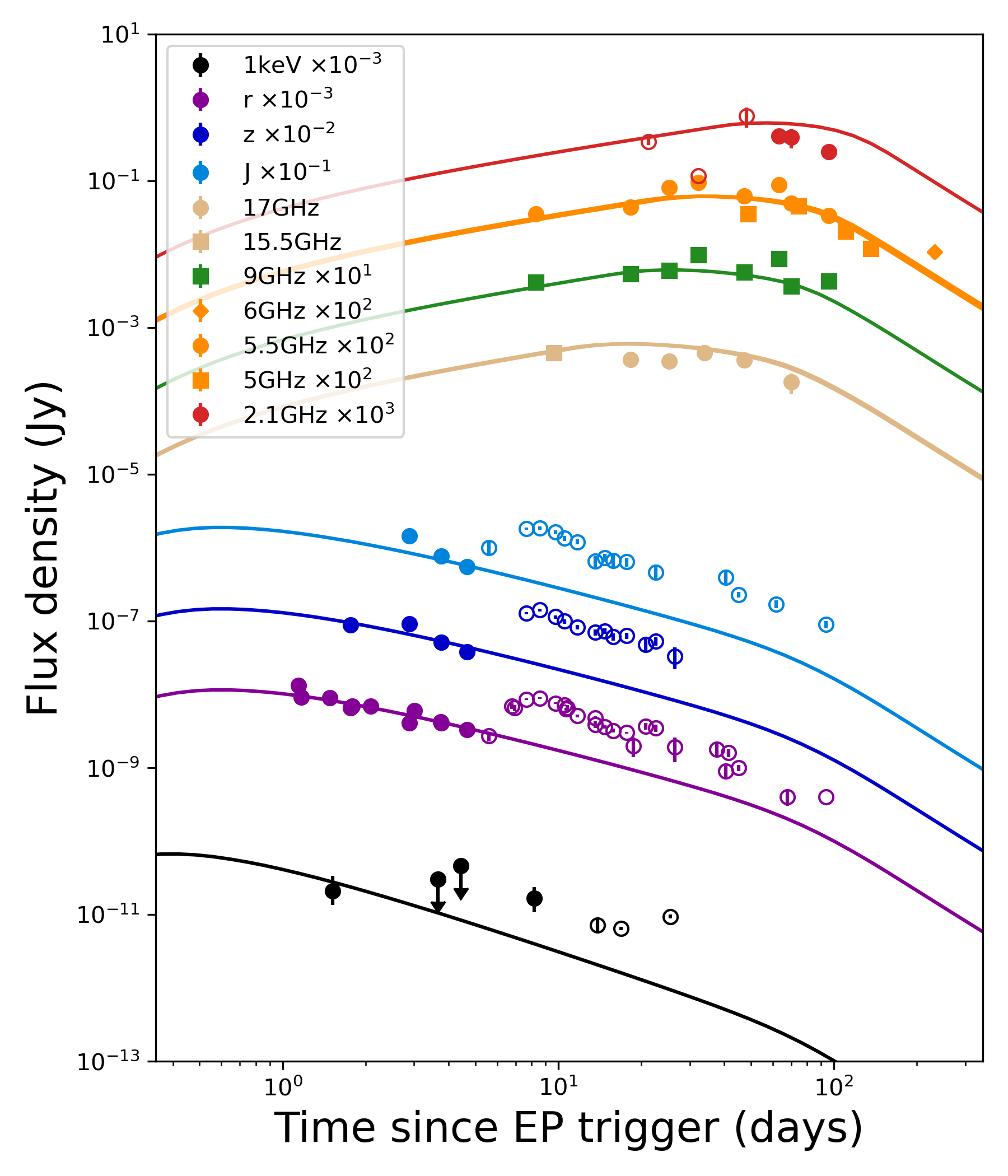}
        \caption{\textbf{Broadband modeling results.} Best-fit afterglow light curves for an on-axis jet (model Top-Hat) compared with observations at multiple frequencies.
    The X-ray (\textit{Swift}/XRT; 1 keV) and optical and near-infrared observations in the r, z and J filters are from \citet{Busmann2025}. Radio observations at 15.5~GHz from AMI-LA \citep{Carotenuto2024GCN}. Downward arrows indicate $3\sigma$ upper limits. Empty symbols show data not used for model fitting. A scaling factor has been applied to different frequencies for plotting purposes.}
        \label{fig:lc}
    \end{minipage}
    \hfill
    \begin{minipage}[t]{0.5\textwidth}
        \centering
        \includegraphics[width=\textwidth]{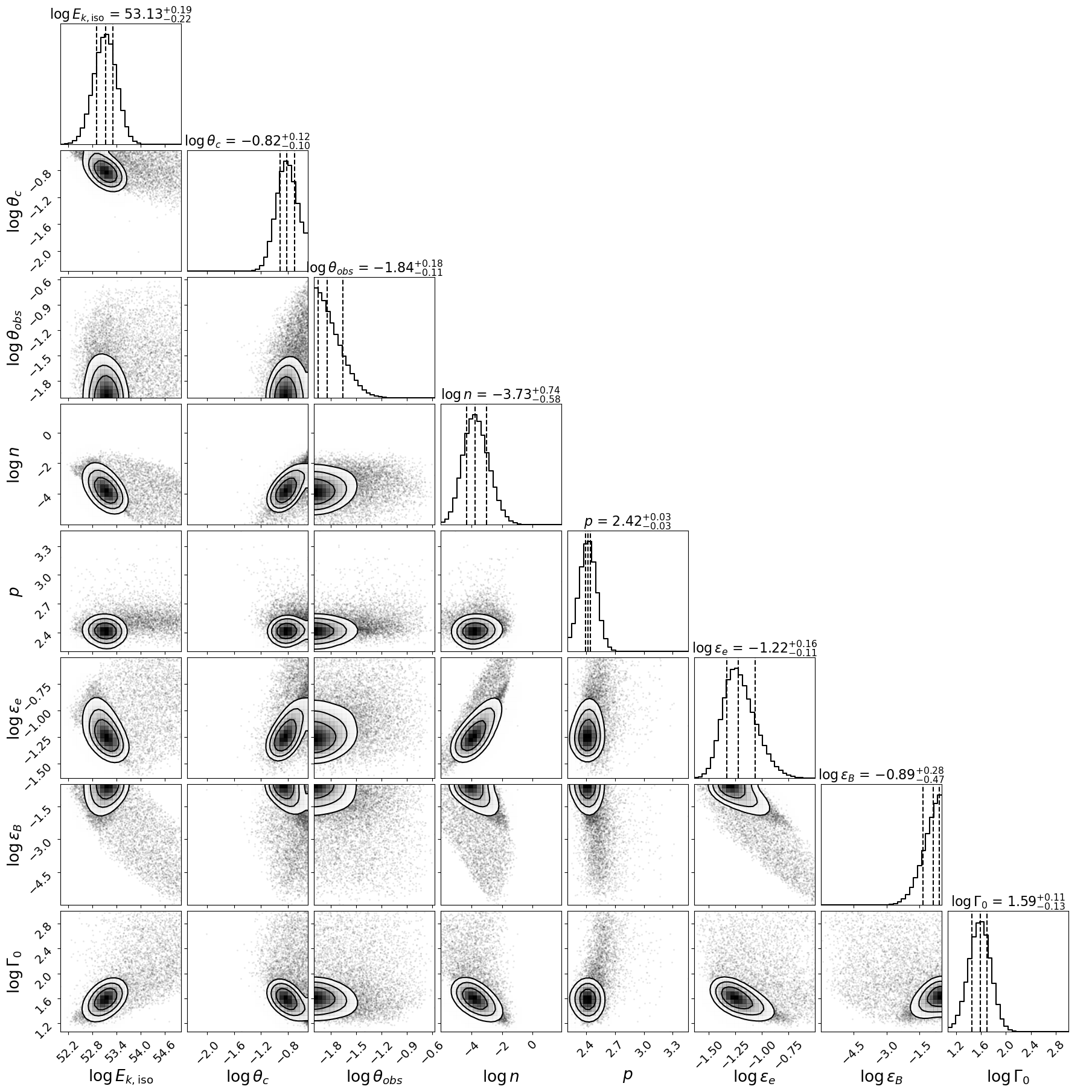}
        \caption{\textbf{Broadband modeling results.} Corner plot showing the posterior distributions of the fitted model parameters using a uniform (top-hat) jet model.}
        \label{fig:corner}
    \end{minipage}
\end{figure*}

To derive physical constraints on the explosion, we model the afterglow using the open-source Python package {\sc Afterglowpy v.~0.8.0} \citep{Ryan2020,Ryan2024}, which implements the single-shell approximation \citep{vanEerten2010,vanEerten2018}. In this framework, the ejecta and forward shock are treated as a single radially uniform fluid element. The model incorporates prescriptions for jet spreading and uses an equation of state that smoothly transitions between the ultra-relativistic and non-relativistic regimes \citep{vanEerten2013,Nava2013}. \mbox{\sc Afterglowpy} supports various jet structures, including Gaussian and power-law jets, by decomposing the jet into multiple top-hat segments. The dynamical evolution of each segment is governed by coupled ordinary differential equations that describe the shock radius, dimensionless four-velocity, and evolving jet opening angle. These equations are integrated over time to generate synthetic afterglow light curves.

\begin{table*}
    \centering
    \caption{\textbf{Table of afterglow parameters derived from afterglow model fitting.} Posterior uncertainties represent 1$\sigma$ from the marginal distributions and limits are at $2\sigma$ confidence level.}
    \renewcommand{\arraystretch}{2}
    \setlength{\tabcolsep}{0.5cm}
    \begin{tabular}{l c c c c}
        \toprule
        Parameter & Prior & \multicolumn{3}{c}{Posterior} \\
        \cmidrule(lr){3-5}
         & & Top-Hat & Gaussian & Gaussian + WXT \\
        \midrule
        $\log E_{K,\mathrm{iso}}$ (erg) & (49, 55) & $53.1^{+0.2}_{-0.2}$ & $53.7^{+0.2}_{-0.2}$ & $53.1^{+0.2}_{-0.1}$ \\
        $\log \theta_c$ (rad) & (-3, -0.5) & $-0.8^{+0.1}_{-0.1}$ & $-1.2^{+0.2}_{-0.1}$ & $-0.9^{+0.02}_{-0.02}$ \\
        $\log \theta_{\rm obs}$ (rad) & (-2, -0.5) & $<-1.5$ & $-0.9^{+0.2}_{-0.2}$ & $-0.5$ (Fixed) \\
        $\log n$ (cm$^{-3}$) & (-6, 2) & $-3.7^{+0.7}_{-0.6}$ & $-2.6^{+1.1}_{-0.9}$ & $-0.3^{+0.3}_{-0.2}$ \\
        $p$ & (2, 3.5) & $2.42^{+0.03}_{-0.03}$ & $2.38^{+0.04}_{-0.04}$ & $2.32^{+0.07}_{-0.13}$ \\
        $\log \epsilon_e$ & (-6, -0.5) & $-1.2^{+0.2}_{-0.1}$ & $-1.1^{+0.2}_{-0.2}$ & $-0.62^{+0.08}_{-0.09}$ \\
        $\log \epsilon_B$ & (-6, -0.5) & $>-2.2$ & $-1.6^{+0.6}_{-0.8}$ & $-2.6^{+0.7}_{-0.7}$ \\
        $\log \Gamma_0$ & (0.1, 5) & $1.6^{+0.1}_{-0.1}$ & $>2.1$ & $> 2.5 $ \\   
        $E_K$ ($10^{51}$ erg) & -- & $1.6^{+0.9}_{-0.6}$ & $1.0^{+1.5}_{-0.4}$ & $0.8^{+0.5}_{-0.3}$ \\
        \bottomrule
    \end{tabular}
    \label{tab:case1}
\end{table*}

We adopt a simple uniform (top-hat) jet profile to fit the afterglow data (see Fig.\ref{fig:lc}) and treat the initial Lorentz factor $\Gamma_0$ as a free parameter\footnote{A finite initial Lorentz factor breaks the assumption of instantaneous deceleration and necessitates disabling jet spreading, as standard spreading prescriptions are invalid under finite-$\Gamma_0$ conditions \citep{Ryan2020}. Accordingly, we set \texttt{spread=False}. For further discussion on the implications of finite initial Lorentz factors in {\sc Afterglowpy} modeling, see \citet{Kaur2024}.}. The other free parameters include the isotropic-equivalent kinetic energy $E_{K,\mathrm{iso}}$, ambient density $n$, viewing angle $\theta_{\mathrm{obs}}$, jet core half-opening angle $\theta_c$, electron power-law index $p$, and the microphysical energy fractions in electrons ($\varepsilon_e$) and magnetic field ($\varepsilon_B$).

The posterior distributions of these parameters are summarized in Table~\ref{tab:case1} and visualized in Fig.~\ref{fig:corner}. 
This model favors an on-axis jet ($\theta_{\rm obs} < \theta_{\rm c}$) with moderate initial Lorentz factor $\Gamma_0\sim$30-50. 
As our line of sight intercepts the jet core, we are not particularly sensitive to the jet structure, and the simple top-hat profile provides an adequate, although not unique, description. 
Adopting the median values of $E_{K,\mathrm{iso}} \sim 1.3 \times 10^{53}\,\mathrm{erg}$ and the jet half opening angle $\theta_c \approx 0.16$ rad  ($\sim 9^\circ$; see Fig.~\ref{fig:corner}), the beaming corrected kinetic energy is $E_K = (1 - \cos \theta_c) E_{k,\mathrm{iso}} \approx 1.6 \times 10^{51} \,\text{ erg}$,
consistent with relativistic TDEs and typical GRB energy scales \citep{Wang2018,Beniamini2023}. 

We also tested the model of a structured jet with a Gaussian angular profile. This model provides an adequate description of the data, but points to a highly relativistic outflow with $\Gamma$\,$>$130 seen off-axis ($\theta_{\rm obs} > \theta_{\rm c}$).
Adopting the median values of \( E_{K,\mathrm{iso}} \sim 5 \times 10^{53}~\mathrm{erg} \) and a jet half-opening angle \( \theta_{\rm c} \approx 0.06~\mathrm{rad} \) (\( \sim 3.4^\circ \)), the inferred beaming-corrected kinetic energy is \( E_K \approx 10^{51}~\mathrm{erg} \). In this model, the jet core has a narrower opening angle than in the top-hat fit, but is surrounded by a wider angle structure from which emission is seen.
Using this model, we can also reproduce the X-ray luminosity of the initial WXT detection (col. Gaussian+WXT). In this scenario, the early WXT flare would mark the afterglow onset rather than the time of the explosion. 

In all cases, the derived blastwave kinetic energy is much higher than the upper limit on $E_{\gamma, \rm iso} \lesssim 8\times10^{50}$ erg (Sect.~\ref{sec:sec1}). This results in a radiative efficiency which is unusually low for classical GRBs, $\eta_{\gamma} = E_{\gamma, \rm iso} / (E_{\gamma, \rm iso} + E_{\rm K,iso}) \lesssim$1\%, although sporadically observed in some events (e.g., GRB190829A, \citealt{Dichiara2022}; and the candidate off-axis GRB150101B, \citealt{Troja2018})

\section{Discussion}
\label{sec:sec6}

Our observations of EP241021a place it within a growing population of FXRTs that show compelling evidence for relativistic outflows despite the absence of prompt gamma-ray detections \citep[e.g.,][]{Cenko2011,Lipunov2022, Bright2024, Perley2025, Srinivasaragavan2025, Busmann2025, LiZhu2025}. In Fig.~\ref{fig:prompt_flux} we show that most of these gamma-ray dark FXRTs require a soft photon index larger than 1 to explain the lack of gamma-rays, and identify a sub-class of events which implies unusually soft spectra with index $\gtrsim$1.5. EP241021a lies within the bulk of the FXRTs population, in a region that is also occupied by canonical GRBs. Therefore, the lack of gamma-ray signal does not immediately single it out as an outlier of the GRB population. 
Its moderate redshift ($z = 0.75$) indicates that the lack of gamma-rays cannot be attributed to cosmological redshift effects, as might be the case for higher-redshift FXRTs  \citep[e.g., EP240315a at $z = 4.9$;][]{Liu2025NatAs,Ricci2025}. Instead, the emission of EP241021a is intrinsically soft, likely reflecting a low spectral peak or a suppressed gamma-ray output. 
The luminosity of its afterglow signal points to a highly energetic explosion with $E_{\rm K,iso} \gtrsim 10^{53}$ erg (Sect.~\ref{sec:sec5}), which requires an unusually low radiative efficiency $\eta_{\gamma} \lesssim 1\%$, at the lower end of the distribution found in typical GRBs \citep{BeniaminiPiran2013,Dichiara2022}.

The presence of both DISS and RISS in its radio afterglow reveals a compact emitting region, with angular size constraints of $\theta_{\text{src}} \gtrsim 1.5~\mu\mathrm{as}$ and a corresponding Lorentz factor of $\Gamma \approx 4$ at 1.5 d. 
Constraints coming from the brightness temperature (Sect.~\ref{sec:sec4.3}) and the optically thin radio spectrum
(Sect.~\ref{sec:sec5}) provide independent support to this value. 
This evidence disfavors scenarios involving sub-relativistic ejecta, such as supernovae shock breakout and classical TDEs, and instead points to a relativistically expanding source, such as GRBs and relativistic TDEs. 
These observations demonstrate that EP241021a launched an outflow that is at least mildly relativistic, regardless of the lack of detectable gamma-ray emission. The absence of gamma rays thus does not preclude relativistic dynamics, and radio scintillation can serve as a powerful diagnostic for uncovering otherwise hidden jets.

Further support for this interpretation comes from our broadband afterglow modeling. The full radio data set, along with the early optical and X-ray data, can be reproduced by a standard forward shock model describing the interaction of a moderately relativistic jet with a low-density \( n \approx 10^{-4}-10^{-1}~\mathrm{cm}^{-3} \)
uniform medium. 
Two different scenarios are consistent with our dataset. The former scenario invokes a moderately relativistic ($\Gamma \approx$40) jet seen close to its axis (Top-Hat model in Table~\ref{tab:case1}), in which the gamma-ray emission is likely suppressed by the compactness of the source \citep{Lithwick2001,Matsumoto2019}. In this model, EP241021a differs from classical GRBs and resembles the properties of a dirty fireball \citep{Dermer1999}, where a baryon-rich outflow  fails to produce bright $\gamma$-rays. This could point to a non-standard central engine, such as a magnetar or a relativistic TDE, launching an outflow with atypical properties.

The latter model invokes a highly relativistic structured jet seen off-axis (Gaussian model in Table~\ref{tab:case1}). In this case, our line of sight is still dominated by the emission produced by ejecta expanding with moderate velocities. However, a high Lorentz factor ($\Gamma \gtrsim$130) jet exists and is not detected due to viewing angle effects. In this model, EP241021a could very well be a standard GRB ``orphan" of its prompt emission
\citep{Rhoads1999}. 
Within this off-axis framework, the origin of the initial WXT flare remains unclear. If our line of sight does not intercept the jet energetic core, then we might not be able to detect the traditional prompt emission phase. Thus, the luminous X-ray flare discovered by EP would be produced by the low-$\Gamma$ ejecta. 
Its onset could mark the break-out of the jet from the progenitor's stellar envelope \citep{Matzner2003,Bromberg2012}, thus favoring the core-collapse of a massive star rather than a TDE. 
Alternatively, the WXT flare could represent the onset of the afterglow \citep{Kobayashi2007}, produced as the relativistic ejecta begin to decelerate upon interacting with the ambient medium. The early X-ray signal would thus not require a separate emission component, but rather reflects the natural transition from the coasting to the deceleration phase in the jet’s evolution (Gaussian+WXT model, Table~\ref{tab:case1}).

Our findings do not depend sensitively on the origin of the unusual optical rebrightening \citep{Busmann2025}, which substantially diverges from traditional afterglow models. 
Multiple interpretations were put forward to describe this atypical feature, such as refreshed shock \citep{Busmann2025}, multi-component jets \citep{Gianfagna2025}, or a reactivation of a long-lived central engine \citep{Shu2025}. 
Our analysis shows that the radio data are mostly consistent with a standard synchrotron radiation from a decelerating, moderately relativistic, wide-angled fireball, and are not substantially affected by any additional emission components probed at higher frequencies. To test the robustness of our results, we included the optical re-brightening and modeled the afterglow with a two-component Top-Hat jet, where both jets share the same $n$, $\theta_c$ and $\theta_{\mathrm{obs}}$ and have independent $E_{K,\mathrm{iso}}$, microphysical parameters and $\Gamma_0$. This shifts some of the inferred physical parameters (e.g., isotropic energy, circumburst density and viewing angle), but not the main inference of a wide-angled relativistic fireball (see Appendix).

\begin{figure}
    \centering
    \includegraphics[width=0.7\textwidth]{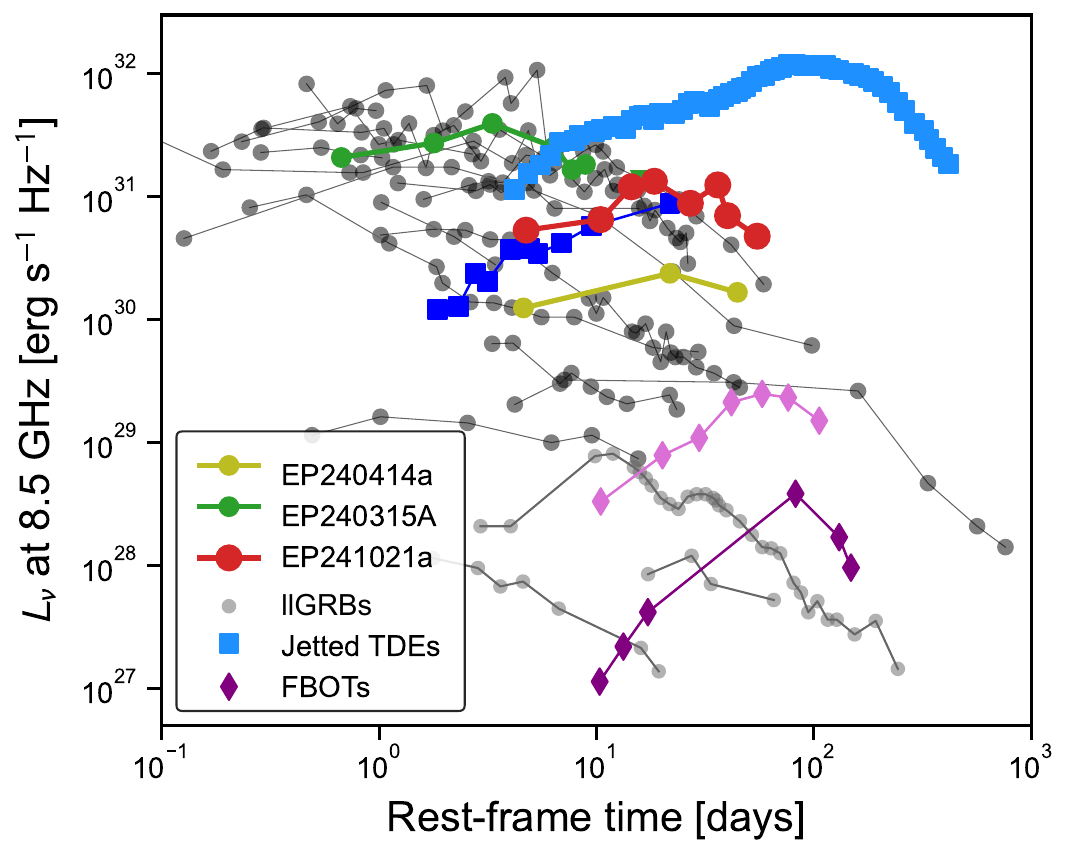}
    \caption{\textbf{Radio luminosity lightcurves.} EP241021A (red circles) is compared to other FXRTs (e.g., EP240315A, \citealt{Ricci2025}; EP240414A, \citealt{Bright2024}) as well as other high-energy transients, including GRBs (circles; \citealt{Chandra2012}), relativistic TDEs (squares; \citealt{Zauderer2011,Andreoni2022}) and FBOTs (diamonds; \citealt{Perley2019,Margutti2019,Perley2021}). }
    \label{fig:eight}
\end{figure}

In Fig.~\ref{fig:eight} we place the radio counterpart of EP241021a in the broader context of high-energy transients. Its luminosity is substantially higher than FBOTs and low-luminosity GRBs (llGRBs), while the observed timescales differ from the population of merger-driven short GRBs. The radio light curve fits well within the distribution of cosmological GRBs from collapsars (dark grey circles) and relativistic jetted TDEs (squares). Its luminosity and timescale appear indeed similar to AT2022cmc (blue squares; \citealt{Andreoni2022,Pasham2023}).  

Additional observations of gamma-ray dark FXRTs are essential to distinguish between different models and to understand the physical origin of this new population of high-energy transients. For instance, the total energetics, temporal profile and variability of the early X-ray emission detected by WXT would help clarify whether it is similar to erratic and highly variable prompt emission flares from an on-axis jet or consistent with other emission mechanisms visible from off-axis observers. 
Important clues might also come from the study of the environment, which would help differentiate between TDEs, core-collapse and other types of stellar explosions \citep[e.g.,][]{Fruchter2006,vanVelzen2021,OConnor2025}.

\section{Conclusion}
\label{sec:sec7}

EP241021a is a luminous ($L_X \approx$ $ 9\times 10^{47}$ erg s$^{-1}$) and relatively nearby ($z\approx$0.75) fast X-ray transient with no associated gamma-ray counterpart down to deep limits. 
We present long-term radio observations and afterglow modeling of this source, providing robust evidence for an outflow expanding with at least mildly relativistic velocities. Through a comprehensive monitoring campaign with ATCA and e-MERLIN over a period exceeding 100 days, we detect sustained radio emission and signatures of interstellar scintillation (ISS) at early times. This observed variability enables us to place constraints on the angular size and Lorentz factor of the emitting region, $\Gamma \approx$4 at 1.5 d.

Although we cannot rule out the presence of a highly relativistic jet oriented far from our line of sight, our modeling suggests that the observed emission comes from material moving toward us with moderately low initial Lorentz factor. 
This rules out progenitor models associated with sub-relativistic ejecta, and points to either relativistic TDEs or GRB-like explosions. 
Our findings reinforce the emerging view that some FXRTs may belong to a new population of energetic and at least moderately relativistic jets, lacking prompt gamma-ray emission due to geometric orientation or intrinsically low radiative inefficiency.

\begin{acknowledgments}
We thank David Williams for promptly scheduling, correlating and making the eMERLIN calibrated data available to the Authors. This work was supported by the European Research Council through the Consolidator grant BHianca (grant agreement ID~101002761). BO acknowledges useful discussions with Paz Beniamini, Ramandeep Gill, and Michael Moss. BO is supported by the McWilliams Fellowship at Carnegie Mellon University. YK is supported by the China Scholarship Council (CSC). 

\end{acknowledgments}





\newpage
\appendix

\section{Gamma-ray and X-ray flux values for Fast X-ray Transients}

\label{sec:appA}

\renewcommand{\arraystretch}{1.2}
\setlength{\tabcolsep}{0.1cm}{\begin{longtable}{cccccc}
\caption{Explicit Gamma and X-ray flux values used in Figure~\ref{fig:prompt_flux} along with their uncertainties.\label{tab:flux_explicit} For gamma-ray dark FXRTs, the gamma-ray fluxes represent 3$\sigma$ upper limits based on 8.192~s time-averaged \textit{Fermi}-GBM data, while detections are reported with 1$\sigma$ uncertainty. X-ray fluxes correspond to time-averaged, unabsorbed 0.5--4~keV fluxes derived from WXT data, with uncertainties reported at the 90\% confidence level. All fluxes are reported in the observer frame. Gamma-ray fluxes reported by the Konus-Wind team are converted to the GBM energy range using the best-fit spectral model.}\\
\toprule
ID&Name&$\gamma$-ray flux&Ref$_\gamma$&X-ray flux&Ref$_X$\\
&&($10^{-8}$ erg cm$^{-2}$  s$^{-1}$)&&($10^{-11}$ erg cm$^{-2}$  s$^{-1}$)\\
\hline
1 & EP240305a &  $<3.8$ &This work&  $83_{-14}^{+20}$ & \citealt{ATel16514} \\
2 & EP240315a &  $118_{-36.8}^{+47.3}$ & \citealt{2024GCN35972} &  $53_{-7}^{+10}$ & \citealt{2024GCN35931} \\
3 & EP240413a &  $<2$ &This work&  $110\pm10$ & \citealt{2024GCN36086} \\
4 & EP240417a &  $<3.5$ &This work&  $9.8_{-8}^{+0.8}$ & \citealt{Hu2024GCN36161} \\
5 & EP240426b &  $<4.0$ & This work& $36\pm6$ & \citealt{2024GCN36330} \\
6 & EP240506a &  $<6.2$ & This work& $110_{-20}^{+30}$ & \citealt{2024GCN36405} \\
7 & EP240617a &  $<4.1$ & This work& $350\pm30$ & \citealt{2024GCN36691} \\
8 & EP240618a &  $<0.85$ & \citealt{2024GCN36725} & $290_{-60}^{+70}$ & \citealt{2024GCN36690} \\
9 & EP240625a & $<5.7$ & This work& $29_{-5}^{+7}$ & \citealt{2024GCN36757} \\
10 & EP240626a &  $<4.8$ & This work& $170_{-50}^{+73}$ & \citealt{2024GCN36766} \\
11 & EP240702a &  $<4.3$ & This work& $540_{-120}^{+150}$ & \citealt{2024GCN36801} \\
12 & EP240703a &  $39.4\pm5.63$ & \citealt{2024GCN36809} & $250_{-30}^{+40}$ & \citealt{2024GCN36807} \\
13 & EP240801a &  $<3.6$ & This work& $48\pm31$ & \citealt{2024GCN36997} \\
14 & EP240802a &  $1249_{-40.0}^{+47.0}$ & \citealt{2024GCN37079} & $170_{-30}^{+40}$ & \citealt{2024GCN37019} \\
15 & EP240804a &  $22\pm9.45$ & \citealt{2024GCN37071} & $61_{-16}^{+26}$ & \citealt{2024GCN37034} \\
16 & EP240807a &  $18.45_{-3.48}^{+2.32}$ & \citealt{2024GCN37130} & $17_{-6}^{+7}$ & \citealt{2024GCN37088} \\
17 & EP240816a &  $<9.3$ & This work& $30\pm5$ & \citealt{2024GCN37188} \\
18 & EP240820a &  $<3.1$ & This work& $12_{-5}^{+7}$ & \citealt{2024GCN37214} \\
19 & EP240913a &  $37\pm0.2$ &  \citealt{2024GCN37481} &  $11_{-4}^{+5}$ & \citealt{2024GCN37492} \\
20 & EP240918b &  $<3.7$ & This work& $26_{-12}^{+41}$ & \citealt{2024GCN37555} \\
21 & EP240918c &  $<3.8$ & This work& $150_{-20}^{+60}$ & \citealt{2024GCN37555} \\
22 & EP240919a &  $4.5\pm0.5$ & \citealt{2024GCN37580} & $70_{-15}^{+18}$ & \citealt{2024GCN37561} \\
23 & EP240930a &  $26\pm1$ &  \citealt{2024GCN37660} & $380\pm50$ & \citealt{2024GCN37648} \\
24 & EP241021a &  $<3.2$ & This work& $33_{-16}^{+48}$ & \citealt{2024GCN37834} \\
25 & EP241025a &  $11.8\pm0.5$ & \citealt{2024GCN37886}  & $16\pm7$ & \citealt{2024GCN37872} \\
26 & EP241026a &  $41\pm2$ &  \citealt{2024GCN37917} & $160\pm30$ & \citealt{2024GCN37909} \\
27 & EP241026b &  $<4.8$ & This work& $12_{-3}^{+4}$ & \citealt{2024GCN37902} \\
28 & EP241030a &  $37.6\pm0.57$ & \citealt{2024GCN38015} &  $7.5_{-2.4}^{+3}$ & \citealt{2024GCN37997} \\
29 & EP241101a &  $<3.2$ & This work& $120_{-40}^{+50}$ & \citealt{2024GCN38039} \\
30 & EP241103a &  $<5.5$ & This work&  $380_{-90}^{+110}$ & \citealt{2024GCN38058} \\
31 & EP241104a &  $32\pm7.6$ & \citealt{2024GCN38094} &  $20_{-11}^{+10}$ & \citealt{2024GCN38081} \\
32 & EP241107a &  $<3.8$ & This work& $10\pm0.5$ & \citealt{2024GCN38171} \\
33 & EP241113a &  $<4.0$ & This work& $55_{-7.6}^{+12.6}$ & \citealt{2024GCN38211} \\
34 & EP241115a &  $15\pm2$ & \cite{2024GCN38323} & $14.2_{-4}^{+6}$ & \citealt{2024GCN38239} \\
35 & EP241202b &  $<4.5$ & This work& $54_{-16}^{+20}$ & \citealt{2024GCN38426} \\
36 & EP241206a &  $<4.8$ & This work& $49.2_{-12.4}^{+11.9}$ & \citealt{2024GCN38457} \\
37 & EP241217a &  $<4.1$ & This work& $73\pm27$ & \citealt{2024GCN38624} \\
38 & EP241217b &  $5.3\pm0.2$ &  \citealt{2024GCN38625} & $119\pm10$ & \citealt{2024GCN38606} \\
39 & EP250101a &  $<8.0$ & This work& $4.4_{-1.7}^{+2.3}$ & \citealt{2025GCN38778} \\
40 & EP250108a &  $<2.6$ & \citealt{2025GCN39146} & $4.2_{-0.9}^{+1.2}$ & \citealt{2025GCN38861} \\
41 & EP250109a &  $13\pm1$ & \citealt{2025GCN38887} & $250_{-120}^{+460}$ & \citealt{2025GCN38889} \\
42 & EP250111a &  $<8.0$ & This work& $139_{-40}^{+53}$ & \citealt{2025GCN38905} \\
43 & EP250125a &  $<5.1$ & This work& $180_{-50}^{+70}$ & \citealt{2025GCN39028} \\
44 & EP250205a &  $6.2\pm0.2$ & \citealt{2025GCN39171} & $42\pm11$ & \citealt{2025GCN39165} \\
45 & EP250207a &  $<2.2$ & This work & $29_{-7}^{+9}$ & \citealt{2025GCN39224} \\
46 & EP250207b &  $<5.4$ & This work& $61_{-25}^{+42}$ & \citealt{2025GCN39266} \\
47 & EP250212a &  $<3.8$ & This work& $90_{-30}^{+100}$ & \citealt{2025GCN39308} \\
48 & EP250223a &  $<4.7$ & This work & $44_{-11}^{+14}$ & \citealt{2025GCN39448} \\
49 & EP250225a &  $<3.7$ & This work& $4.3_{-1.3}^{+2.1}$ & \citealt{2025GCN39475} \\
50 & EP250226a &  $32\pm1$ & \citealt{2025GCN39530} &  $980$ & \citealt{2025GCN39513} \\
51 & EP250227a &  $<2.9$ & This work& $210_{-120}^{+260}$ & \citealt{2025GCN39532} \\
52 & EP250302a &  $<2.6$ & This work& $700_{-160}^{+200}$ & \citealt{2025GCN39556} \\
53 & EP250304a &  $<3.9$ & This work & $53\pm4$ & \citealt{2025GCN39591} \\
54 & EP250321a &  $<3.7$ & This work& $173_{-19}^{+21}$ & \citealt{2025GCN39833} \\
55 & EP250321b &  $<5.7$ & This work& $7.6_{-3}^{+5.3}$ & \citealt{ATel17103} \\
\hline
\end{longtable}}
\clearpage
\section{Two-component TopHat Jet Fit (Including Optical Bump)}

\begin{figure*}[t]
    \centering
    \begin{minipage}[t]{0.45\textwidth}
        \centering
        \includegraphics[width=\textwidth]{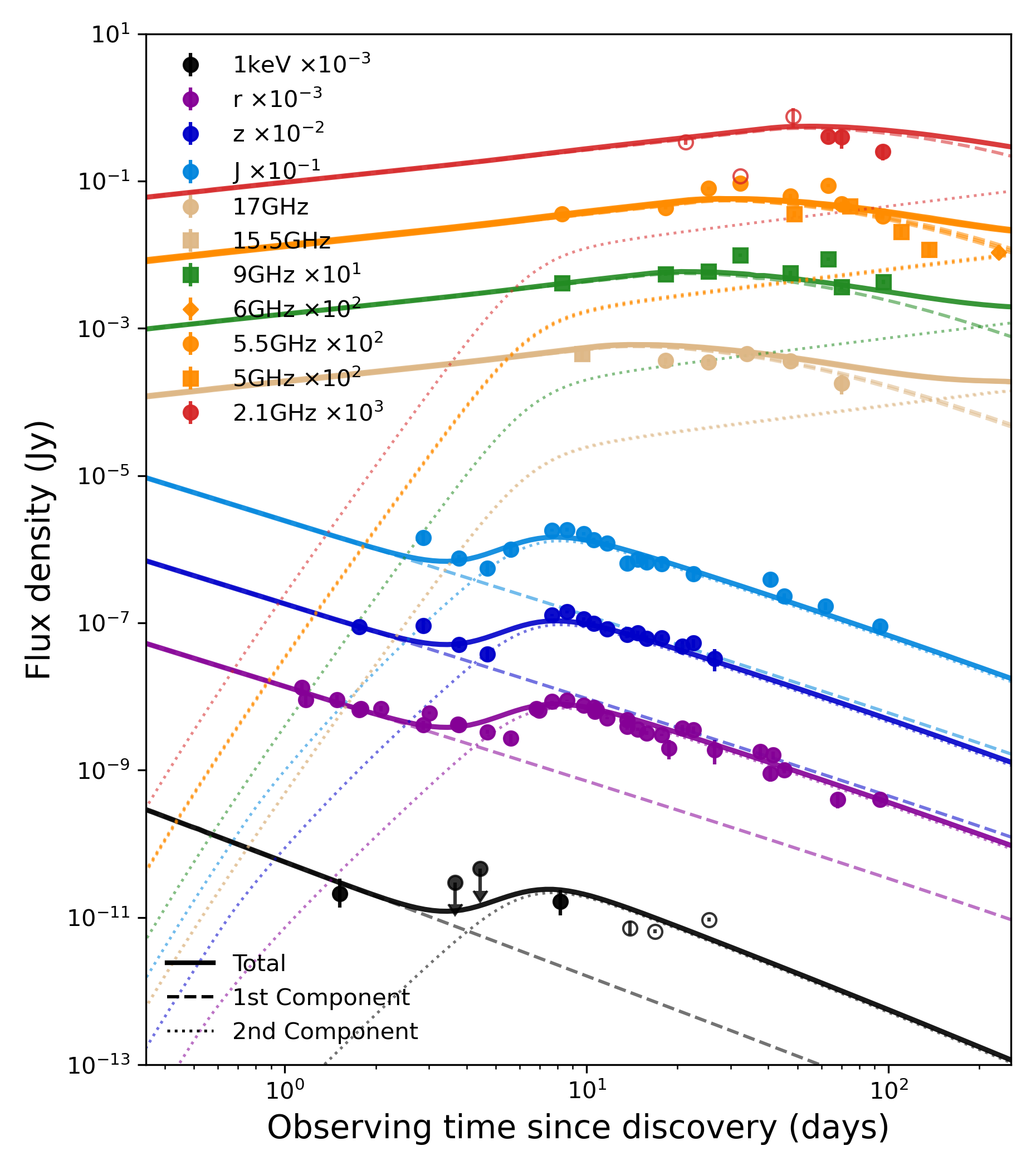}
        \caption{\textbf{Broadband modeling including optical re-brightening results.} Multi-band light curves fitted with a two-component uniform (Top-Hat) jet model.
X-ray (\textit{Swift}/XRT; 1 keV), optical/NIR ($r$, $z$, $J$; \citealt{Busmann2025}), and 15.5~GHz radio (AMI-LA; \citealt{Carotenuto2024GCN}) data are shown.
Downward arrows mark $3\sigma$ upper limits; open symbols indicate points excluded in the fit. Solid lines show the best-fitting two-component jet model, with contributions from component 1(dashed) and second component(dotted) indicated separately. Band-dependent scaling factors are applied for clarity. }
        \label{fig:lc}
    \end{minipage}
    \hfill
    \begin{minipage}[t]{0.5\textwidth}
        \centering
        \includegraphics[width=\textwidth]{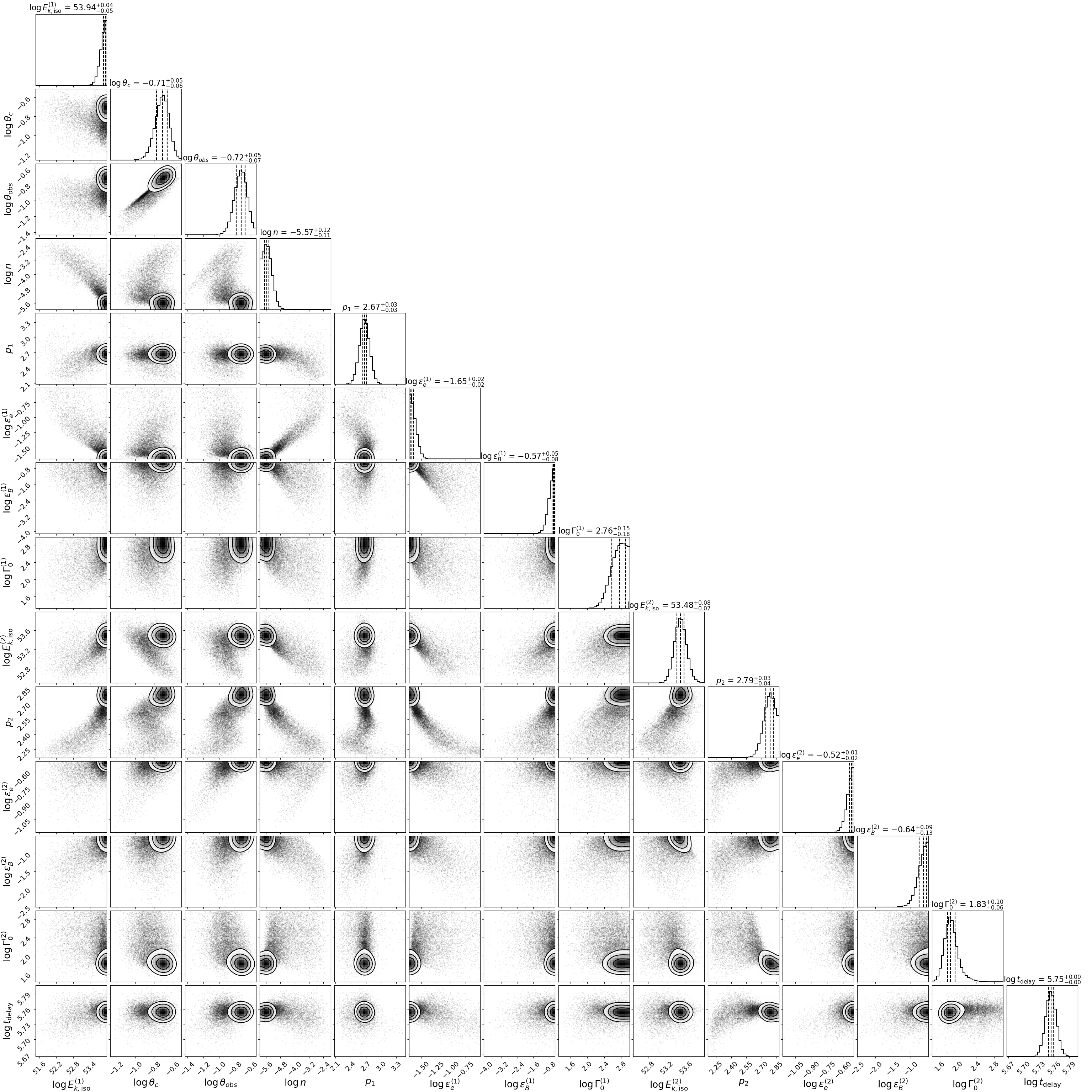}
        \caption{\textbf{Broadband modeling including optical re-brightening results.} Posterior distributions of the fitted parameters for the uniform (Top-Hat) jet model.}
        \label{fig:corner}
    \end{minipage}
\end{figure*}


\bibliography{sample7}{}

\begin{thebibliography}{}
\expandafter\ifx\csname natexlab\endcsname\relax\def\natexlab#1{#1}\fi
\providecommand{\url}[1]{\href{#1}{#1}}
\providecommand{\dodoi}[1]{doi:~\href{http://doi.org/#1}{\nolinkurl{#1}}}
\providecommand{\doeprint}[1]{\href{http://ascl.net/#1}{\nolinkurl{http://ascl.net/#1}}}
\providecommand{\doarXiv}[1]{\href{https://arxiv.org/abs/#1}{\nolinkurl{https://arxiv.org/abs/#1}}}

\bibitem[{S. {Ai} \& B. {Zhang}(2021){Ai} \& {Zhang}}]{Ai2021}
{Ai}, S., \& {Zhang}, B. 2021, \bibinfo{title}{{On the Binary Neutron Star Post-merger Magnetar Origin of XRT 210423},} \apjl, 915, L11, \dodoi{10.3847/2041-8213/ac097d}

\bibitem[{G.~E. {Anderson} {et~al.}(2014){Anderson}, {van der Horst}, {Staley}, {Fender}, {Wijers}, {Scaife}, {Rumsey}, {Titterington}, {Rowlinson}, \& {Saunders}}]{Anderson2014}
{Anderson}, G.~E., {van der Horst}, A.~J., {Staley}, T.~D., {et~al.} 2014, \bibinfo{title}{{Probing the bright radio flare and afterglow of GRB 130427A with the Arcminute Microkelvin Imager},} \mnras, 440, 2059, \dodoi{10.1093/mnras/stu478}

\bibitem[{I. {Andreoni} {et~al.}(2022){Andreoni}, {Coughlin}, {Perley}, {Yao}, {Lu}, {Cenko}, {Kumar}, {Anand}, {Ho}, {Kasliwal}, {de Ugarte Postigo}, {Sagu{\'e}s-Carracedo}, {Schulze}, {Kann}, {Kulkarni}, {Sollerman}, {Tanvir}, {Rest}, {Izzo}, {Somalwar}, {Kaplan}, {Ahumada}, {Anupama}, {Auchettl}, {Barway}, {Bellm}, {Bhalerao}, {Bloom}, {Bremer}, {Bulla}, {Burns}, {Campana}, {Chandra}, {Charalampopoulos}, {Cooke}, {D'Elia}, {Das}, {Dobie}, {Ag{\"u}{\'\i} Fern{\'a}ndez}, {Freeburn}, {Fremling}, {Gezari}, {Goode}, {Graham}, {Hammerstein}, {Karambelkar}, {Kilpatrick}, {Kool}, {Krips}, {Laher}, {Leloudas}, {Levan}, {Lundquist}, {Mahabal}, {Medford}, {Miller}, {M{\"o}ller}, {Mooley}, {Nayana}, {Nir}, {Pang}, {Paraskeva}, {Perley}, {Petitpas}, {Pursiainen}, {Ravi}, {Ridden-Harper}, {Riddle}, {Rigault}, {Rodriguez}, {Rusholme}, {Sharma}, {Smith}, {Stein}, {Th{\"o}ne}, {Tohuvavohu}, {Valdes}, {van Roestel}, {Vergani}, {Wang}, \& {Zhang}}]{Andreoni2022}
{Andreoni}, I., {Coughlin}, M.~W., {Perley}, D.~A., {et~al.} 2022, \bibinfo{title}{{A very luminous jet from the disruption of a star by a massive black hole},} \nat, 612, 430, \dodoi{10.1038/s41586-022-05465-8}

\bibitem[{R.~L. {Aptekar} {et~al.}(1995){Aptekar}, {Frederiks}, {Golenetskii}, {Ilynskii}, {Mazets}, {Panov}, {Sokolova}, {Terekhov}, {Sheshin}, {Cline}, \& {Stilwell}}]{Aptekar1995}
{Aptekar}, R.~L., {Frederiks}, D.~D., {Golenetskii}, S.~V., {et~al.} 1995, \bibinfo{title}{{Konus-W Gamma-Ray Burst Experiment for the GGS Wind Spacecraft},} \ssr, 71, 265, \dodoi{10.1007/BF00751332}

\bibitem[{V.~A. {Arefiev} {et~al.}(2003){Arefiev}, {Priedhorsky}, \& {Borozdin}}]{Arefiev2003}
{Arefiev}, V.~A., {Priedhorsky}, W.~C., \& {Borozdin}, K.~N. 2003, \bibinfo{title}{{Fast X-Ray Transients and Their Connection to Gamma-Ray Bursts},} \apj, 586, 1238, \dodoi{10.1086/367761}

\bibitem[{J.~W. {Armstrong}(1984){Armstrong}}]{Armstrong1984}
{Armstrong}, J.~W. 1984, \bibinfo{title}{{Interstellar scintillation and ultra-low-frequency gravitational wave observations},} \nat, 307, 527, \dodoi{10.1038/307527a0}

\bibitem[{D. {Band} {et~al.}(1993){Band}, {Matteson}, {Ford}, {Schaefer}, {Palmer}, {Teegarden}, {Cline}, {Briggs}, {Paciesas}, {Pendleton}, {Fishman}, {Kouveliotou}, {Meegan}, {Wilson}, \& {Lestrade}}]{Band1993}
{Band}, D., {Matteson}, J., {Ford}, L., {et~al.} 1993, \bibinfo{title}{{BATSE Observations of Gamma-Ray Burst Spectra. I. Spectral Diversity},} \apj, 413, 281, \dodoi{10.1086/172995}

\bibitem[{R. {Barniol Duran} {et~al.}(2013){Barniol Duran}, {Nakar}, \& {Piran}}]{BarniolDuran2013}
{Barniol Duran}, R., {Nakar}, E., \& {Piran}, T. 2013, \bibinfo{title}{{Radius Constraints and Minimal Equipartition Energy of Relativistically Moving Synchrotron Sources},} \apj, 772, 78, \dodoi{10.1088/0004-637X/772/1/78}

\bibitem[{F.~E. {Bauer} {et~al.}(2017){Bauer}, {Treister}, {Schawinski}, {Schulze}, {Luo}, {Alexander}, {Brandt}, {Comastri}, {Forster}, {Gilli}, {Kann}, {Maeda}, {Nomoto}, {Paolillo}, {Ranalli}, {Schneider}, {Shemmer}, {Tanaka}, {Tolstov}, {Tominaga}, {Tozzi}, {Vignali}, {Wang}, {Xue}, \& {Yang}}]{Bauer2017}
{Bauer}, F.~E., {Treister}, E., {Schawinski}, K., {et~al.} 2017, \bibinfo{title}{{A new, faint population of X-ray transients},} \mnras, 467, 4841, \dodoi{10.1093/mnras/stx417}

\bibitem[{P. {Beniamini} \& T. {Piran}(2013){Beniamini} \& {Piran}}]{BeniaminiPiran2013}
{Beniamini}, P., \& {Piran}, T. 2013, \bibinfo{title}{{Constraints on the Synchrotron Emission Mechanism in Gamma-Ray Bursts},} \apj, 769, 69, \dodoi{10.1088/0004-637X/769/1/69}

\bibitem[{P. {Beniamini} {et~al.}(2023){Beniamini}, {Piran}, \& {Matsumoto}}]{Beniamini2023}
{Beniamini}, P., {Piran}, T., \& {Matsumoto}, T. 2023, \bibinfo{title}{{Swift J1644+57 as an off-axis Jet},} \mnras, 524, 1386, \dodoi{10.1093/mnras/stad1950}

\bibitem[{J.~S. {Bright} {et~al.}(2024){Bright}, {Carotenuto}, {Fender}, {Choza}, {Mummery}, {Jonker}, {Smartt}, {DeBoer}, {Farah}, {Matthews}, {Pollak}, {Rhodes}, \& {Siemion}}]{Bright2024}
{Bright}, J.~S., {Carotenuto}, F., {Fender}, R., {et~al.} 2024, \bibinfo{title}{{The Radio Counterpart to the Fast X-ray Transient EP240414a},} arXiv e-prints, arXiv:2409.19055, \dodoi{10.48550/arXiv.2409.19055}

\bibitem[{J.~S. {Bright} {et~al.}(2025){Bright}, {Carotenuto}, {Fender}, {Choza}, {Mummery}, {Jonker}, {Smartt}, {DeBoer}, {Farah}, {Matthews}, {Pollak}, {Rhodes}, \& {Siemion}}]{Bright2025}
{Bright}, J.~S., {Carotenuto}, F., {Fender}, R., {et~al.} 2025, \bibinfo{title}{{The Radio Counterpart to the Fast X-Ray Transient EP240414a},} \apj, 981, 48, \dodoi{10.3847/1538-4357/adaaef}

\bibitem[{O. {Bromberg} {et~al.}(2012){Bromberg}, {Nakar}, {Piran}, \& {Sari}}]{Bromberg2012}
{Bromberg}, O., {Nakar}, E., {Piran}, T., \& {Sari}, R. 2012, \bibinfo{title}{{An Observational Imprint of the Collapsar Model of Long Gamma-Ray Bursts},} \apj, 749, 110, \dodoi{10.1088/0004-637X/749/2/110}

\bibitem[{D.~N. {Burrows} {et~al.}(2005){Burrows}, {Hill}, {Nousek}, {Kennea}, {Wells}, {Osborne}, {Abbey}, {Beardmore}, {Mukerjee}, {Short}, {Chincarini}, {Campana}, {Citterio}, {Moretti}, {Pagani}, {Tagliaferri}, {Giommi}, {Capalbi}, {Tamburelli}, {Angelini}, {Cusumano}, {Br{\"a}uninger}, {Burkert}, \& {Hartner}}]{Burrows2005}
{Burrows}, D.~N., {Hill}, J.~E., {Nousek}, J.~A., {et~al.} 2005, \bibinfo{title}{{The Swift X-Ray Telescope},} \ssr, 120, 165, \dodoi{10.1007/s11214-005-5097-2}

\bibitem[{M. {Busmann} {et~al.}(2025){Busmann}, {O'Connor}, {Sommer}, {Gruen}, {Beniamini}, {Gill}, {Moss}, {Palmese}, {Riffeser}, {Yang}, {Troja}, {Dichiara}, {Ricci}, {Klingler}, {G{\"o}ssl}, {Hu}, {Rau}, {Ries}, {Ryan}, {Schmidt}, {Yadav}, \& {Zeimann}}]{Busmann2025}
{Busmann}, M., {O'Connor}, B., {Sommer}, J., {et~al.} 2025, \bibinfo{title}{{The curious case of EP241021a: Unraveling the mystery of its exceptional rebrightening},} arXiv e-prints, arXiv:2503.14588, \dodoi{10.48550/arXiv.2503.14588}

\bibitem[{F. {Carotenuto} {et~al.}(2024){Carotenuto}, {Bright}, \& {Fender}}]{Carotenuto2024GCN}
{Carotenuto}, F., {Bright}, J., \& {Fender}, R. 2024, \bibinfo{title}{{EP241021a: AMI-LA radio detection},} GRB Coordinates Network, 38014, 1

\bibitem[{J. {Castaneda Jaimes} {et~al.}(2024){Castaneda Jaimes}, {Brightman}, \& {Harrison}}]{Castaneda2024}
{Castaneda Jaimes}, J., {Brightman}, M., \& {Harrison}, F. 2024, in American Astronomical Society Meeting Abstracts, Vol. 243, American Astronomical Society Meeting Abstracts, 359.35

\bibitem[{S.~B. {Cenko} {et~al.}(2008){Cenko}, {Fox}, {Penprase}, {Cucchiara}, {Price}, {Berger}, {Kulkarni}, {Harrison}, {Gal-Yam}, {Ofek}, {Rau}, {Chandra}, {Frail}, {Kasliwal}, {Schmidt}, {Soderberg}, {Cameron}, \& {Roth}}]{Cenko2008}
{Cenko}, S.~B., {Fox}, D.~B., {Penprase}, B.~E., {et~al.} 2008, \bibinfo{title}{{GRB 070125: The First Long-Duration Gamma-Ray Burst in a Halo Environment},} \apj, 677, 441, \dodoi{10.1086/526491}

\bibitem[{S.~B. {Cenko} {et~al.}(2011){Cenko}, {Frail}, {Harrison}, {Haislip}, {Reichart}, {Butler}, {Cobb}, {Cucchiara}, {Berger}, {Bloom}, {Chandra}, {Fox}, {Perley}, {Prochaska}, {Filippenko}, {Glazebrook}, {Ivarsen}, {Kasliwal}, {Kulkarni}, {LaCluyze}, {Lopez}, {Morgan}, {Pettini}, \& {Rana}}]{Cenko2011}
{Cenko}, S.~B., {Frail}, D.~A., {Harrison}, F.~A., {et~al.} 2011, \bibinfo{title}{{Afterglow Observations of Fermi Large Area Telescope Gamma-ray Bursts and the Emerging Class of Hyper-energetic Events},} \apj, 732, 29, \dodoi{10.1088/0004-637X/732/1/29}

\bibitem[{P. {Chandra} \& D.~A. {Frail}(2012){Chandra} \& {Frail}}]{Chandra2012}
{Chandra}, P., \& {Frail}, D.~A. 2012, \bibinfo{title}{{A Radio-selected Sample of Gamma-Ray Burst Afterglows},} \apj, 746, 156, \dodoi{10.1088/0004-637X/746/2/156}

\bibitem[{P. {Chandra} {et~al.}(2008){Chandra}, {Cenko}, {Frail}, {Chevalier}, {Macquart}, {Kulkarni}, {Bock}, {Bertoldi}, {Kasliwal}, {Fox}, {Price}, {Berger}, {Soderberg}, {Harrison}, {Gal-Yam}, {Ofek}, {Rau}, {Schmidt}, {Cameron}, {Cowie}, {Cowie}, {Roth}, {Dopita}, {Peterson}, \& {Penprase}}]{Chandra2008}
{Chandra}, P., {Cenko}, S.~B., {Frail}, D.~A., {et~al.} 2008, \bibinfo{title}{{A Comprehensive Study of GRB 070125, A Most Energetic Gamma-Ray Burst},} \apj, 683, 924, \dodoi{10.1086/589807}

\bibitem[{W. {Chen} {et~al.}(2024){Chen}, {Dai}, {Wen}, {Ling}, {Yuan}, {Liu}, {Zhang}, {Jin}, {Cheng}, {Cui}, {Fan}, {Hu}, {Hu}, {Huang}, {Li}, {Liu}, {Liu}, {Lv}, {Lian}, {Mao}, {Pan}, {Pan}, {Sun}, {Wang}, {Wang}, {Wu}, {Xu}, {Xu}, {Yang}, {Zhang}, {Zhang}, {Zhang}, {Zhang}, {Zhao}, {Chen}, {Jia}, {Zhang}, {Kuulkers}, {Santovincenzo}, {O'Brien}, {Nandra}, {Rau}, {Cordier}, \& {Einstein Probe Team}}]{2024GCN36801}
{Chen}, W., {Dai}, C.~Y., {Wen}, S.~X., {et~al.} 2024, \bibinfo{title}{{EP240702a: EP-WXT detection of a fast X-ray transient},} GRB Coordinates Network, 36801, 1

\bibitem[{H.~Q. {Cheng} {et~al.}(2025){Cheng}, {Hu}, {Zhao}, {Yang}, {Chen}, \& {Zhang}}]{ATel17103}
{Cheng}, H.~Q., {Hu}, D.~F., {Zhao}, Q.~C., {et~al.} 2025, \bibinfo{title}{{Einstein Probe detection of an X-ray transient EP250321b},} The Astronomer's Telegram, 17103, 1

\bibitem[{A. {Ciaramella} {et~al.}(2004){Ciaramella}, {Bongardo}, {Aller}, {Aller}, {De Zotti}, {L{\"a}hteenmaki}, {Longo}, {Milano}, {Tagliaferri}, {Ter{\"a}sranta}, {Tornikoski}, \& {Urpo}}]{Ciaramella2004}
{Ciaramella}, A., {Bongardo}, C., {Aller}, H.~D., {et~al.} 2004, \bibinfo{title}{{A multifrequency analysis of radio variability of blazars},} \aap, 419, 485, \dodoi{10.1051/0004-6361:20035771}

\bibitem[{A. {Connors}(1988){Connors}}]{ConnorsAlanna1988}
{Connors}, A. 1988, PhD thesis, University of Maryland, College Park

\bibitem[{J.~M. {Cordes} \& T.~J.~W. {Lazio}(2002){Cordes} \& {Lazio}}]{Cordes2002}
{Cordes}, J.~M., \& {Lazio}, T.~J.~W. 2002, \bibinfo{title}{{NE2001.I. A New Model for the Galactic Distribution of Free Electrons and its Fluctuations},} arXiv e-prints, astro, \dodoi{10.48550/arXiv.astro-ph/0207156}

\bibitem[{C.~Y. {Dai} {et~al.}(2025){Dai}, {Zhao}, {Mao}, {Wen}, {Yuan}, \& {Einstein Probe Team}}]{2025GCN39556}
{Dai}, C.~Y., {Zhao}, Y.~Q., {Mao}, X., {et~al.} 2025, \bibinfo{title}{{EP250302a: Einstein Probe detection of an X-ray transient},} GRB Coordinates Network, 39556, 1

\bibitem[{V. {D'Alessio} {et~al.}(2006){D'Alessio}, {Piro}, \& {Rossi}}]{Alessio2006}
{D'Alessio}, V., {Piro}, L., \& {Rossi}, E.~M. 2006, \bibinfo{title}{{Properties of X-ray rich gamma ray bursts and X-ray flashes detected with BeppoSAX and Hete-2},} \aap, 460, 653, \dodoi{10.1051/0004-6361:20054501}

\bibitem[{C. {de Barra} {et~al.}(2024){de Barra}, {Meegan}, \& {Fermi GBM Team}}]{2024GCN38015}
{de Barra}, C., {Meegan}, C., \& {Fermi GBM Team}. 2024, \bibinfo{title}{{GRB 241030A: Fermi GBM Observation},} GRB Coordinates Network, 38015, 1

\bibitem[{J. {Dennett-Thorpe} \& A.~G. {de Bruyn}(2002){Dennett-Thorpe} \& {de Bruyn}}]{DennettThorpe2002Natur}
{Dennett-Thorpe}, J., \& {de Bruyn}, A.~G. 2002, \bibinfo{title}{{Interstellar scintillation as the origin of the rapid radio variability of the quasar J1819+3845},} \nat, 415, 57, \dodoi{10.1038/415057a}

\bibitem[{C.~D. {Dermer}(1999){Dermer}}]{Dermer1999}
{Dermer}, C.~D. 1999, \bibinfo{title}{{Gamma-ray burst phenomenology explained through the blast wave model},} \aaps, 138, 519, \dodoi{10.1051/aas:1999335}

\bibitem[{S. {Dichiara} {et~al.}(2022){Dichiara}, {Troja}, {Lipunov}, {Ricci}, {Oates}, {Butler}, {Liuzzo}, {Ryan}, {O'Connor}, {Cenko}, {Cosentino}, {Lien}, {Gorbovskoy}, {Tyurina}, {Balanutsa}, {Vlasenko}, {Gorbunov}, {Podesta}, {Podesta}, {Rebolo}, {Serra}, \& {Buckley}}]{Dichiara2022}
{Dichiara}, S., {Troja}, E., {Lipunov}, V., {et~al.} 2022, \bibinfo{title}{{The early afterglow of GRB 190829A},} \mnras, 512, 2337, \dodoi{10.1093/mnras/stac454}

\bibitem[{ {e-MERLIN pipeline}({2019}){e-MERLIN pipeline}}]{emerlin_casa_pipeline}
{e-MERLIN pipeline}. {2019}, \bibinfo{title}{{CASA pipeline},}, \url{https://github.com/e-merlin/eMERLIN_CASA_pipeline}

\bibitem[{ {Fermi GBM Team}(2024){Fermi GBM Team}}]{2024GCN37481}
{Fermi GBM Team}. 2024, \bibinfo{title}{{GRB 240913A: Fermi GBM Final Real-time Localization},} GRB Coordinates Network, 37481, 1

\bibitem[{C. {Ferrigno} {et~al.}(2019){Ferrigno}, {Bozzo}, {Sanna}, {Jaisawal}, {Girard}, {Di Salvo}, \& {Burderi}}]{Ferrigno2019}
{Ferrigno}, C., {Bozzo}, E., {Sanna}, A., {et~al.} 2019, \bibinfo{title}{{IGR J17503-2636: a candidate supergiant fast X-ray transient},} \aap, 624, A142, \dodoi{10.1051/0004-6361/201935185}

\bibitem[{D.~A. {Frail} {et~al.}(1997){Frail}, {Kulkarni}, {Nicastro}, {Feroci}, \& {Taylor}}]{Frail1997Natur}
{Frail}, D.~A., {Kulkarni}, S.~R., {Nicastro}, L., {Feroci}, M., \& {Taylor}, G.~B. 1997, \bibinfo{title}{{The radio afterglow from the {\ensuremath{\gamma}}-ray burst of 8 May 1997},} \nat, 389, 261, \dodoi{10.1038/38451}

\bibitem[{D.~A. {Frail} {et~al.}(2000){Frail}, {Waxman}, \& {Kulkarni}}]{Frail2000}
{Frail}, D.~A., {Waxman}, E., \& {Kulkarni}, S.~R. 2000, \bibinfo{title}{{A 450 Day Light Curve of the Radio Afterglow of GRB 970508: Fireball Calorimetry},} \apj, 537, 191, \dodoi{10.1086/309024}

\bibitem[{D. {Frederiks} {et~al.}(2024{\natexlab{a}}){Frederiks}, {Lysenko}, {Ridnaia}, {Svinkin}, {Tsvetkova}, {Ulanov}, {Cline}, \& {Konus-Wind Team}}]{2024GCN36809}
{Frederiks}, D., {Lysenko}, A., {Ridnaia}, A., {et~al.} 2024{\natexlab{a}}, \bibinfo{title}{{Konus-Wind detection of GRB 240703A (a counterpart of EP240703a)},} GRB Coordinates Network, 36809, 1

\bibitem[{D. {Frederiks} {et~al.}(2024{\natexlab{b}}){Frederiks}, {Lysenko}, {Ridnaia}, {Svinkin}, {Tsvetkova}, {Ulanov}, {Cline}, \& {Konus-Wind Team}}]{2024GCN37079}
{Frederiks}, D., {Lysenko}, A., {Ridnaia}, A., {et~al.} 2024{\natexlab{b}}, \bibinfo{title}{{Konus-Wind detection of GRB 240802A (EP240802a)},} GRB Coordinates Network, 37079, 1

\bibitem[{D. {Frederiks} {et~al.}(2024{\natexlab{c}}){Frederiks}, {Lysenko}, {Ridnaia}, {Svinkin}, {Tsvetkova}, {Ulanov}, {Cline}, \& {Konus-Wind Team}}]{2024GCN37071}
{Frederiks}, D., {Lysenko}, A., {Ridnaia}, A., {et~al.} 2024{\natexlab{c}}, \bibinfo{title}{{Konus-Wind detection of GRB 240804B (a counterpart of EP240804a)},} GRB Coordinates Network, 37071, 1

\bibitem[{A.~S. {Fruchter} {et~al.}(2006){Fruchter}, {Levan}, {Strolger}, {Vreeswijk}, {Thorsett}, {Bersier}, {Burud}, {Castro Cer{\'o}n}, {Castro-Tirado}, {Conselice}, {Dahlen}, {Ferguson}, {Fynbo}, {Garnavich}, {Gibbons}, {Gorosabel}, {Gull}, {Hjorth}, {Holland}, {Kouveliotou}, {Levay}, {Livio}, {Metzger}, {Nugent}, {Petro}, {Pian}, {Rhoads}, {Riess}, {Sahu}, {Smette}, {Tanvir}, {Wijers}, \& {Woosley}}]{Fruchter2006}
{Fruchter}, A.~S., {Levan}, A.~J., {Strolger}, L., {et~al.} 2006, \bibinfo{title}{{Long {\ensuremath{\gamma}}-ray bursts and core-collapse supernovae have different environments},} \nat, 441, 463, \dodoi{10.1038/nature04787}

\bibitem[{Y.~C. {Fu} {et~al.}(2024){Fu}, {Jiang}, {Hu}, {Ling}, {Zhang}, {Liu}, {Jin}, {Zhang}, {Cheng}, {Chen}, {Cui}, {Fan}, {Hu}, {Huang}, {Li}, {Liu}, {Liu}, {Lv}, {Lian}, {Mao}, {Pan}, {Pan}, {Sun}, {Wang}, {Wang}, {Wu}, {Xu}, {Xu}, {Yang}, {Yuan}, {Zhang}, {Zhang}, {Zhang}, {Zhao}, {Chen}, {Jia}, {Cui}, {Han}, {Li}, {Song}, {Zhao}, {Zhang}, {Zhang}, {Kuulkers}, {Santovincenzo}, {O'Brien}, {Nandra}, {Rau}, {Cordier}, \& {Einstein Probe Team}}]{2024GCN37088}
{Fu}, Y.~C., {Jiang}, S.~Q., {Hu}, J.~W., {et~al.} 2024, \bibinfo{title}{{EP240807a: EP-WXT detection of a fast X-ray transient},} GRB Coordinates Network, 37088, 1

\bibitem[{T.~J. {Galama} {et~al.}(2003){Galama}, {Frail}, {Sari}, {Berger}, {Taylor}, \& {Kulkarni}}]{Galama2003}
{Galama}, T.~J., {Frail}, D.~A., {Sari}, R., {et~al.} 2003, \bibinfo{title}{{Continued Radio Monitoring of the Gamma-Ray Burst 991208},} \apj, 585, 899, \dodoi{10.1086/346083}

\bibitem[{T.~J. {Galama} {et~al.}(1999){Galama}, {Briggs}, {Wijers}, {Vreeswijk}, {Rol}, {Band}, {van Paradijs}, {Kouveliotou}, {Preece}, {Bremer}, {Smith}, {Tilanus}, {de Bruyn}, {Strom}, {Pooley}, {Castro-Tirado}, {Tanvir}, {Robinson}, {Hurley}, {Heise}, {Telting}, {Rutten}, {Packham}, {Swaters}, {Davies}, {Fassia}, {Green}, {Foster}, {Sagar}, {Pandey}, {Nilakshi}, {Yadav}, {Ofek}, {Leibowitz}, {Ibbetson}, {Rhoads}, {Falco}, {Petry}, {Impey}, {Geballe}, \& {Bhattacharya}}]{Galama1999Natur}
{Galama}, T.~J., {Briggs}, M.~S., {Wijers}, R.~A.~M.~J., {et~al.} 1999, \bibinfo{title}{{The effect of magnetic fields on {\ensuremath{\gamma}}-ray bursts inferred from multi-wavelength observations of the burst of 23 January 1999},} \nat, 398, 394, \dodoi{10.1038/18828}

\bibitem[{G. {Gianfagna} {et~al.}(2025){Gianfagna}, {Piro}, {Bruni}, {Linesh Thakur}, {Van Eerten}, {Castro-Tirado}, {Chen}, {Cheng}, {He}, {Jia}, {Ling}, {Maiorano}, {Paladino}, {Tripodi}, {Rossi}, {Yang}, {Yuan}, {Yuan}, \& {Zhang}}]{Gianfagna2025}
{Gianfagna}, G., {Piro}, L., {Bruni}, G., {et~al.} 2025, \bibinfo{title}{{The soft X-ray transient EP241021a: a cosmic explosion with a complex off-axis jet and cocoon from a massive progenitor},} arXiv e-prints, arXiv:2505.05444.
\newblock \doarXiv{2505.05444}

\bibitem[{J.~H. {Gillanders} {et~al.}(2024){Gillanders}, {Rhodes}, {Srivastav}, {Carotenuto}, {Bright}, {Huber}, {Stevance}, {Smartt}, {Chambers}, {Chen}, {Fender}, {Andersson}, {Cooper}, {Jonker}, {Cowie}, {de Boer}, {Erasmus}, {Fulton}, {Gao}, {Herman}, {Lin}, {Lowe}, {Magnier}, {Miao}, {Minguez}, {Moore}, {Ngeow}, {Nicholl}, {Pan}, {Pignata}, {Rest}, {Sheng}, {Smith}, {Smith}, {Tonry}, {Wainscoat}, {Weston}, {Yang}, \& {Young}}]{Gillanders2024}
{Gillanders}, J.~H., {Rhodes}, L., {Srivastav}, S., {et~al.} 2024, \bibinfo{title}{{Discovery of the Optical and Radio Counterpart to the Fast X-Ray Transient EP 240315a},} \apjl, 969, L14, \dodoi{10.3847/2041-8213/ad55cd}

\bibitem[{A. {Glennie} {et~al.}(2015){Glennie}, {Jonker}, {Fender}, {Nagayama}, \& {Pretorius}}]{Glennie2015}
{Glennie}, A., {Jonker}, P.~G., {Fender}, R.~P., {Nagayama}, T., \& {Pretorius}, M.~L. 2015, \bibinfo{title}{{Two fast X-ray transients in archival Chandra data},} \mnras, 450, 3765, \dodoi{10.1093/mnras/stv801}

\bibitem[{M. {Godwin} \&  {Fermi GBM Team}(2024){Godwin} \& {Fermi GBM Team}}]{2024GCN37886}
{Godwin}, M., \& {Fermi GBM Team}. 2024, \bibinfo{title}{{GRB 241025A: Fermi GBM Observation},} GRB Coordinates Network, 37886, 1

\bibitem[{A. Goldstein {et~al.}(2024)Goldstein, Cleveland, \& Kocevski}]{GDT-Core}
Goldstein, A., Cleveland, W.~H., \& Kocevski, D. 2024, \bibinfo{title}{Gamma-ray Data Tools Core Package: v2.0.4,} \url{https://github.com/USRA-STI/gdt-core}

\bibitem[{J. {Goodman}(1997){Goodman}}]{Goodman1997}
{Goodman}, J. 1997, \bibinfo{title}{{Radio scintillation of gamma-ray-burst afterglows},} \na, 2, 449, \dodoi{10.1016/S1384-1076(97)00031-6}

\bibitem[{J. {Granot}(2008){Granot}}]{Granot2008}
{Granot}, J. 2008, \bibinfo{title}{{Critical Review of Basic Afterglow Concepts},} arXiv e-prints, arXiv:0811.1657, \dodoi{10.48550/arXiv.0811.1657}

\bibitem[{J. {Granot} \& R. {Sari}(2002){Granot} \& {Sari}}]{Granot2002}
{Granot}, J., \& {Sari}, R. 2002, \bibinfo{title}{{The Shape of Spectral Breaks in Gamma-Ray Burst Afterglows},} \apj, 568, 820, \dodoi{10.1086/338966}

\bibitem[{J. {Granot} \& A.~J. {van der Horst}(2014){Granot} \& {van der Horst}}]{Granot2014}
{Granot}, J., \& {van der Horst}, A.~J. 2014, \bibinfo{title}{{Gamma-Ray Burst Jets and their Radio Observations},} \pasa, 31, e008, \dodoi{10.1017/pasa.2013.44}

\bibitem[{D.~S. {Heeschen} \& B.~J. {Rickett}(1987){Heeschen} \& {Rickett}}]{Heeschen1987}
{Heeschen}, D.~S., \& {Rickett}, B.~J. 1987, \bibinfo{title}{{The Galactic Latitude Dependence of Centimeter-Wavelength Flicker},} \aj, 93, 589, \dodoi{10.1086/114340}

\bibitem[{J. {Heise}(2003){Heise}}]{Heise2003}
{Heise}, J. 2003, in American Institute of Physics Conference Series, Vol. 662, Gamma-Ray Burst and Afterglow Astronomy 2001: A Workshop Celebrating the First Year of the HETE Mission, ed. G.~R. {Ricker} \& R.~K. {Vanderspek} (AIP), 229--236, \dodoi{10.1063/1.1579346}

\bibitem[{T.~W.~S. {Holoien} {et~al.}(2016){Holoien}, {Kochanek}, {Prieto}, {Grupe}, {Chen}, {Godoy-Rivera}, {Stanek}, {Shappee}, {Dong}, {Brown}, {Basu}, {Beacom}, {Bersier}, {Brimacombe}, {Carlson}, {Falco}, {Johnston}, {Madore}, {Pojmanski}, \& {Seibert}}]{Holoien2016}
{Holoien}, T.~W.~S., {Kochanek}, C.~S., {Prieto}, J.~L., {et~al.} 2016, \bibinfo{title}{{ASASSN-15oi: a rapidly evolving, luminous tidal disruption event at 216 Mpc},} \mnras, 463, 3813, \dodoi{10.1093/mnras/stw2272}

\bibitem[{D.~F. {Hu} {et~al.}(2025){Hu}, {Cheng}, {Yang}, {ZHAO}, {Chen}, {Zhang}, \& {Einstein Probe Team}}]{2025GCN39833}
{Hu}, D.~F., {Cheng}, H.~Q., {Yang}, Z.~H., {et~al.} 2025, \bibinfo{title}{{EP250321a: refined analysis of the EP-WXT and EP-FXT observations},} GRB Coordinates Network, 39833, 1

\bibitem[{J.~W. {Hu} {et~al.}(2024{\natexlab{a}}){Hu}, {Huang}, {Liu}, {Dai}, {Zhang}, {Zhang}, {Liu}, \& {Einstein Probe Team}}]{2024GCN38239}
{Hu}, J.~W., {Huang}, M.~Q., {Liu}, Z.~Y., {et~al.} 2024{\natexlab{a}}, \bibinfo{title}{{EP241115a: EP detection of a fast X-ray transient},} GRB Coordinates Network, 38239, 1

\bibitem[{J.~W. {Hu} {et~al.}(2024{\natexlab{b}}){Hu}, {Wang}, {He}, {Yang}, {Cheng}, {Yuan}, \& {Einstein Probe Team}}]{Hu2024GCN}
{Hu}, J.~W., {Wang}, Y., {He}, H., {et~al.} 2024{\natexlab{b}}, \bibinfo{title}{{EP241021a: EP detection of a fast X-ray transient},} GRB Coordinates Network, 37834, 1

\bibitem[{J.~W. {Hu} {et~al.}(2024{\natexlab{c}}){Hu}, {Wang}, {He}, {Yang}, {Cheng}, {Yuan}, \& {Einstein Probe Team}}]{2024GCN37834}
{Hu}, J.~W., {Wang}, Y., {He}, H., {et~al.} 2024{\natexlab{c}}, \bibinfo{title}{{EP241021a: EP detection of a fast X-ray transient},} GRB Coordinates Network, 37834, 1

\bibitem[{J.~W. {Hu} {et~al.}(2024{\natexlab{d}}){Hu}, {Chen}, {Jiang}, {Jin}, {Liu}, {Ling}, {Zhang}, {Cheng}, {Cui}, {Fan}, {Hu}, {Huang}, {Li}, {Lian}, {Liu}, {Liu}, {Lv}, {Mao}, {Pan}, {Pan}, {Sun}, {Wang}, {Wang}, {Wu}, {Xu}, {Xu}, {Yang}, {Yuan}, {Zhang}, {Zhang}, {Zhang}, {Zhang}, {Zhao}, {Kuulkers}, {Santovincenzo}, {O'Brien}, {Nandra}, {Rau}, {Cordier}, \& {Einstein Probe Team.}}]{Hu2024GCN36161}
{Hu}, J.~W., {Chen}, W., {Jiang}, S.~Q., {et~al.} 2024{\natexlab{d}}, \bibinfo{title}{{EP240417a: EP-WXT detection of a fast X-ray transient},} GRB Coordinates Network, 36161, 1

\bibitem[{P.~A. {Hughes} {et~al.}(1992){Hughes}, {Aller}, \& {Aller}}]{Hughes1992}
{Hughes}, P.~A., {Aller}, H.~D., \& {Aller}, M.~F. 1992, \bibinfo{title}{{The University of Michigan Radio Astronomy Data Base. I. Structure Function Analysis and the Relation between BL Lacertae Objects and Quasi-stellar Objects},} \apj, 396, 469, \dodoi{10.1086/171734}

\bibitem[{F. {Jansen} {et~al.}(2001){Jansen}, {Lumb}, {Altieri}, {Clavel}, {Ehle}, {Erd}, {Gabriel}, {Guainazzi}, {Gondoin}, {Much}, {Munoz}, {Santos}, {Schartel}, {Texier}, \& {Vacanti}}]{Jansen2001}
{Jansen}, F., {Lumb}, D., {Altieri}, B., {et~al.} 2001, \bibinfo{title}{{XMM-Newton observatory. I. The spacecraft and operations},} \aap, 365, L1, \dodoi{10.1051/0004-6361:20000036}

\bibitem[{S.~Q. {Jiang} {et~al.}(2025{\natexlab{a}}){Jiang}, {Wang}, {Wu}, {Zhao}, {Song}, {Liu}, \& {Einstein Probe Team}}]{2025GCN39475}
{Jiang}, S.~Q., {Wang}, B.~T., {Wu}, H.~Z., {et~al.} 2025{\natexlab{a}}, \bibinfo{title}{{EP250225a: Einstein Probe detected of a fast X-ray transient},} GRB Coordinates Network, 39475, 1

\bibitem[{S.~Q. {Jiang} {et~al.}(2025{\natexlab{b}}){Jiang}, {Wang}, {Wu}, {Zhao}, {Song}, {Liu}, \& {Einstein Probe Team}}]{2025GCN39513}
{Jiang}, S.~Q., {Wang}, B.~T., {Wu}, H.~Z., {et~al.} 2025{\natexlab{b}}, \bibinfo{title}{{EP250226a/GRB 250226A: refined analysis of the EP-WXT and FXT observations},} GRB Coordinates Network, 39513, 1

\bibitem[{S.-Q. {Jiang} {et~al.}(2025){Jiang}, {Xu}, {van Hoof}, {Lei}, {Liu}, {Zhou}, {Chen}, {Fu}, {Yang}, {Liu}, {Zhu}, {Filippenko}, {Jonker}, {Pozanenko}, {Gao}, {Wu}, {Zhang}, {Lamb}, {De Pasquale}, {Kobayashi}, {Bauer}, {Sun}, {Pugliese}, {An}, {D'Elia}, {Fynbo}, {Zheng}, {Tirado}, {Yin}, {Zou}, {Deller}, {Pankov}, {Volnova}, {Moskvitin}, {Spiridonova}, {Oparin}, {Rumyantsev}, {Burkhonov}, {Egamberdiyev}, {Kim}, {Krugov}, {Tatarnikov}, {Inasaridze}, {Levan}, {Bj{\o}rn Malesani}, {Ravasio}, {Quirola-V{\'a}squez}, {van Dalen}, {S{\'a}nchez-Sierras}, {Mata S{\'a}nchez}, {Littlefair}, {Chac{\'o}n}, {Torres}, {Chrimes}, {Sarin}, {Martin-Carrillo}, {Dhillon}, {Yang}, {Brink}, {Davies}, {Yang}, {Aryan}, {Chen}, {Kong}, {Li}, {Li}, {Mao}, {P{\'e}rez-Garc{\'\i}a}, {Fern{\'a}ndez-Garc{\'\i}a}, {Andrews}, {Farah}, {Fan}, {Padilla Gonzalez}, {Howell}, {Hartmann}, {Hu}, {Jakobsson}, {Li}, {Ling}, {McCully}, {Newsome}, {Schneider}, {Samaporn Tinyanont}, {Sun}, {Terreran}, {Tang}, {Wang}, {Xu}, {Yuan}, {Zhang},
  {Zhao}, \& {Zhang}}]{Jiang2025}
{Jiang}, S.-Q., {Xu}, D., {van Hoof}, A. P.~C., {et~al.} 2025, \bibinfo{title}{{EP240801a/XRF 240801B: An X-ray Flash Detected by the Einstein Probe and Implications of its Multiband Afterglow},} arXiv e-prints, arXiv:2503.04306, \dodoi{10.48550/arXiv.2503.04306}

\bibitem[{ {KARMA kvis}({1995}){KARMA kvis}}]{KVIS}
{KARMA kvis}. {1995}, \bibinfo{title}{{CASA pipeline},}, \url{https://www.atnf.csiro.au/computing/software/karma/}

\bibitem[{R. {Kaur} {et~al.}(2024){Kaur}, {O'Connor}, {Palmese}, \& {Kunnumkai}}]{Kaur2024}
{Kaur}, R., {O'Connor}, B., {Palmese}, A., \& {Kunnumkai}, K. 2024, \bibinfo{title}{{Detecting prompt and afterglow jet emission of gravitational wave events from LIGO/Virgo/KAGRA and next generation detectors},} arXiv e-prints, arXiv:2410.10579, \dodoi{10.48550/arXiv.2410.10579}

\bibitem[{S. {Kobayashi} \& B. {Zhang}(2007){Kobayashi} \& {Zhang}}]{Kobayashi2007}
{Kobayashi}, S., \& {Zhang}, B. 2007, \bibinfo{title}{{The Onset of Gamma-Ray Burst Afterglow},} \apj, 655, 973, \dodoi{10.1086/510203}

\bibitem[{C. {Kouveliotou} {et~al.}(1993){Kouveliotou}, {Meegan}, {Fishman}, {Bhat}, {Briggs}, {Koshut}, {Paciesas}, \& {Pendleton}}]{Kouveliotou1993}
{Kouveliotou}, C., {Meegan}, C.~A., {Fishman}, G.~J., {et~al.} 1993, \bibinfo{title}{{Identification of Two Classes of Gamma-Ray Bursts},} \apjl, 413, L101, \dodoi{10.1086/186969}

\bibitem[{L.~C. {Lee} \& J.~R. {Jokipii}(1975){Lee} \& {Jokipii}}]{LeeLC1975}
{Lee}, L.~C., \& {Jokipii}, J.~R. 1975, \bibinfo{title}{{Strong scintillations in astrophysics. I. The Markov approximation, its validity and application to angular broadening.},} \apj, 196, 695, \dodoi{10.1086/153458}

\bibitem[{W. {Lewandowski} {et~al.}(2011){Lewandowski}, {Kijak}, {Gupta}, \& {Krzeszowski}}]{Lewandowski2011}
{Lewandowski}, W., {Kijak}, J., {Gupta}, Y., \& {Krzeszowski}, K. 2011, \bibinfo{title}{{Diffractive and refractive timescales at 4.8 GHz in PSR B0329+54},} \aap, 534, A66, \dodoi{10.1051/0004-6361/201116850}

\bibitem[{D.~Y. {Li} {et~al.}(2024{\natexlab{a}}){Li}, {Lian}, {Wang}, {Wen}, {Yuan}, \& {Einstein Probe Team}}]{2024GCN37909}
{Li}, D.~Y., {Lian}, T.~Y., {Wang}, Y.~L., {et~al.} 2024{\natexlab{a}}, \bibinfo{title}{{EP241026a (GRB 241026A): EP on-board trigger and follow-up observation},} GRB Coordinates Network, 37909, 1

\bibitem[{D.~Y. {Li} {et~al.}(2024{\natexlab{b}}){Li}, {Liu}, {Huang}, {Peng}, {Jin}, \& {Einstein Probe Team}}]{2024GCN37872}
{Li}, D.~Y., {Liu}, Z.~Y., {Huang}, M.~Q., {et~al.} 2024{\natexlab{b}}, \bibinfo{title}{{EP241025a (GRB 241025A): EP observations update},} GRB Coordinates Network, 37872, 1

\bibitem[{D.~Y. {Li} {et~al.}(2025{\natexlab{a}}){Li}, {Zhang}, {Fu}, {Yuan}, \& {Einstein Probe Team}}]{2025GCN39308}
{Li}, D.~Y., {Zhang}, Y.~J.~T., {Fu}, Y.~C., {Yuan}, W., \& {Einstein Probe Team}. 2025{\natexlab{a}}, \bibinfo{title}{{EP250212a: EP-WXT detection and FXT follow-up observation of a fast X-ray transient},} GRB Coordinates Network, 39308, 1

\bibitem[{D.~Y. {Li} {et~al.}(2024{\natexlab{c}}){Li}, {Yang}, {Li}, {Yuan}, {Ling}, {Liu}, {Zhang}, {Cheng}, {Chen}, {Cui}, {Fan}, {Hu}, {Hu}, {Huang}, {Liu}, {Liu}, {Lv}, {Lian}, {Mao}, {Pan}, {Sun}, {Wang}, {Wang}, {Wu}, {Xu}, {Xu}, {Yang}, {Zhang}, {Zhang}, {Zhang}, {Zhang}, {Chen}, {Jia}, {Zhang}, {Kuulkers}, {Santovincenzo}, {O'Brien}, {Nandra}, {Rau}, {Cordier}, \& {Einstein Probe Team}}]{2024GCN36405}
{Li}, D.~Y., {Yang}, J., {Li}, A., {et~al.} 2024{\natexlab{c}}, \bibinfo{title}{{EP240506a: EP-WXT detection of a new fast X-ray transient},} GRB Coordinates Network, 36405, 1

\bibitem[{D.~Y. {Li} {et~al.}(2024{\natexlab{d}}){Li}, {Xu}, {Wang}, {Zhang}, {Hu}, {Jin}, {Ling}, {Zhang}, {Liu}, {Zhang}, {Cheng}, {Chen}, {Cui}, {Fan}, {Hu}, {Hu}, {Huang}, {Liu}, {Liu}, {Lv}, {Lian}, {Mao}, {Pan}, {Pan}, {Sun}, {Wang}, {Wang}, {Wu}, {Xu}, {Yang}, {Yuan}, {Zhang}, {Zhang}, {Zhang}, {Zhao}, {Chen}, {Jia}, {Cui}, {Han}, {Li}, {Song}, {Zhao}, {Zhang}, {Zhang}, {Kuulkers}, {Santovincenzo}, {O'Brien}, {Nandra}, {Rau}, {Cordier}, \& {Einstein Probe Team}}]{2024GCN37492}
{Li}, D.~Y., {Xu}, X.~P., {Wang}, B.~T., {et~al.} 2024{\natexlab{d}}, \bibinfo{title}{{EP240913a: EP-WXT detection of a fast X-ray transient},} GRB Coordinates Network, 37492, 1

\bibitem[{D.~Y. {Li} {et~al.}(2025{\natexlab{b}}){Li}, {Liu}, {An}, {Fu}, {Jiang}, {Liu}, {Xu}, {Zhu}, {Zhang}, {Shui}, {Zhang}, \& {Einstein Probe Team}}]{2025GCN39224}
{Li}, D.~Y., {Liu}, M.~J., {An}, J., {et~al.} 2025{\natexlab{b}}, \bibinfo{title}{{EP250207a: EP-WXT detection and XRT follow-up observation},} GRB Coordinates Network, 39224, 1

\bibitem[{R.~Z. {Li} {et~al.}(2024){Li}, {Chen}, {Chatterjee}, {Chen}, {Zhou}, {Zhao}, {Pan}, \& {Einstein Probe Team}}]{2024GCN38171}
{Li}, R.~Z., {Chen}, W., {Chatterjee}, K., {et~al.} 2024, \bibinfo{title}{{EP241107a: EP-FXT follow-up observations},} GRB Coordinates Network, 38171, 1

\bibitem[{R.~Z. {Li} {et~al.}(2025{\natexlab{a}}){Li}, {Chen}, {Chatterjee}, {Hua}, {Liu}, \& {Einstein Probe Team}}]{2025GCN38861}
{Li}, R.~Z., {Chen}, X.~L., {Chatterjee}, K., {et~al.} 2025{\natexlab{a}}, \bibinfo{title}{{EP250108a: EP-WXT detection of an X-ray transient},} GRB Coordinates Network, 38861, 1

\bibitem[{R.~Z. {Li} {et~al.}(2025{\natexlab{b}}){Li}, {Chen}, {Chatterjee}, {Hua}, {Sun}, {Liu}, \& {Einstein Probe Team}}]{2025GCN38889}
{Li}, R.~Z., {Chen}, X.~L., {Chatterjee}, K., {et~al.} 2025{\natexlab{b}}, \bibinfo{title}{{EP250109a / GRB 250109A: refined EP-WXT analysis and EP-FXT follow-up observation},} GRB Coordinates Network, 38889, 1

\bibitem[{W.~X. {Li} {et~al.}(2025){Li}, {Zhu}, {Zou}, {Geng}, {Liu}, {Wang}, {Li}, {Xu}, {Sun}, {Wang}, {Yu}, {Zhang}, {Wu}, {Yang}, {Filippenko}, {Liu}, {Yuan}, {Aguado}, {An}, {An}, {Buckley}, {Castro-Tirado}, {Fu}, {Fynbo}, {Howell}, {Hu}, {Jiang}, {Kumar}, {Mao}, {Maund}, {Liu}, {Mockler}, {Moskvitin}, {Andrews}, {Bom}, {Brink}, {Chatterjee}, {Chen}, {Cheng}, {Cooke}, {Dai}, {Du}, {Erasmus}, {Fang}, {Farah}, {Goranskij}, {Gritsevich}, {Gu}, {Guo}, {Hsiao}, {Hu}, {Hua}, {Jacobson-Gal{\'a}n}, {Jia}, {Jin}, {Kasliwal}, {Kilpatrick}, {Kumar}, {Lei}, {Li}, {Li}, {Li}, {Ling}, {Liu}, {Liu}, {Liu}, {L{\'o}pez-Oramas}, {Maslennikova}, {McCully}, {Monageng}, {Newsone}, {Padilla Gonzalez}, {Pan}, {Peng}, {Pignata}, {Poidevin}, {Potter}, {P{\'e}rez-Fournon}, {Santana-Silva}, {Santos}, {Song}, {Song}, {Spiridonova}, {Sun}, {Sun}, {Terreran}, {Wang}, {Wang}, {Wang}, {Wang}, {Wu}, {Xiang}, {Xiao}, {Xu}, {Xue}, {Yan}, {Yang}, {Yu}, {Zhang}, {Zhang}, {Zhang}, {Zhang}, {Zhang}, {Zheng}, \& {Zou}}]{LiZhu2025}
{Li}, W.~X., {Zhu}, Z.~P., {Zou}, X.~Z., {et~al.} 2025, \bibinfo{title}{{An extremely soft and weak fast X-ray transient associated with a luminous supernova},} arXiv e-prints, arXiv:2504.17034, \dodoi{10.48550/arXiv.2504.17034}

\bibitem[{T.~Y. {Lian} {et~al.}(2024{\natexlab{a}}){Lian}, {Li}, {Wang}, {Yuan}, \& {Einstein Probe Team}}]{2024GCN37902}
{Lian}, T.~Y., {Li}, D.~Y., {Wang}, Y.~L., {Yuan}, S.~X. W.~W., \& {Einstein Probe Team}. 2024{\natexlab{a}}, \bibinfo{title}{{EP241026b: EP detection of a fast X-ray transient},} GRB Coordinates Network, 37902, 1

\bibitem[{T.~Y. {Lian} {et~al.}(2024{\natexlab{b}}){Lian}, {Pan}, {Ling}, {Liu}, {Ling}, {Zhang}, {Jin}, {Cheng}, {Chen}, {Cui}, {Fan}, {Hu}, {Hu}, {Huang}, {Li}, {Liu}, {Liu}, {Lv}, {Mao}, {Pan}, {Sun}, {Wang}, {Wang}, {Wu}, {Xu}, {Xu}, {Yang}, {Yuan}, {Zhang}, {Zhang}, {Zhang}, {Zhang}, {Zhao}, {Yang}, {Dai}, {Fang}, {Chen}, {Jia}, {Zhang}, {Kuulkers}, {Santovincenzo}, {O'Brien}, {Nandra}, {Rau}, {Cordier}, \& {Einstein Probe Team}}]{2024GCN36086}
{Lian}, T.~Y., {Pan}, X., {Ling}, Z.~X., {et~al.} 2024{\natexlab{b}}, \bibinfo{title}{{EP240413a: EP-WXT detection of a fast X-ray transient},} GRB Coordinates Network, 36086, 1

\bibitem[{Y.~F. {Liang} {et~al.}(2025){Liang}, {Li}, {Peng}, {Wen}, {Zhang}, \& {Einstein Probe Team}}]{2025GCN38778}
{Liang}, Y.~F., {Li}, A., {Peng}, H.~L., {et~al.} 2025, \bibinfo{title}{{EP241231b and EP250101a: EP-WXT detection of two X-ray transients},} GRB Coordinates Network, 38778, 1

\bibitem[{Y.~F. {Liang} {et~al.}(2024{\natexlab{a}}){Liang}, {Liu}, {Wu}, {Zhao}, {Yang}, {Yuan}, \& {Einstein Probe Team}}]{2024GCN38039}
{Liang}, Y.~F., {Liu}, Q.~C., {Wu}, Q.~Y., {et~al.} 2024{\natexlab{a}}, \bibinfo{title}{{EP241101a: EP-WXT detection of a fast X-ray transient},} GRB Coordinates Network, 38039, 1

\bibitem[{Y.~F. {Liang} {et~al.}(2024{\natexlab{b}}){Liang}, {Liu}, {Mao}, {Jin}, {Liu}, {Ling}, {Zhang}, {Cheng}, {Chen}, {Cui}, {Fan}, {Hu}, {W.}, {Huang}, {Li}, {Liu}, {Lv}, {Lian}, {Pan}, {Pan}, {Sun}, {Wang}, {Wang}, {Xu}, {Xu}, {Yang}, {Yuan}, {Zhang}, {Zhang}, {Zhang}, {Zhang}, {Zhao}, {Chen}, {Jia}, {Cui}, {Han}, {Li}, {Song}, {Zhao}, {Zhang}, {Zhang}, {Kuulkers}, {Santovincenzo}, {O'Brien}, {Nandra}, {Rau}, {Cordier}, \& {Einstein Probe Team}}]{2024GCN37214}
{Liang}, Y.~F., {Liu}, H.~Y., {Mao}, X., {et~al.} 2024{\natexlab{b}}, \bibinfo{title}{{EP240820a: EP detection of a fast X-ray transient},} GRB Coordinates Network, 37214, 1

\bibitem[{Y.~F. {Liang} {et~al.}(2024{\natexlab{c}}){Liang}, {Zhang}, {Liu}, {Xu}, {Jin}, {Ling}, {Yuan}, {Liu}, {Zhang}, {Chen}, {Cheng}, {Cui}, {Fan}, {Hu}, {Hu}, {Huang}, {Li}, {Liu}, {Lv}, {Lian}, {Mao}, {Pan}, {Pan}, {Sun}, {Wang}, {Wang}, {Xu}, {Yang}, {Zhang}, {Zhang}, {Zhang}, {Zhang}, {Zhao}, {Chen}, {Jia}, {Cui}, {Han}, {Li}, {Song}, {Zhao}, {Zhang}, {Zhang}, {Kuulkers}, {Santovincenzo}, {O'Brien}, {Nandra}, {Rau}, {Cordier}, \& {Einstein Probe Team}}]{2024GCN37555}
{Liang}, Y.~F., {Zhang}, Z.~J., {Liu}, M.~J., {et~al.} 2024{\natexlab{c}}, \bibinfo{title}{{EP240918b and EP240918c: EP-WXT detection of two fast X-ray transients},} GRB Coordinates Network, 37555, 1

\bibitem[{Y.~F. {Liang} {et~al.}(2024{\natexlab{d}}){Liang}, {Cheng}, {Liu}, {Pan}, {Xu}, {Yang(NAOC}, {CAS)}, {Zhang}, {Dai}, {Zhang}, {Jin}, {Pan}, {Ling}, {Yuan}, {Liu}, {Zhang}, {Chen}, {Cui}, {Fan}, {Hu}, {Hu}, {Huang}, {Li}, {Liu}, {Lv}, {Lian}, {Mao}, {Sun}, {Wang}, {Wang}, {Xu}, {Zhang}, {Zhang}, {Zhang}, {Zhang}, {Zhao}, {Chen}, {Jia}, {Cui}, {Han}, {Li}, {Song}, {Zhao}, {Zhang}, {Zhang}, {Kuulkers}, {Santovincenzo}, {O'Brien}, {Nandra}, {Rau}, {Cordier}, \& {Einstein Probe Team}}]{2024GCN37561}
{Liang}, Y.~F., {Cheng}, H.~Q., {Liu}, M.~J., {et~al.} 2024{\natexlab{d}}, \bibinfo{title}{{EP240919a: EP-WXT detection of a fast X-ray transient},} GRB Coordinates Network, 37561, 1

\bibitem[{D. {Lin} {et~al.}(2022){Lin}, {Irwin}, {Berger}, \& {Nguyen}}]{Lin2022}
{Lin}, D., {Irwin}, J.~A., {Berger}, E., \& {Nguyen}, R. 2022, \bibinfo{title}{{Discovery of Three Candidate Magnetar-powered Fast X-Ray Transients from Chandra Archival Data},} \apj, 927, 211, \dodoi{10.3847/1538-4357/ac4fc6}

\bibitem[{D. {Lin} {et~al.}(2020){Lin}, {Strader}, {Romanowsky}, {Irwin}, {Godet}, {Barret}, {Webb}, {Homan}, \& {Remillard}}]{Lin2020}
{Lin}, D., {Strader}, J., {Romanowsky}, A.~J., {et~al.} 2020, \bibinfo{title}{{Multiwavelength Follow-up of the Hyperluminous Intermediate-mass Black Hole Candidate 3XMM J215022.4-055108},} \apjl, 892, L25, \dodoi{10.3847/2041-8213/ab745b}

\bibitem[{V.~M. {Lipunov} {et~al.}(2022){Lipunov}, {Kornilov}, {Topolev}, {Tyurina}, {Gorbovskoy}, {Simakov}, {Zhirkov}, {Vlasenko}, {Francile}, {Podesta}, {Podesta}, {Svinkin}, {Budnev}, {Balanutsa}, {Cheryasov}, {Chasovnikov}, {Rebolo}, {Serra-Ricart}, {Gress}, {Ershova}, {Yurkov}, {Gabovich}, {Tlatov}, {Minkina}, {Vladimirov}, {Kuznetsov}, {Antipov}, {Svertilov}, {Tselik}, \& {Kechin}}]{Lipunov2022}
{Lipunov}, V.~M., {Kornilov}, V.~G., {Topolev}, V.~V., {et~al.} 2022, \bibinfo{title}{{The First Detection of an Orphan Burst at the Rise Phase},} Astronomy Letters, 48, 623, \dodoi{10.1134/S1063773722110093}

\bibitem[{Y. {Lithwick} \& R. {Sari}(2001){Lithwick} \& {Sari}}]{Lithwick2001}
{Lithwick}, Y., \& {Sari}, R. 2001, \bibinfo{title}{{Lower Limits on Lorentz Factors in Gamma-Ray Bursts},} \apj, 555, 540, \dodoi{10.1086/321455}

\bibitem[{M.~J. {Liu} {et~al.}(2024){Liu}, {Li}, {Liu}, {Cheng}, {Zhang}, {Zhang}, {Ling}, {Jin}, {Chen}, {Cui}, {Fan}, {Hu}, {Hu}, {Huang}, {Lian}, {Liu}, {Lv}, {Mao}, {Pan}, {Pan}, {Sun}, {Wang}, {Wang}, {Wu}, {Xu}, {Xu}, {Yang}, {Zhang}, {Zhang}, {Zhang}, {Zhang}, {Zhao}, {Kuulkers}, {O'Brien}, \& {Yuan}}]{ATel16514}
{Liu}, M.~J., {Li}, D.~Y., {Liu}, Y., {et~al.} 2024, \bibinfo{title}{{Swift follow-up observation of the EP X-ray transient EP240305a (EPW20240305aa)},} The Astronomer's Telegram, 16514, 1

\bibitem[{Y. {Liu} {et~al.}(2022{\natexlab{a}}){Liu}, {Verbiest}, {Main}, {Wu}, {Ambalappat}, {Champion}, {Cognard}, {Guillemot}, {Gaikwad}, {Janssen}, {Kramer}, {Keith}, {Karuppusamy}, {K{\"u}nkel}, {Liu}, {McKee}, {Mickaliger}, {Stappers}, {Shaifullah}, \& {Theureau}}]{LiuYulan2022}
{Liu}, Y., {Verbiest}, J. P.~W., {Main}, R.~A., {et~al.} 2022{\natexlab{a}}, \bibinfo{title}{{Long-term scintillation studies of EPTA pulsars. I. Observations and basic results},} \aap, 664, A116, \dodoi{10.1051/0004-6361/202142552}

\bibitem[{Y. {Liu} {et~al.}(2022{\natexlab{b}}){Liu}, {Verbiest}, {Main}, {Wu}, {Ambalappat}, {Champion}, {Cognard}, {Guillemot}, {Gaikwad}, {Janssen}, {Kramer}, {Keith}, {Karuppusamy}, {K{\"u}nkel}, {Liu}, {McKee}, {Mickaliger}, {Stappers}, {Shaifullah}, \& {Theureau}}]{Liu2022}
{Liu}, Y., {Verbiest}, J. P.~W., {Main}, R.~A., {et~al.} 2022{\natexlab{b}}, \bibinfo{title}{{Long-term scintillation studies of EPTA pulsars. I. Observations and basic results},} \aap, 664, A116, \dodoi{10.1051/0004-6361/202142552}

\bibitem[{Y. {Liu} {et~al.}(2025){Liu}, {Sun}, {Xu}, {Svinkin}, {Delaunay}, {Tanvir}, {Gao}, {Zhang}, {Chen}, {Wu}, {Zhang}, {Yuan}, {An}, {Bruni}, {Frederiks}, {Ghirlanda}, {Hu}, {Li}, {Li}, {Li}, {Malesani}, {Piro}, {Raman}, {Ricci}, {Troja}, {Vergani}, {Wu}, {Yang}, {Zhang}, {Zhu}, {de Ugarte Postigo}, {Demin}, {Dobie}, {Fan}, {Fu}, {Fynbo}, {Geng}, {Gianfagna}, {Hu}, {Huang}, {Jiang}, {Jonker}, {Julakanti}, {Kennea}, {Kokomov}, {Kuulkers}, {Lei}, {Leung}, {Levan}, {Li}, {Li}, {Littlefair}, {Liu}, {Lysenko}, {Ma}, {Martin-Carrillo}, {O'Brien}, {Parsotan}, {Quirola-V{\'a}squez}, {Ridnaia}, {Ronchini}, {Rossi}, {Mata-S{\'a}nchez}, {Schneider}, {Shen}, {Thakur}, {Tohuvavohu}, {Torres}, {Tsvetkova}, {Ulanov}, {Wei}, {Xiao}, {Yin}, {Bai}, {Burwitz}, {Cai}, {Chen}, {Chen}, {Chen}, {Chen}, {Chen}, {Chen}, {Cheng}, {Cordier}, {Cui}, {Cui}, {Dai}, {Dai}, {Eder}, {Eyles-Ferris}, {Fan}, {Feldman}, {Feng}, {Feng}, {Friedrich}, {Gao}, {Gonzalez}, {Guan}, {Han}, {Han}, {Hou}, {Hu}, {Hu}, {Huang}, {Huo}, {Hutchinson},
  {Ji}, {Jia}, {Jia}, {Jiang}, {Jin}, {Jin}, {Jin}, {Keereman}, {Lerman}, {Li}, {Li}, {Li}, {Li}, {Li}, {Lian}, {Liang}, {Ling}, {Liu}, {Liu}, {Liu}, {Liu}, {Liu}, {Lu}, {L{\"u}}, {Luo}, {Ma}, {Ma}, {Mao}, {Mao}, {McHugh}, {Meidinger}, {Nandra}, {Osborne}, {Pan}, {Pan}, {Ravasio}, {Rau}, {Rea}, {Rehman}, {Sanders}, {Santovincenzo}, {Song}, {Su}, {Sun}, {Sun}, {Sun}, {Tan}, {Tang}, {Tao}, {Tong}, {Wang}, {Wang}, {Wang}, {Wang}, {Wang}, {Wang}, {Wang}, {Wang}, {Wang}, {Wei}, {Willingale}, {Xiong}, {Xu}, {Xu}, {Xu}, {Xu}, {Xu}, {Xue}, {Xue}, {Yan}, {Yang}, {Yang}, {Yang}, {Yang}, {Yu}, {Zhang}, {Zhang}, {Zhang}, {Zhang}, {Zhang}, {Zhang}, {Zhang}, {Zhang}, {Zhang}, {Zhao}, {Zhao}, {Zhao}, {Zhao}, {Zhou}, {Zhou}, {Zhu}, {Zhu}, \& {Zuo}}]{Liu2025NatAs}
{Liu}, Y., {Sun}, H., {Xu}, D., {et~al.} 2025, \bibinfo{title}{{Soft X-ray prompt emission from the high-redshift gamma-ray burst EP240315a},} Nature Astronomy, \dodoi{10.1038/s41550-024-02449-8}

\bibitem[{Z.~Y. {Liu} {et~al.}(2024){Liu}, {Huang}, {Dai}, {Xu}, {Zhang}, {Yuan}, \& {Einstein Probe Team}}]{2024GCN38211}
{Liu}, Z.~Y., {Huang}, M.~Q., {Dai}, C.~Y., {et~al.} 2024, \bibinfo{title}{{EP241113a: EP on-board trigger and autonomous follow-up observation},} GRB Coordinates Network, 38211, 1

\bibitem[{Z.~Y. {Liu} {et~al.}(2025){Liu}, {Zhang}, {Liu}, {Yang}, {Yuan}, \& {Einstein Probe Team}}]{2025GCN39165}
{Liu}, Z.~Y., {Zhang}, M.~H., {Liu}, M.~J., {et~al.} 2025, \bibinfo{title}{{EP250205a/GRB 250205A: Einstein Probe observation},} GRB Coordinates Network, 39165, 1

\bibitem[{M.~S. {Longair}(2011){Longair}}]{Longair2011}
{Longair}, M.~S. 2011, {High Energy Astrophysics} (Cambridge University Press)

\bibitem[{J.~E.~J. {Lovell} {et~al.}(2008){Lovell}, {Rickett}, {Macquart}, {Jauncey}, {Bignall}, {Kedziora-Chudczer}, {Ojha}, {Pursimo}, {Dutka}, {Senkbeil}, \& {Shabala}}]{Lovell2008}
{Lovell}, J.~E.~J., {Rickett}, B.~J., {Macquart}, J.~P., {et~al.} 2008, \bibinfo{title}{{The Micro-Arcsecond Scintillation-Induced Variability (MASIV) Survey. II. The First Four Epochs},} \apj, 689, 108, \dodoi{10.1086/592485}

\bibitem[{R. {Margutti} {et~al.}(2019){Margutti}, {Metzger}, {Chornock}, {Vurm}, {Roth}, {Grefenstette}, {Savchenko}, {Cartier}, {Steiner}, {Terreran}, {Margalit}, {Migliori}, {Milisavljevic}, {Alexander}, {Bietenholz}, {Blanchard}, {Bozzo}, {Brethauer}, {Chilingarian}, {Coppejans}, {Ducci}, {Ferrigno}, {Fong}, {G{\"o}tz}, {Guidorzi}, {Hajela}, {Hurley}, {Kuulkers}, {Laurent}, {Mereghetti}, {Nicholl}, {Patnaude}, {Ubertini}, {Banovetz}, {Bartel}, {Berger}, {Coughlin}, {Eftekhari}, {Frederiks}, {Kozlova}, {Laskar}, {Svinkin}, {Drout}, {MacFadyen}, \& {Paterson}}]{Margutti2019}
{Margutti}, R., {Metzger}, B.~D., {Chornock}, R., {et~al.} 2019, \bibinfo{title}{{An Embedded X-Ray Source Shines through the Aspherical AT 2018cow: Revealing the Inner Workings of the Most Luminous Fast-evolving Optical Transients},} \apj, 872, 18, \dodoi{10.3847/1538-4357/aafa01}

\bibitem[{T. {Matsumoto} {et~al.}(2019){Matsumoto}, {Nakar}, \& {Piran}}]{Matsumoto2019}
{Matsumoto}, T., {Nakar}, E., \& {Piran}, T. 2019, \bibinfo{title}{{Generalized compactness limit from an arbitrary viewing angle},} \mnras, 486, 1563, \dodoi{10.1093/mnras/stz923}

\bibitem[{C.~D. {Matzner}(2003){Matzner}}]{Matzner2003}
{Matzner}, C.~D. 2003, \bibinfo{title}{{Supernova hosts for gamma-ray burst jets: dynamical constraints},} \mnras, 345, 575, \dodoi{10.1046/j.1365-8711.2003.06969.x}

\bibitem[{P.~A. {Mazzali} {et~al.}(2008){Mazzali}, {Valenti}, {Della Valle}, {Chincarini}, {Sauer}, {Benetti}, {Pian}, {Piran}, {D'Elia}, {Elias-Rosa}, {Margutti}, {Pasotti}, {Antonelli}, {Bufano}, {Campana}, {Cappellaro}, {Covino}, {D'Avanzo}, {Fiore}, {Fugazza}, {Gilmozzi}, {Hunter}, {Maguire}, {Maiorano}, {Marziani}, {Masetti}, {Mirabel}, {Navasardyan}, {Nomoto}, {Palazzi}, {Pastorello}, {Panagia}, {Pellizza}, {Sari}, {Smartt}, {Tagliaferri}, {Tanaka}, {Taubenberger}, {Tominaga}, {Trundle}, \& {Turatto}}]{Mazzali2008}
{Mazzali}, P.~A., {Valenti}, S., {Della Valle}, M., {et~al.} 2008, \bibinfo{title}{{The Metamorphosis of Supernova SN 2008D/XRF 080109: A Link Between Supernovae and GRBs/Hypernovae},} Science, 321, 1185, \dodoi{10.1126/science.1158088}

\bibitem[{C. {Meegan} {et~al.}(2009){Meegan}, {Lichti}, {Bhat}, {Bissaldi}, {Briggs}, {Connaughton}, {Diehl}, {Fishman}, {Greiner}, {Hoover}, {van der Horst}, {von Kienlin}, {Kippen}, {Kouveliotou}, {McBreen}, {Paciesas}, {Preece}, {Steinle}, {Wallace}, {Wilson}, \& {Wilson-Hodge}}]{Meegan2009}
{Meegan}, C., {Lichti}, G., {Bhat}, P.~N., {et~al.} 2009, \bibinfo{title}{{The Fermi Gamma-ray Burst Monitor},} \apj, 702, 791, \dodoi{10.1088/0004-637X/702/1/791}

\bibitem[{P. {M{\'e}sz{\'a}ros} {et~al.}(1999){M{\'e}sz{\'a}ros}, {Rees}, \& {Wijers}}]{Wijers1999}
{M{\'e}sz{\'a}ros}, P., {Rees}, M.~J., \& {Wijers}, R.~A.~M.~J. 1999, \bibinfo{title}{{Energetics and beaming of gamma ray burst triggers},} \na, 4, 303, \dodoi{10.1016/S1384-1076(99)00013-5}

\bibitem[{P. {M{\'e}sz{\'a}ros} \& E. {Waxman}(2001){M{\'e}sz{\'a}ros} \& {Waxman}}]{MeszarosWaxman2001}
{M{\'e}sz{\'a}ros}, P., \& {Waxman}, E. 2001, \bibinfo{title}{{TeV Neutrinos from Successful and Choked Gamma-Ray Bursts},} \prl, 87, 171102, \dodoi{10.1103/PhysRevLett.87.171102}

\bibitem[{M. {Modjaz} {et~al.}(2008){Modjaz}, {Kewley}, {Kirshner}, {Stanek}, {Challis}, {Garnavich}, {Greene}, {Kelly}, \& {Prieto}}]{Modjaz2008}
{Modjaz}, M., {Kewley}, L., {Kirshner}, R.~P., {et~al.} 2008, in IAU Symposium, Vol. 250, Massive Stars as Cosmic Engines, ed. F.~{Bresolin}, P.~A. {Crowther}, \& J.~{Puls}, 503--508, \dodoi{10.1017/S1743921308020887}

\bibitem[{J. {Moldon}(2021){Moldon}}]{Moldon2021}
{Moldon}, J. 2021, \bibinfo{title}{{eMCP: e-MERLIN CASA pipeline},}, Astrophysics Source Code Library, record ascl:2109.006

\bibitem[{E. {Molinari} {et~al.}(2007){Molinari}, {Vergani}, {Malesani}, {Covino}, {D'Avanzo}, {Chincarini}, {Zerbi}, {Antonelli}, {Conconi}, {Testa}, {Tosti}, {Vitali}, {D'Alessio}, {Malaspina}, {Nicastro}, {Palazzi}, {Guetta}, {Campana}, {Goldoni}, {Masetti}, {Meurs}, {Monfardini}, {Norci}, {Pian}, {Piranomonte}, {Rizzuto}, {Stefanon}, {Stella}, {Tagliaferri}, {Ward}, {Ihle}, {Gonzalez}, {Pizarro}, {Sinclaire}, \& {Valenzuela}}]{Molinari2007}
{Molinari}, E., {Vergani}, S.~D., {Malesani}, D., {et~al.} 2007, \bibinfo{title}{{REM observations of GRB 060418 and GRB 060607A: the onset of the afterglow and the initial fireball Lorentz factor determination},} \aap, 469, L13, \dodoi{10.1051/0004-6361:20077388}

\bibitem[{O. {Mukherjee} {et~al.}(2025){Mukherjee}, {Meegan}, \& {Fermi GBM Team}}]{2025GCN39171}
{Mukherjee}, O., {Meegan}, C., \& {Fermi GBM Team}. 2025, \bibinfo{title}{{GRB 250205A: Fermi GBM Observation},} GRB Coordinates Network, 39171, 1

\bibitem[{R. {Narayan}(1992){Narayan}}]{Narayan1992}
{Narayan}, R. 1992, \bibinfo{title}{{The Physics of Pulsar Scintillation},} Philosophical Transactions of the Royal Society of London Series A, 341, 151, \dodoi{10.1098/rsta.1992.0090}

\bibitem[{L. {Nava} {et~al.}(2013){Nava}, {Sironi}, {Ghisellini}, {Celotti}, \& {Ghirlanda}}]{Nava2013}
{Nava}, L., {Sironi}, L., {Ghisellini}, G., {Celotti}, A., \& {Ghirlanda}, G. 2013, \bibinfo{title}{{Afterglow emission in gamma-ray bursts - I. Pair-enriched ambient medium and radiative blast waves},} \mnras, 433, 2107, \dodoi{10.1093/mnras/stt872}

\bibitem[{G. {Novara} {et~al.}(2020){Novara}, {Esposito}, {Tiengo}, {Vianello}, {Salvaterra}, {Belfiore}, {De Luca}, {D'Avanzo}, {Greiner}, {Scodeggio}, {Rosen}, {Delvaux}, {Pian}, {Campana}, {Lisini}, {Mereghetti}, \& {Israel}}]{Novara2020}
{Novara}, G., {Esposito}, P., {Tiengo}, A., {et~al.} 2020, \bibinfo{title}{{A Supernova Candidate at z = 0.092 in XMM-Newton Archival Data},} \apj, 898, 37, \dodoi{10.3847/1538-4357/ab98f8}

\bibitem[{B. {O'Connor} {et~al.}(2025){O'Connor}, {Pasham}, {Andreoni}, {Hare}, {Beniamini}, {Troja}, {Ricci}, {Dobie}, {Chakraborty}, {Ng}, {Klingler}, {Karambelkar}, {Rose}, {Schulze}, {Ryan}, {Dichiara}, {Monageng}, {Buckley}, {Hu}, {Srinivasaragavan}, {Bruni}, {Cabrera}, {Cenko}, {van Eerten}, {Freeburn}, {Hammerstein}, {Kasliwal}, {Kouveliotou}, {Kunnumkai}, {Leung}, {Lien}, {Palmese}, \& {Sakamoto}}]{OConnor2025}
{O'Connor}, B., {Pasham}, D., {Andreoni}, I., {et~al.} 2025, \bibinfo{title}{{Characterization of a Peculiar Einstein Probe Transient EP240408a: An Exotic Gamma-Ray Burst or an Abnormal Jetted Tidal Disruption Event?},} \apjl, 979, L30, \dodoi{10.3847/2041-8213/ada7f5}

\bibitem[{B. {Paczy{\'n}ski}(1998){Paczy{\'n}ski}}]{Paczynski1998}
{Paczy{\'n}ski}, B. 1998, \bibinfo{title}{{Are Gamma-Ray Bursts in Star-Forming Regions?},} \apjl, 494, L45, \dodoi{10.1086/311148}

\bibitem[{X. {Pan} {et~al.}(2024{\natexlab{a}}){Pan}, {Zhao}, {Peng}, {Jin}, {Ling}, {Liu}, {Ling}, {Zhang}, {Cheng}, {Chen}, {Cui}, {Fan}, {Hu}, {Hu}, {Huang}, {Li}, {Liu}, {Liu}, {Lv}, {Lian}, {Mao}, {Pan}, {Sun}, {Wang}, {Wang}, {Wu}, {Xu}, {Xu}, {Yang}, {Yuan}, {Zhang}, {Zhang}, {Zhang}, {Zhang}, {Chen}, {Jia}, {Zhang}, {Kuulkers}, {Santovincenzo}, {O'Brien}, {Nandra}, {Rau}, {Cordier}, \& {Einstein Probe Team}}]{2024GCN36330}
{Pan}, X., {Zhao}, D.~H., {Peng}, J.~Q., {et~al.} 2024{\natexlab{a}}, \bibinfo{title}{{EP240426b: EP-WXT detection of a new fast X-ray transient},} GRB Coordinates Network, 36330, 1

\bibitem[{X. {Pan} {et~al.}(2024{\natexlab{b}}){Pan}, {Lv}, {Fu}, {Liu}, {Ling}, {Zhang}, {Jin}, {Chen}, {Cheng}, {Cui}, {Fan}, {Hu}, {Hu}, {Huang}, {Li}, {Liu}, {Liu}, {Lian}, {Mao}, {Pan}, {Sun}, {Wang}, {Wang}, {Wu}, {Xu}, {Xu}, {Yang}, {Yuan}, {Zhang}, {Zhang}, {Zhang}, {Zhang}, {Zhao}, {Chen}, {Jia}, {Zhang}, {Kuulkers}, {Santovincenzo}, {O'Brien}, {Nandra}, {Rau}, {Cordier}, \& {Einstein Probe Team}}]{2024GCN36757}
{Pan}, X., {Lv}, Z.~Z., {Fu}, Y.~C., {et~al.} 2024{\natexlab{b}}, \bibinfo{title}{{EP240625a: EP-WXT detection of a fast X-ray transient},} GRB Coordinates Network, 36757, 1

\bibitem[{D.~R. {Pasham} {et~al.}(2023){Pasham}, {Lucchini}, {Laskar}, {Gompertz}, {Srivastav}, {Nicholl}, {Smartt}, {Miller-Jones}, {Alexander}, {Fender}, {Smith}, {Fulton}, {Dewangan}, {Gendreau}, {Coughlin}, {Rhodes}, {Horesh}, {van Velzen}, {Sfaradi}, {Guolo}, {Castro Segura}, {Aamer}, {Anderson}, {Arcavi}, {Brennan}, {Chambers}, {Charalampopoulos}, {Chen}, {Clocchiatti}, {de Boer}, {Dennefeld}, {Ferrara}, {Galbany}, {Gao}, {Gillanders}, {Goodwin}, {Gromadzki}, {Huber}, {Jonker}, {Joshi}, {Kara}, {Killestein}, {Kosec}, {Kocevski}, {Leloudas}, {Lin}, {Margutti}, {Mattila}, {Moore}, {M{\"u}ller-Bravo}, {Ngeow}, {Oates}, {Onori}, {Pan}, {Perez-Torres}, {Rani}, {Remillard}, {Ridley}, {Schulze}, {Sheng}, {Shingles}, {Smith}, {Steiner}, {Wainscoat}, {Wevers}, \& {Yang}}]{Pasham2023}
{Pasham}, D.~R., {Lucchini}, M., {Laskar}, T., {et~al.} 2023, \bibinfo{title}{{The Birth of a Relativistic Jet Following the Disruption of a Star by a Cosmological Black Hole},} Nature Astronomy, 7, 88, \dodoi{10.1038/s41550-022-01820-x}

\bibitem[{U. {Pathak} {et~al.}(2025{\natexlab{a}}){Pathak}, {Meegan}, \& {Fermi GBM Team}}]{2025GCN38887}
{Pathak}, U., {Meegan}, C., \& {Fermi GBM Team}. 2025{\natexlab{a}}, \bibinfo{title}{{GRB 250109A: Fermi GBM Observation},} GRB Coordinates Network, 38887, 1

\bibitem[{U. {Pathak} {et~al.}(2025{\natexlab{b}}){Pathak}, {Meegan}, \& {Fermi GBM Team}}]{2025GCN39530}
{Pathak}, U., {Meegan}, C., \& {Fermi GBM Team}. 2025{\natexlab{b}}, \bibinfo{title}{{GRB 250226A: Fermi GBM Observation},} GRB Coordinates Network, 39530, 1

\bibitem[{I. {P{\'e}rez-Fournon} {et~al.}(2024){P{\'e}rez-Fournon}, {Sun}, {Li}, {Poidevin}, {Aguado}, {Wang}, {Cabrera-Lavers}, {Acosta-Pulido}, {L{\'o}pez-Oramas}, {Nespral}, {Niu}, \& {Acero}}]{PerezFournon2024GCN}
{P{\'e}rez-Fournon}, I., {Sun}, N.~C., {Li}, W., {et~al.} 2024, \bibinfo{title}{{EP241021a: GTC OSIRIS+ spectroscopy of the optical counterpart},} GRB Coordinates Network, 37858, 1

\bibitem[{D.~A. {Perley} {et~al.}(2019){Perley}, {Mazzali}, {Yan}, {Cenko}, {Gezari}, {Taggart}, {Blagorodnova}, {Fremling}, {Mockler}, {Singh}, {Tominaga}, {Tanaka}, {Watson}, {Ahumada}, {Anupama}, {Ashall}, {Becerra}, {Bersier}, {Bhalerao}, {Bloom}, {Butler}, {Copperwheat}, {Coughlin}, {De}, {Drake}, {Duev}, {Frederick}, {Gonz{\'a}lez}, {Goobar}, {Heida}, {Ho}, {Horst}, {Hung}, {Itoh}, {Jencson}, {Kasliwal}, {Kawai}, {Khanam}, {Kulkarni}, {Kumar}, {Kumar}, {Kutyrev}, {Lee}, {Maeda}, {Mahabal}, {Murata}, {Neill}, {Ngeow}, {Penprase}, {Pian}, {Quimby}, {Ramirez-Ruiz}, {Richer}, {Rom{\'a}n-Z{\'u}{\~n}iga}, {Sahu}, {Srivastav}, {Socia}, {Sollerman}, {Tachibana}, {Taddia}, {Tinyanont}, {Troja}, {Ward}, {Wee}, \& {Yu}}]{Perley2019}
{Perley}, D.~A., {Mazzali}, P.~A., {Yan}, L., {et~al.} 2019, \bibinfo{title}{{The fast, luminous ultraviolet transient AT2018cow: extreme supernova, or disruption of a star by an intermediate-mass black hole?},} \mnras, 484, 1031, \dodoi{10.1093/mnras/sty3420}

\bibitem[{D.~A. {Perley} {et~al.}(2021){Perley}, {Ho}, {Yao}, {Fremling}, {Anderson}, {Schulze}, {Kumar}, {Anupama}, {Barway}, {Bellm}, {Bhalerao}, {Chen}, {Duev}, {Galbany}, {Graham}, {Gromadzki}, {Guti{\'e}rrez}, {Ihanec}, {Inserra}, {Kasliwal}, {Kool}, {Kulkarni}, {Laher}, {Masci}, {Neill}, {Nicholl}, {Pursiainen}, {van Roestel}, {Sharma}, {Sollerman}, {Walters}, \& {Wiseman}}]{Perley2021}
{Perley}, D.~A., {Ho}, A. Y.~Q., {Yao}, Y., {et~al.} 2021, \bibinfo{title}{{Real-time discovery of AT2020xnd: a fast, luminous ultraviolet transient with minimal radioactive ejecta},} \mnras, 508, 5138, \dodoi{10.1093/mnras/stab2785}

\bibitem[{D.~A. {Perley} {et~al.}(2025){Perley}, {Ho}, {Fausnaugh}, {Lamb}, {Kasliwal}, {Ahumada}, {Anand}, {Andreoni}, {Bellm}, {Bhalerao}, {Bolin}, {Brink}, {Burns}, {Cenko}, {Corsi}, {Filippenko}, {Frederiks}, {Goldstein}, {Hamburg}, {Jayaraman}, {Jonker}, {Kool}, {Kulkarni}, {Kumar}, {Laher}, {Levan}, {Lysenko}, {Perley}, {Ricker}, {Riddle}, {Ridnaia}, {Rusholme}, {Smith}, {Svinkin}, {Ulanov}, {Vanderspek}, {Waratkar}, \& {Yao}}]{Perley2025}
{Perley}, D.~A., {Ho}, A. Y.~Q., {Fausnaugh}, M., {et~al.} 2025, \bibinfo{title}{{The luminous, slow-rising orphan afterglow AT2019pim as a candidate moderately relativistic outflow},} \mnras, 537, 1, \dodoi{10.1093/mnras/staf125}

\bibitem[{T. {Piran}(1999){Piran}}]{Piran1999}
{Piran}, T. 1999, \bibinfo{title}{{Gamma-ray bursts and the fireball model},} \physrep, 314, 575, \dodoi{10.1016/S0370-1573(98)00127-6}

\bibitem[{ {Planck Collaboration} {et~al.}(2020){Planck Collaboration}, {Aghanim}, {Akrami}, {Ashdown}, {Aumont}, {Baccigalupi}, {Ballardini}, {Banday}, {Barreiro}, {Bartolo}, {Basak}, {Battye}, {Benabed}, {Bernard}, {Bersanelli}, {Bielewicz}, {Bock}, {Bond}, {Borrill}, {Bouchet}, {Boulanger}, {Bucher}, {Burigana}, {Butler}, {Calabrese}, {Cardoso}, {Carron}, {Challinor}, {Chiang}, {Chluba}, {Colombo}, {Combet}, {Contreras}, {Crill}, {Cuttaia}, {de Bernardis}, {de Zotti}, {Delabrouille}, {Delouis}, {Di Valentino}, {Diego}, {Dor{\'e}}, {Douspis}, {Ducout}, {Dupac}, {Dusini}, {Efstathiou}, {Elsner}, {En{\ss}lin}, {Eriksen}, {Fantaye}, {Farhang}, {Fergusson}, {Fernandez-Cobos}, {Finelli}, {Forastieri}, {Frailis}, {Fraisse}, {Franceschi}, {Frolov}, {Galeotta}, {Galli}, {Ganga}, {G{\'e}nova-Santos}, {Gerbino}, {Ghosh}, {Gonz{\'a}lez-Nuevo}, {G{\'o}rski}, {Gratton}, {Gruppuso}, {Gudmundsson}, {Hamann}, {Handley}, {Hansen}, {Herranz}, {Hildebrandt}, {Hivon}, {Huang}, {Jaffe}, {Jones}, {Karakci}, {Keih{\"a}nen},
  {Keskitalo}, {Kiiveri}, {Kim}, {Kisner}, {Knox}, {Krachmalnicoff}, {Kunz}, {Kurki-Suonio}, {Lagache}, {Lamarre}, {Lasenby}, {Lattanzi}, {Lawrence}, {Le Jeune}, {Lemos}, {Lesgourgues}, {Levrier}, {Lewis}, {Liguori}, {Lilje}, {Lilley}, {Lindholm}, {L{\'o}pez-Caniego}, {Lubin}, {Ma}, {Mac{\'\i}as-P{\'e}rez}, {Maggio}, {Maino}, {Mandolesi}, {Mangilli}, {Marcos-Caballero}, {Maris}, {Martin}, {Martinelli}, {Mart{\'\i}nez-Gonz{\'a}lez}, {Matarrese}, {Mauri}, {McEwen}, {Meinhold}, {Melchiorri}, {Mennella}, {Migliaccio}, {Millea}, {Mitra}, {Miville-Desch{\^e}nes}, {Molinari}, {Montier}, {Morgante}, {Moss}, {Natoli}, {N{\o}rgaard-Nielsen}, {Pagano}, {Paoletti}, {Partridge}, {Patanchon}, {Peiris}, {Perrotta}, {Pettorino}, {Piacentini}, {Polastri}, {Polenta}, {Puget}, {Rachen}, {Reinecke}, {Remazeilles}, {Renzi}, {Rocha}, {Rosset}, {Roudier}, {Rubi{\~n}o-Mart{\'\i}n}, {Ruiz-Granados}, {Salvati}, {Sandri}, {Savelainen}, {Scott}, {Shellard}, {Sirignano}, {Sirri}, {Spencer}, {Sunyaev}, {Suur-Uski}, {Tauber}, {Tavagnacco},
  {Tenti}, {Toffolatti}, {Tomasi}, {Trombetti}, {Valenziano}, {Valiviita}, {Van Tent}, {Vibert}, {Vielva}, {Villa}, {Vittorio}, {Wandelt}, {Wehus}, {White}, {White}, {Zacchei}, \& {Zonca}}]{Planck2020}
{Planck Collaboration}, {Aghanim}, N., {Akrami}, Y., {et~al.} 2020, \bibinfo{title}{{Planck 2018 results. VI. Cosmological parameters},} \aap, 641, A6, \dodoi{10.1051/0004-6361/201833910}

\bibitem[{S. {Poolakkil} {et~al.}(2021){Poolakkil}, {Preece}, {Fletcher}, {Goldstein}, {Bhat}, {Bissaldi}, {Briggs}, {Burns}, {Cleveland}, {Giles}, {Hui}, {Kocevski}, {Lesage}, {Mailyan}, {Malacaria}, {Paciesas}, {Roberts}, {Veres}, {von Kienlin}, \& {Wilson-Hodge}}]{Poolakkil2021}
{Poolakkil}, S., {Preece}, R., {Fletcher}, C., {et~al.} 2021, \bibinfo{title}{{The Fermi-GBM Gamma-Ray Burst Spectral Catalog: 10 yr of Data},} \apj, 913, 60, \dodoi{10.3847/1538-4357/abf24d}

\bibitem[{P. {Predehl} {et~al.}(2021){Predehl}, {Andritschke}, {Arefiev}, {Babyshkin}, {Batanov}, {Becker}, {B{\"o}hringer}, {Bogomolov}, {Boller}, {Borm}, {Bornemann}, {Br{\"a}uninger}, {Br{\"u}ggen}, {Brunner}, {Brusa}, {Bulbul}, {Buntov}, {Burwitz}, {Burkert}, {Clerc}, {Churazov}, {Coutinho}, {Dauser}, {Dennerl}, {Doroshenko}, {Eder}, {Emberger}, {Eraerds}, {Finoguenov}, {Freyberg}, {Friedrich}, {Friedrich}, {F{\"u}rmetz}, {Georgakakis}, {Gilfanov}, {Granato}, {Grossberger}, {Gueguen}, {Gureev}, {Haberl}, {H{\"a}lker}, {Hartner}, {Hasinger}, {Huber}, {Ji}, {Kienlin}, {Kink}, {Korotkov}, {Kreykenbohm}, {Lamer}, {Lomakin}, {Lapshov}, {Liu}, {Maitra}, {Meidinger}, {Menz}, {Merloni}, {Mernik}, {Mican}, {Mohr}, {M{\"u}ller}, {Nandra}, {Nazarov}, {Pacaud}, {Pavlinsky}, {Perinati}, {Pfeffermann}, {Pietschner}, {Ramos-Ceja}, {Rau}, {Reiffers}, {Reiprich}, {Robrade}, {Salvato}, {Sanders}, {Santangelo}, {Sasaki}, {Scheuerle}, {Schmid}, {Schmitt}, {Schwope}, {Shirshakov}, {Steinmetz}, {Stewart}, {Str{\"u}der},
  {Sunyaev}, {Tenzer}, {Tiedemann}, {Tr{\"u}mper}, {Voron}, {Weber}, {Wilms}, \& {Yaroshenko}}]{Predehl2021}
{Predehl}, P., {Andritschke}, R., {Arefiev}, V., {et~al.} 2021, \bibinfo{title}{{The eROSITA X-ray telescope on SRG},} \aap, 647, A1, \dodoi{10.1051/0004-6361/202039313}

\bibitem[{S.~J. {Prentice} {et~al.}(2018){Prentice}, {Maguire}, {Smartt}, {Magee}, {Schady}, {Sim}, {Chen}, {Clark}, {Colin}, {Fulton}, {McBrien}, {O'Neill}, {Smith}, {Ashall}, {Chambers}, {Denneau}, {Flewelling}, {Heinze}, {Holoien}, {Huber}, {Kochanek}, {Mazzali}, {Prieto}, {Rest}, {Shappee}, {Stalder}, {Stanek}, {Stritzinger}, {Thompson}, \& {Tonry}}]{2018ApJ...865L...3P}
{Prentice}, S.~J., {Maguire}, K., {Smartt}, S.~J., {et~al.} 2018, \bibinfo{title}{{The Cow: Discovery of a Luminous, Hot, and Rapidly Evolving Transient},} \apjl, 865, L3, \dodoi{10.3847/2041-8213/aadd90}

\bibitem[{G. {Pugliese} {et~al.}(2024){Pugliese}, {Xu}, {Izzo}, {Levan}, {Zhu}, {Malesani}, {D'Elia}, {Jonker}, {Martin-Carrillo}, {Gompertz}, {Rossi}, {Saccardi}, \& {Stargate Collaboration}}]{Pugliese2024GCN}
{Pugliese}, G., {Xu}, D., {Izzo}, L., {et~al.} 2024, \bibinfo{title}{{EP241021a: VLT/FORS2 redshift z = 0.75},} GRB Coordinates Network, 37852, 1

\bibitem[{J. {Quirola-V{\'a}squez} {et~al.}(2022){Quirola-V{\'a}squez}, {Bauer}, {Jonker}, {Brandt}, {Yang}, {Levan}, {Xue}, {Eappachen}, {Zheng}, \& {Luo}}]{QuirolaVasquez2022}
{Quirola-V{\'a}squez}, J., {Bauer}, F.~E., {Jonker}, P.~G., {et~al.} 2022, \bibinfo{title}{{Extragalactic fast X-ray transient candidates discovered by Chandra (2000-2014)},} \aap, 663, A168, \dodoi{10.1051/0004-6361/202243047}

\bibitem[{J. {Quirola-V{\'a}squez} {et~al.}(2024){Quirola-V{\'a}squez}, {Bauer}, {Jonker}, {Levan}, {Brandt}, {Ravasio}, {Eappachen}, {Xue}, \& {Zheng}}]{QuirolaVasquez2024}
{Quirola-V{\'a}squez}, J., {Bauer}, F.~E., {Jonker}, P.~G., {et~al.} 2024, \bibinfo{title}{{New JWST redshifts for the host galaxies of CDF-S XT1 and XT2: understanding their nature},} arXiv e-prints, arXiv:2410.10015, \dodoi{10.48550/arXiv.2410.10015}

\bibitem[{M.~E. {Ravasio} {et~al.}(2025){Ravasio}, {Burns}, {Wilson-Hodge}, {Jonker}, \& {Fermi-GBM Team}}]{2025GCN39146}
{Ravasio}, M.~E., {Burns}, E., {Wilson-Hodge}, C., {Jonker}, P.~G., \& {Fermi-GBM Team}. 2025, \bibinfo{title}{{EP 250108A: Upper limits from Fermi-GBM Observations},} GRB Coordinates Network, 39146, 1

\bibitem[{M.~E. {Ravasio} \& P.~G. {Jonker}(2024){Ravasio} \& {Jonker}}]{2024GCN36725}
{Ravasio}, M.~E., \& {Jonker}, P.~G. 2024, \bibinfo{title}{{X-ray transient EP240618a: Fermi/GBM non-detection},} GRB Coordinates Network, 36725, 1

\bibitem[{M.~E. {Ravasio} {et~al.}(2024){Ravasio}, {Veres}, {Hamburg}, {Burns}, {Jonker}, \& {Fermi-GBM Team}}]{2024GCN38625}
{Ravasio}, M.~E., {Veres}, P., {Hamburg}, R., {et~al.} 2024, \bibinfo{title}{{GRB 241217A/EP241217B: Fermi-GBM Sub-Threshold Detection},} GRB Coordinates Network, 38625, 1

\bibitem[{R.~J. {Reynolds}(1989){Reynolds}}]{Reynolds1989ApJ}
{Reynolds}, R.~J. 1989, \bibinfo{title}{{The Column Density and Scale Height of Free Electrons in the Galactic Disk},} \apjl, 339, L29, \dodoi{10.1086/185412}

\bibitem[{J.~E. {Rhoads}(1999){Rhoads}}]{Rhoads1999}
{Rhoads}, J.~E. 1999, \bibinfo{title}{{The Dynamics and Light Curves of Beamed Gamma-Ray Burst Afterglows},} \apj, 525, 737, \dodoi{10.1086/307907}

\bibitem[{J.~E. {Rhoads}(2003){Rhoads}}]{2003ApJ...591.1097R}
{Rhoads}, J.~E. 2003, \bibinfo{title}{{Dirty Fireballs and Orphan Afterglows: A Tale of Two Transients},} \apj, 591, 1097, \dodoi{10.1086/368125}

\bibitem[{R. {Ricci} {et~al.}(2025){Ricci}, {Troja}, {Yang}, {Yadav}, {Liu}, {Sun}, {Wu}, {Gao}, {Zhang}, \& {Yuan}}]{Ricci2025}
{Ricci}, R., {Troja}, E., {Yang}, Y.-H., {et~al.} 2025, \bibinfo{title}{{Long-term Radio Monitoring of the Fast X-Ray Transient EP 240315a: Evidence for a Relativistic Jet},} \apjl, 979, L28, \dodoi{10.3847/2041-8213/ad8b3f}

\bibitem[{B.~J. {Rickett}(1990){Rickett}}]{Rickett1990}
{Rickett}, B.~J. 1990, \bibinfo{title}{{Radio propagation through the turbulent interstellar plasma.},} \araa, 28, 561, \dodoi{10.1146/annurev.aa.28.090190.003021}

\bibitem[{B.~J. {Rickett} {et~al.}(1995){Rickett}, {Quirrenbach}, {Wegner}, {Krichbaum}, \& {Witzel}}]{Rickett1995}
{Rickett}, B.~J., {Quirrenbach}, A., {Wegner}, R., {Krichbaum}, T.~P., \& {Witzel}, A. 1995, \bibinfo{title}{{Interstellar scintillation of the radio source 0917+624.},} \aap, 293, 479

\bibitem[{A. {Ridnaia} {et~al.}(2024){Ridnaia}, {Frederiks}, {Lysenko}, {Svinkin}, {Tsvetkova}, {Ulanov}, {Cline}, \& {Konus-Wind Team}}]{2024GCN37130}
{Ridnaia}, A., {Frederiks}, D., {Lysenko}, A., {et~al.} 2024, \bibinfo{title}{{Konus-Wind detection of GRB 240807A (a counterpart of EP240807a)},} GRB Coordinates Network, 37130, 1

\bibitem[{O.~J. {Roberts} {et~al.}(2024){Roberts}, {Wood}, {Goldstein}, {Hui}, {Hamburg}, \& {Fermi-GBM Team}}]{2024GCN37580}
{Roberts}, O.~J., {Wood}, J., {Goldstein}, A., {et~al.} 2024, \bibinfo{title}{{Fermi GBM Sub-Threshold Detection of GRB240919A / EP240919a},} GRB Coordinates Network, 37580, 1

\bibitem[{P. {Rossi} {et~al.}(1993){Rossi}, {Bodo}, {Massaglia}, \& {Ferrari}}]{Rossi1993}
{Rossi}, P., {Bodo}, G., {Massaglia}, S., \& {Ferrari}, A. 1993, \bibinfo{title}{{Radiative Instability in Synchrotron-emitting Plasmas},} \apj, 414, 112, \dodoi{10.1086/173061}

\bibitem[{G. {Ryan} {et~al.}(2020){Ryan}, {van Eerten}, {Piro}, \& {Troja}}]{Ryan2020}
{Ryan}, G., {van Eerten}, H., {Piro}, L., \& {Troja}, E. 2020, \bibinfo{title}{{Gamma-Ray Burst Afterglows in the Multimessenger Era: Numerical Models and Closure Relations},} \apj, 896, 166, \dodoi{10.3847/1538-4357/ab93cf}

\bibitem[{G. {Ryan} {et~al.}(2024){Ryan}, {van Eerten}, {Troja}, {Piro}, {O'Connor}, \& {Ricci}}]{Ryan2024}
{Ryan}, G., {van Eerten}, H., {Troja}, E., {et~al.} 2024, \bibinfo{title}{{Modeling of Long-term Afterglow Counterparts to Gravitational Wave Events: The Full View of GRB 170817A},} \apj, 975, 131, \dodoi{10.3847/1538-4357/ad6a14}

\bibitem[{R. {Sari} {et~al.}(1999){Sari}, {Piran}, \& {Halpern}}]{Sari1999}
{Sari}, R., {Piran}, T., \& {Halpern}, J.~P. 1999, \bibinfo{title}{{Jets in Gamma-Ray Bursts},} \apjl, 519, L17, \dodoi{10.1086/312109}

\bibitem[{R. {Sari} {et~al.}(1998){Sari}, {Piran}, \& {Narayan}}]{Sari1998}
{Sari}, R., {Piran}, T., \& {Narayan}, R. 1998, \bibinfo{title}{{Spectra and Light Curves of Gamma-Ray Burst Afterglows},} \apjl, 497, L17, \dodoi{10.1086/311269}

\bibitem[{R.~J. {Sault} {et~al.}(1995){Sault}, {Teuben}, \& {Wright}}]{SaultTeuben1995}
{Sault}, R.~J., {Teuben}, P.~J., \& {Wright}, M.~C.~H. 1995, in Astronomical Society of the Pacific Conference Series, Vol.~77, Astronomical Data Analysis Software and Systems IV, ed. R.~A. {Shaw}, H.~E. {Payne}, \& J.~J.~E. {Hayes}, 433, \dodoi{10.48550/arXiv.astro-ph/0612759}

\bibitem[{V. {Sguera} {et~al.}(2016){Sguera}, {Sidoli}, {Paizis}, \& {Bird}}]{Sguera2016}
{Sguera}, V., {Sidoli}, L., {Paizis}, A., \& {Bird}, A.~J. 2016, \bibinfo{title}{{Discovery of two new fast X-ray transients with INTEGRAL: IGR J03346+4414 and IGR J20344+3913},} \mnras, 463, 2885, \dodoi{10.1093/mnras/stw2183}

\bibitem[{X. Shu {et~al.}(2025)Shu, Yang, Yang, Xu, Chen, Eyles-Ferris, Dai, Yu, Shen, Sun, Ding, Jiang, Li, Sun, Xu, Zheng, Zhang, Jin, Rau, Wang, Wu, Yuan, Zhang, Nandra, Aguado, An, An, An, Andrews, Anutarawiramkul, Baldini, Brink, Butpa, Cai, Castro-Tirado, Cheng, Cui, Farah, Filippenko, Fu, Fynbo, Gao, Han, Han, Howell, Hu, Jiang, Kumar, Lei, Li, Li, Liu, Liu, Liu, López-Oramas, López Fernández-Nespral, Maund, McCully, Niu, Newsome, O'Brien, Pan, Padilla~Gonzalez, Pérez-Fournon, Poidevin, Silima, Soria, Sun, Sun, Sun, Terreran, Tinyanont, Wang, Wang, Wang, Wiersema, Xu, Xue, Yang, Zhang, Zhang, Zhang, Zhang, Zhang, Zhao, \& Zhu}]{Shu2025}
Shu, X., Yang, L., Yang, H., {et~al.} 2025, \bibinfo{title}{EP241021a: a months-duration X-ray transient with luminous optical and radio emission,} arXiv e-prints, arXiv:2505.07665.
\newblock \doarXiv{2505.07665}

\bibitem[{J.~H. {Simonetti} {et~al.}(1985){Simonetti}, {Cordes}, \& {Heeschen}}]{Simonetti1985}
{Simonetti}, J.~H., {Cordes}, J.~M., \& {Heeschen}, D.~S. 1985, \bibinfo{title}{{Flicker of extragalactic radio sources at two frequencies.},} \apj, 296, 46, \dodoi{10.1086/163418}

\bibitem[{D.~M. {Smith} {et~al.}(2006){Smith}, {Heindl}, {Markwardt}, {Swank}, {Negueruela}, {Harrison}, \& {Huss}}]{Smith2006}
{Smith}, D.~M., {Heindl}, W.~A., {Markwardt}, C.~B., {et~al.} 2006, \bibinfo{title}{{XTE J1739-302 as a Supergiant Fast X-Ray Transient},} \apj, 638, 974, \dodoi{10.1086/498936}

\bibitem[{J. {Smith} {et~al.}(2024){Smith}, {Meegan}, \& {Fermi GBM Team}}]{2024GCN37660}
{Smith}, J., {Meegan}, C., \& {Fermi GBM Team}. 2024, \bibinfo{title}{{GRB 240930B: Fermi GBM Observation},} GRB Coordinates Network, 37660, 1

\bibitem[{ {Smithsonian Astrophysical Observatory}(2000){Smithsonian Astrophysical Observatory}}]{SAO2000}
{Smithsonian Astrophysical Observatory}. 2000, \bibinfo{title}{{SAOImage DS9: A utility for displaying astronomical images in the X11 window environment},}, Astrophysics Source Code Library, record ascl:0003.002

\bibitem[{A.~M. {Soderberg} {et~al.}(2008){Soderberg}, {Berger}, {Page}, {Schady}, {Parrent}, {Pooley}, {Wang}, {Ofek}, {Cucchiara}, {Rau}, {Waxman}, {Simon}, {Bock}, {Milne}, {Page}, {Barentine}, {Barthelmy}, {Beardmore}, {Bietenholz}, {Brown}, {Burrows}, {Burrows}, {Byrngelson}, {Cenko}, {Chandra}, {Cummings}, {Fox}, {Gal-Yam}, {Gehrels}, {Immler}, {Kasliwal}, {Kong}, {Krimm}, {Kulkarni}, {Maccarone}, {M{\'e}sz{\'a}ros}, {Nakar}, {O'Brien}, {Overzier}, {de Pasquale}, {Racusin}, {Rea}, \& {York}}]{Soderberg2008Natur}
{Soderberg}, A.~M., {Berger}, E., {Page}, K.~L., {et~al.} 2008, \bibinfo{title}{{An extremely luminous X-ray outburst at the birth of a supernova},} \nat, 453, 469, \dodoi{10.1038/nature06997}

\bibitem[{R. {Sonawane} {et~al.}(2024){Sonawane}, {Pathak}, {Mukherjee}, {Bala}, {Meegan}, \& {Fermi GBM Team}}]{2024GCN38094}
{Sonawane}, R., {Pathak}, U., {Mukherjee}, O., {et~al.} 2024, \bibinfo{title}{{GRB 241104A: Fermi GBM Observation},} GRB Coordinates Network, 38094, 1

\bibitem[{G.~P. {Srinivasaragavan} {et~al.}(2025){Srinivasaragavan}, {Hamidani}, {Schroeder}, {Sarin}, {Ho}, {Piro}, {Cenko}, {Anand}, {Sollerman}, {Perley}, {Maeda}, {O'Connor}, {Kuncarayakti}, {Miller}, {Ahumada}, {Vail}, {Duffell}, {Ghosh Dastidar}, {Andreoni}, {Bochenek}, {Brennan}, {Carney}, {Chen}, {Freeburn}, {Gal-Yam}, {Jacobson-Gal{\'a}n}, {Kasliwal}, {Li}, {Li}, {Sravan}, \& {Warshofsky}}]{Srinivasaragavan2025}
{Srinivasaragavan}, G.~P., {Hamidani}, H., {Schroeder}, G., {et~al.} 2025, \bibinfo{title}{{EP250108a/SN 2025kg: A Broad-Line Type Ic Supernova Associated with a Fast X-ray Transient Showing Evidence of Extended CSM Interaction},} arXiv e-prints, arXiv:2504.17516, \dodoi{10.48550/arXiv.2504.17516}

\bibitem[{S. {Srivastav} {et~al.}(2025){Srivastav}, {Chen}, {Gillanders}, {Rhodes}, {Smartt}, {Huber}, {Aryan}, {Yang}, {Beri}, {Cooper}, {Nicholl}, {Smith}, {Stevance}, {Carotenuto}, {Chambers}, {Aamer}, {Angus}, {Fulton}, {Moore}, {Smith}, {Young}, {de Boer}, {Gao}, {Lin}, {Lowe}, {Magnier}, {Minguez}, {Pan}, \& {Wainscoat}}]{Srivastav2025}
{Srivastav}, S., {Chen}, T.~W., {Gillanders}, J.~H., {et~al.} 2025, \bibinfo{title}{{Identification of the Optical Counterpart of the Fast X-Ray Transient EP240414a},} \apjl, 978, L21, \dodoi{10.3847/2041-8213/ad9c75}

\bibitem[{H. {Sun} {et~al.}(2024{\natexlab{a}}){Sun}, {Li}, {Liu}, {Gao}, {Wang}, {Yuan}, {Zhang}, {Filippenko}, {Xu}, {An}, {Ai}, {Brink}, {Liu}, {Liu}, {Wang}, {Wu}, {Wu}, {Yang}, {Zhang}, {Zheng}, {Ahumada}, {Dai}, {Delaunay}, {Elias-Rosa}, {Benetti}, {Fu}, {Howell}, {Huang}, {Kasliwal}, {Karambelkar}, {Stein}, {Lei}, {Lian}, {Peng}, {Ridnaia}, {Svinkin}, {Wang}, {Wang}, {Wei}, {An}, {Andrews}, {Bai}, {Dai}, {Ehgamberdiev}, {Fan}, {Farah}, {Feng}, {Fynbo}, {Guo}, {Guo}, {Hu}, {Hu}, {Jiang}, {Jin}, {Li}, {Li}, {Li}, {Liang}, {Ling}, {Liu}, {Mao}, {McCully}, {Mirzaqulov}, {Newsome}, {Padilla Gonzalez}, {Pan}, {Terreran}, {Tinyanont}, {Wang}, {Wang}, {Wen}, {Xiang}, {Xue}, {Yang}, {Zhu}, {Cai}, {Castro-Tirado}, {Chen}, {Chen}, {Chen}, {Chen}, {Chen}, {Chen}, {Chen}, {Cheng}, {Cordier}, {Cui}, {Cui}, {Dai}, {Fan}, {Feng}, {Guan}, {Han}, {Hou}, {Hu}, {Huang}, {Huo}, {Jia}, {Jia}, {Jiang}, {Jin}, {Jin}, {Kuulkers}, {Li}, {Li}, {Li}, {Li}, {Li}, {Li}, {Li}, {Liu}, {Liu}, {Liu}, {Liu}, {Lu}, {Luo}, {Ma}, {Mao},
  {Nandra}, {O'Brien}, {Pan}, {Rau}, {Rea}, {Sanders}, {Song}, {Sun}, {Sun}, {Tan}, {Tang}, {Tao}, {Wang}, {Wang}, {Wang}, {Wang}, {Wang}, {Wang}, {Xiong}, {Xu}, {Xu}, {Xu}, {Xu}, {Xu}, {Xue}, {Xue}, {Yan}, {Yang}, {Yang}, {Yang}, {Zhang}, {Zhang}, {Zhang}, {Zhang}, {Zhang}, {Zhang}, {Zhang}, {Zhang}, {Zhang}, {Zhang}, {Zhao}, {Zhao}, {Zhao}, {Zhao}, {Zhou}, {Zhu}, \& {Zhu}}]{Sun2024}
{Sun}, H., {Li}, W.~X., {Liu}, L.~D., {et~al.} 2024{\natexlab{a}}, \bibinfo{title}{{Extragalactic fast X-ray transient from a weak relativistic jet associated with a Type Ic-BL supernova},} arXiv e-prints, arXiv:2410.02315, \dodoi{10.48550/arXiv.2410.02315}

\bibitem[{H. {Sun} {et~al.}(2024{\natexlab{b}}){Sun}, {Chen}, {Zhou}, {Zhang}, {Hu}, {Li}, {Liu}, {Ling}, {Zhang}, {Jin}, {Cheng}, {Cui}, {Fan}, {Hu}, {Huang}, {Liu}, {Liu}, {Lv}, {Lian}, {Mao}, {Pan}, {Pan}, {Wang}, {Wang}, {Wu}, {Xu}, {Xu}, {Yang}, {Yuan}, {Zhang}, {Zhang}, {Zhang}, {Zhang}, {Zhao}, {Yang}, {Dai}, {Liang}, {Chen}, {Jia}, {Zhang}, {Kuulkers}, {Santovincenzo}, {O'Brien}, {Nandra}, {Rau}, {Cordier}, \& {Einstein Probe Team}}]{2024GCN36690}
{Sun}, H., {Chen}, W., {Zhou}, H., {et~al.} 2024{\natexlab{b}}, \bibinfo{title}{{EP240618a: EP-WXT detection of a fast X-ray transient and EP-FXT quick follow-up},} GRB Coordinates Network, 36690, 1

\bibitem[{D. {Svinkin} {et~al.}(2024{\natexlab{a}}){Svinkin}, {Frederiks}, {Lysenko}, {Ridnaia}, {Tsvetkova}, {Ulanov}, {Cline}, \& {Konus-Wind team}}]{Svinkin2024GCN}
{Svinkin}, D., {Frederiks}, D., {Lysenko}, A., {et~al.} 2024{\natexlab{a}}, \bibinfo{title}{{EP241021a: EP detection of a fast X-ray transient},} GRB Coordinates Network, 38034, 1

\bibitem[{D. {Svinkin} {et~al.}(2024{\natexlab{b}}){Svinkin}, {Frederiks}, {Lysenko}, {Ridnaia}, {Tsvetkova}, {Ulanov}, {Cline}, \& {Konus-Wind Team}}]{2024GCN35972}
{Svinkin}, D., {Frederiks}, D., {Lysenko}, A., {et~al.} 2024{\natexlab{b}}, \bibinfo{title}{{Konus-Wind detection of GRB 240315C (possible counterpart of EP240315a)},} GRB Coordinates Network, 35972, 1

\bibitem[{D. {Svinkin} {et~al.}(2024{\natexlab{c}}){Svinkin}, {Frederiks}, {Lysenko}, {Ridnaia}, {Tsvetkova}, {Ulanov}, {Cline}, \& {Konus-Wind Team}}]{2024GCN38323}
{Svinkin}, D., {Frederiks}, D., {Lysenko}, A., {et~al.} 2024{\natexlab{c}}, \bibinfo{title}{{Konus-Wind detection of GRB 241115D / EP 241115a},} GRB Coordinates Network, 38323, 1

\bibitem[{X. {Tian} {et~al.}(2024){Tian}, {Peng}, {Mao}, {Jin}, {Liu}, {Liu}, {Ling}, {Zhang}, {Cheng}, {Chen}, {Cui}, {Fan}, {Hu}, {W.}, {Huang}, {Li}, {Liu}, {Lv}, {Lian}, {Pan}, {Pan}, {Sun}, {Wang}, {Wang}, {Xu}, {Xu}, {Yang}, {Yuan}, {Zhang}, {Zhang}, {Zhang}, {Zhang}, {Zhao}, {Chen}, {Jia}, {Cui}, {Han}, {Li}, {Song}, {Zhao}, {Zhang}, {Zhang}, {Liang}, {Kuulkers}, {Santovincenzo}, {O'Brien}, {Nandra}, {Rau}, {Cordier}, \& {Einstein Probe Team}}]{2024GCN37648}
{Tian}, X., {Peng}, H.~L., {Mao}, X., {et~al.} 2024, \bibinfo{title}{{EP240930a: EP-WXT detection of a fast X-ray transient},} GRB Coordinates Network, 37648, 1

\bibitem[{A. {Trigg} \&  {Fermi GBM Team}(2024){Trigg} \& {Fermi GBM Team}}]{2024GCN37917}
{Trigg}, A., \& {Fermi GBM Team}. 2024, \bibinfo{title}{{GRB 241026A Fermi GBM Analysis},} GRB Coordinates Network, 37917, 1

\bibitem[{E. {Troja} {et~al.}(2018){Troja}, {Ryan}, {Piro}, {van Eerten}, {Cenko}, {Yoon}, {Lee}, {Im}, {Sakamoto}, {Gatkine}, {Kutyrev}, \& {Veilleux}}]{Troja2018}
{Troja}, E., {Ryan}, G., {Piro}, L., {et~al.} 2018, \bibinfo{title}{{A luminous blue kilonova and an off-axis jet from a compact binary merger at z = 0.1341},} Nature Communications, 9, 4089, \dodoi{10.1038/s41467-018-06558-7}

\bibitem[{E. {Troja} {et~al.}(2019){Troja}, {van Eerten}, {Ryan}, {Ricci}, {Burgess}, {Wieringa}, {Piro}, {Cenko}, \& {Sakamoto}}]{Troja2019}
{Troja}, E., {van Eerten}, H., {Ryan}, G., {et~al.} 2019, \bibinfo{title}{{A year in the life of GW 170817: the rise and fall of a structured jet from a binary neutron star merger},} \mnras, 489, 1919, \dodoi{10.1093/mnras/stz2248}

\bibitem[{J.~N.~D. {van Dalen} {et~al.}(2024){van Dalen}, {Levan}, {Jonker}, {Malesani}, {Izzo}, {Sarin}, {Quirola-V{\'a}squez}, {Mata S{\'a}nchez}, {de Ugarte Postigo}, {van Hoof}, {Torres}, {Schulze}, {Littlefair}, {Chrimes}, {Ravasio}, {Bauer}, {Martin-Carrillo}, {Fraser}, {van der Horst}, {Jakobsson}, {O'Brien}, {De Pasquale}, {Pugliese}, {Sollerman}, {Tanvir}, {Zafar}, {Anderson}, {Galbany}, {Gal-Yam}, {Gromadzki}, {Muller-Bravo}, {Ragosta}, \& {Terwel}}]{vanDalen2024}
{van Dalen}, J. N.~D., {Levan}, A.~J., {Jonker}, P.~G., {et~al.} 2024, \bibinfo{title}{{The Einstein Probe transient EP240414a: Linking Fast X-ray Transients, Gamma-ray Bursts and Luminous Fast Blue Optical Transients},} arXiv e-prints, arXiv:2409.19056, \dodoi{10.48550/arXiv.2409.19056}

\bibitem[{H. {van Eerten}(2013){van Eerten}}]{vanEerten2013}
{van Eerten}, H. 2013, \bibinfo{title}{{Gamma-ray burst afterglow theory},} arXiv e-prints, arXiv:1309.3869, \dodoi{10.48550/arXiv.1309.3869}

\bibitem[{H. {van Eerten}(2018){van Eerten}}]{vanEerten2018}
{van Eerten}, H. 2018, \bibinfo{title}{{Gamma-ray burst afterglow blast waves},} International Journal of Modern Physics D, 27, 1842002, \dodoi{10.1142/S0218271818420026}

\bibitem[{H. {van Eerten} {et~al.}(2010){van Eerten}, {Zhang}, \& {MacFadyen}}]{vanEerten2010}
{van Eerten}, H., {Zhang}, W., \& {MacFadyen}, A. 2010, \bibinfo{title}{{Off-axis Gamma-ray Burst Afterglow Modeling Based on a Two-dimensional Axisymmetric Hydrodynamics Simulation},} \apj, 722, 235, \dodoi{10.1088/0004-637X/722/1/235}

\bibitem[{S. {van Velzen} {et~al.}(2021){van Velzen}, {Pasham}, {Komossa}, {Yan}, \& {Kara}}]{vanVelzen2021}
{van Velzen}, S., {Pasham}, D.~R., {Komossa}, S., {Yan}, L., \& {Kara}, E.~A. 2021, \bibinfo{title}{{Reverberation in Tidal Disruption Events: Dust Echoes, Coronal Emission Lines, Multi-wavelength Cross-correlations, and QPOs},} \ssr, 217, 63, \dodoi{10.1007/s11214-021-00835-6}

\bibitem[{M.~A. {Walker}(1998){Walker}}]{Walker1998}
{Walker}, M.~A. 1998, \bibinfo{title}{{Interstellar scintillation of compact extragalactic radio sources},} \mnras, 294, 307, \dodoi{10.1046/j.1365-8711.1998.01238.x10.1111/j.1365-8711.1998.01238.x}

\bibitem[{X.-G. {Wang} {et~al.}(2018){Wang}, {Zhang}, {Liang}, {Lu}, {Lin}, {Li}, \& {Li}}]{Wang2018}
{Wang}, X.-G., {Zhang}, B., {Liang}, E.-W., {et~al.} 2018, \bibinfo{title}{{Gamma-Ray Burst Jet Breaks Revisited},} \apj, 859, 160, \dodoi{10.3847/1538-4357/aabc13}

\bibitem[{Y. {Wang} {et~al.}(2024{\natexlab{a}}){Wang}, {He}, {Yang}, {Cheng}, {Hu}, {Xu}, {Yuan}, \& {Einstein Probe Team}}]{Wang2024GCN}
{Wang}, Y., {He}, H., {Yang}, S.~K., {et~al.} 2024{\natexlab{a}}, \bibinfo{title}{{EP241021a: EP-FXT follow-up observation update},} GRB Coordinates Network, 37848, 1

\bibitem[{Y. {Wang} {et~al.}(2025){Wang}, {Lian}, {Zhang}, {Tian}, {Li}, {Pan}, \& {Einstein Probe Team}}]{2025GCN39448}
{Wang}, Y., {Lian}, T.~Y., {Zhang}, W.~J., {et~al.} 2025, \bibinfo{title}{{EP250223a: refined analysis of the EP-WXT and FXT observations},} GRB Coordinates Network, 39448, 1

\bibitem[{Y. {Wang} {et~al.}(2024{\natexlab{b}}){Wang}, {Sun}, {Wang}, {Zhang}, {Liu}, {Pan}, {Ling}, {Jin}, {Zhang}, {Cheng}, {Chen}, {Cui}, {Fan}, {Hu}, {Hu}, {Huang}, {Li}, {Liu}, {Liu}, {Lv}, {Lian}, {Mao}, {Pan}, {Wang}, {Wang}, {Wu}, {Xu}, {Xu}, {Yang}, {Yuan}, {Zhang}, {Zhang}, {Zhang}, {Zhao}, {Chen}, {Jia}, {Cui}, {Han}, {Li}, {Song}, {Zhao}, {Zhang}, {Zhang}, {Kuulkers}, {Santovincenzo}, {O'Brien}, {Nandra}, {Rau}, {Cordier}, \& {Einstein Probe Team}}]{2024GCN37034}
{Wang}, Y., {Sun}, H., {Wang}, Y.~L., {et~al.} 2024{\natexlab{b}}, \bibinfo{title}{{EP240804a: EP-WXT detection and EP-FXT follow-up observation of a fast X-ray transient},} GRB Coordinates Network, 37034, 1

\bibitem[{Y.~L. {Wang} {et~al.}(2024{\natexlab{a}}){Wang}, {Dai}, {Peng}, {Shui}, {Jin}, {Ling}, {Yuan}, {Liu}, {Zhang}, {Cheng}, {Cui}, {Fan}, {Hu}, {Hu}, {Huang}, {Li}, {Liu}, {Liu}, {Lv}, {Lian}, {Mao}, {Pan}, {Pan}, {Sun}, {Wang}, {Wang}, {Wen}, {Wu}, {Xu}, {Xu}, {Yang}, {Zhang}, {Zhang}, {Zhang}, {Zhang}, {Zhao}, {Chen}, {Jia}, {Zhang}, {Kuulkers}, {Santovincenzo}, {O'Brien}, {Nandra}, {Rau}, {Cordier}, \& {Einstein Probe Team}}]{2024GCN36807}
{Wang}, Y.~L., {Dai}, C.~Y., {Peng}, J.~Q., {et~al.} 2024{\natexlab{a}}, \bibinfo{title}{{EP240703a: EP-WXT detection of a fast X-ray transient},} GRB Coordinates Network, 36807, 1

\bibitem[{Y.~L. {Wang} {et~al.}(2024{\natexlab{b}}){Wang}, {Li}, {Wang}, {Pan}, {Ling}, {Jin}, {Liu}, {Zhang}, {Cheng}, {Chen}, {Cui}, {Fan}, {Hu}, {Huang}, {Li}, {Liu}, {Liu}, {Lv}, {Lian}, {Mao}, {Pan}, {Sun}, {Wang}, {Wu}, {Xu}, {Xu}, {Yang}, {Yuan}, {Zhang}, {Zhang}, {Zhang}, {Zhang}, {Zhao}, {Chen}, {Jia}, {Cui}, {Han}, {Li}, {Song}, {Zhao}, {Zhang}, {Zhang}, {Kuulkers}, {Santovincenzo}, {O'Brien}, {Nandra}, {Rau}, {Cordier}, \& {Einstein Probe Team}}]{2024GCN37019}
{Wang}, Y.~L., {Li}, A., {Wang}, W.~X., {et~al.} 2024{\natexlab{b}}, \bibinfo{title}{{EP240802a: EP-WXT detection and EP-FXT follow-up observation of a fast X-ray transient},} GRB Coordinates Network, 37019, 1

\bibitem[{M.~C. {Weisskopf} {et~al.}(2002){Weisskopf}, {Brinkman}, {Canizares}, {Garmire}, {Murray}, \& {Van Speybroeck}}]{Weisskopf2002}
{Weisskopf}, M.~C., {Brinkman}, B., {Canizares}, C., {et~al.} 2002, \bibinfo{title}{{An Overview of the Performance and Scientific Results from the Chandra X-Ray Observatory},} \pasp, 114, 1, \dodoi{10.1086/338108}

\bibitem[{W.~F. {Wen} {et~al.}(2025){Wen}, {Wu}, {Wu}, {Wang}, {Ling}, \& {Einstein Probe Team}}]{2025GCN39532}
{Wen}, W.~F., {Wu}, J.~H., {Wu}, H.~Z., {et~al.} 2025, \bibinfo{title}{{EP250227a: Einstein Probe detected of a fast X-ray transient},} GRB Coordinates Network, 39532, 1

\bibitem[{S.~E. {Woosley} \& J.~S. {Bloom}(2006){Woosley} \& {Bloom}}]{WoosleyBloom2006}
{Woosley}, S.~E., \& {Bloom}, J.~S. 2006, \bibinfo{title}{{The Supernova Gamma-Ray Burst Connection},} \araa, 44, 507, \dodoi{10.1146/annurev.astro.43.072103.150558}

\bibitem[{H.~Z. {Wu} {et~al.}(2024){Wu}, {Wang}, {Hu}, {Liang}, {Wu}, {Jiang}, {Yang}, {Jin}, {Ling}, \& {Einstein Probe Team}}]{2024GCN37997}
{Wu}, H.~Z., {Wang}, B.~T., {Hu}, D.~F., {et~al.} 2024, \bibinfo{title}{{EP241030a: EP detection of GRB 241030A X-ray Afterglow},} GRB Coordinates Network, 37997, 1

\bibitem[{Q.~Y. {Wu} {et~al.}(2025){Wu}, {Yang}, {Zhou}, {Jin}, \& {Einstein Probe Team}}]{2025GCN39028}
{Wu}, Q.~Y., {Yang}, J., {Zhou}, X.~Y., {Jin}, C.~C., \& {Einstein Probe Team}. 2025, \bibinfo{title}{{EP250125a: a new X-ray transient detected by Einstein Probe},} GRB Coordinates Network, 39028, 1

\bibitem[{Q.~Y. {Wu} {et~al.}(2024){Wu}, {CAS)}, {Shui}, {Lv}, {Ling}, {Liu}, {Zhang}, {Jin}, {Chen}, {Cheng}, {Cui}, {Fan}, {Hu}, {Hu}, {Huang}, {Li}, {Liu}, {Liu}, {Lian}, {Mao}, {Pan}, {Pan}, {Sun}, {Wang}, {Wang}, {Xu}, {Xu}, {Yang}, {Yuan}, {Zhang}, {Zhang}, {Zhang}, {Zhang}, {Zhao}, {Chen}, {Jia}, {Zhang}, {Kuulkers}, {Santovincenzo}, {O'Brien}, {Nandra}, {Rau}, {Cordier}, \& {Einstein Probe Team}}]{2024GCN36766}
{Wu}, Q.~Y., {CAS)}, {Shui}, Q.~C., {et~al.} 2024, \bibinfo{title}{{EP240626a: EP-WXT detection of a fast X-ray transient},} GRB Coordinates Network, 36766, 1

\bibitem[{Y.~Q. {Xue} {et~al.}(2019){Xue}, {Zheng}, {Li}, {Brandt}, {Zhang}, {Luo}, {Zhang}, {Bauer}, {Sun}, {Lehmer}, {Wu}, {Yang}, {Kong}, {Li}, {Sun}, {Wang}, \& {Vito}}]{Xue2019Natur}
{Xue}, Y.~Q., {Zheng}, X.~C., {Li}, Y., {et~al.} 2019, \bibinfo{title}{{A magnetar-powered X-ray transient as the aftermath of a binary neutron-star merger},} \nat, 568, 198, \dodoi{10.1038/s41586-019-1079-5}

\bibitem[{H.~N. {Yang} {et~al.}(2024){Yang}, {Zhang}, {Wang}, {Yuan}, {Ling}, {Jin}, {Liu}, {Zhang}, {Cheng}, {Chen}, {Cui}, {Fan}, {Hu}, {Hu}, {Huang}, {Li}, {Liu}, {Liu}, {Lv}, {Lian}, {Mao}, {Pan}, {Pan}, {Sun}, {Wang}, {Wu}, {Xu}, {Xu}, {Zhang}, {Zhang}, {Zhang}, {Zhao}, {Chen}, {Jia}, {Cui}, {Han}, {Li}, {Song}, {Zhao}, {Zhang}, {Zhang}, {Kuulkers}, {Santovincenzo}, {O'Brien}, {Nandra}, {Rau}, {Cordier}, \& {Einstein Probe Team}}]{2024GCN37188}
{Yang}, H.~N., {Zhang}, W.~J., {Wang}, W.~X., {et~al.} 2024, \bibinfo{title}{{EP240816a: EP detection of a fast X-ray transient and follow-up observation},} GRB Coordinates Network, 37188, 1

\bibitem[{Y.~H.~I. {Yin} {et~al.}(2024){Yin}, {Wang}, {Zhao}, {Pan}, \& {Einstein Probe Team}}]{2024GCN38457}
{Yin}, Y. H.~I., {Wang}, C.~Y., {Zhao}, D.~H., {Pan}, H.~W., \& {Einstein Probe Team}. 2024, \bibinfo{title}{{EP241206a: EP detection of a fast X-ray transient},} GRB Coordinates Network, 38457, 1

\bibitem[{Y.-H.~I. {Yin} {et~al.}(2024){Yin}, {Zhang}, {Yang}, {Sun}, {Zhang}, {Shao}, {Hu}, {Zhu}, {Xu}, {An}, {Gao}, {Wu}, {Zhang}, {Castro-Tirado}, {Pandey}, {Rau}, {Lei}, {Xie}, {Ghirlanda}, {Piro}, {O'Brien}, {Troja}, {Jonker}, {Yu}, {An}, {Chen}, {Chen}, {Dong}, {Eyles-Ferris}, {Fan}, {Fu}, {Fynbo}, {Gao}, {Huang}, {Jiang}, {Jiang}, {Julakanti}, {Kuulkers}, {Lao}, {Li}, {Ling}, {Liu}, {Liu}, {Mou}, {Pan}, {Wei}, {Wu}, {Yadav}, {Yang}, {Yuan}, \& {Zhang}}]{Yin2024}
{Yin}, Y.-H.~I., {Zhang}, B.-B., {Yang}, J., {et~al.} 2024, \bibinfo{title}{{Triggering the Untriggered: The First Einstein Probe-detected Gamma-Ray Burst 240219A and Its Implications},} \apjl, 975, L27, \dodoi{10.3847/2041-8213/ad8652}

\bibitem[{W. {Yuan} {et~al.}(2022){Yuan}, {Zhang}, {Chen}, \& {Ling}}]{Yuan2022}
{Yuan}, W., {Zhang}, C., {Chen}, Y., \& {Ling}, Z. 2022, in Handbook of X-ray and Gamma-ray Astrophysics, ed. C.~{Bambi} \& A.~{Sangangelo}, 86, \dodoi{10.1007/978-981-16-4544-0_151-1}

\bibitem[{W. {Yuan} {et~al.}(2025){Yuan}, {Dai}, {Feng}, {Jin}, {Jonker}, {Kuulkers}, {Liu}, {Nandra}, {O'Brien}, {Piro}, {Rau}, {Rea}, {Sanders}, {Tao}, {Wang}, {Wu}, {Zhang}, {Zhang}, {Ai}, {Buchner}, {Bulbul}, {Chen}, {Chen}, {Chen}, {Chen}, {Coleiro}, {Zelati}, {Dai}, {Fan}, {Fan}, {Friedrich}, {Gao}, {Ge}, {Ge}, {Geng}, {Ghirlanda}, {Gianfagna}, {Gou}, {Guillot}, {Hou}, {Hu}, {Huang}, {Ji}, {Jia}, {Komossa}, {Kong}, {Lan}, {Li}, {Li}, {Li}, {Li}, {Li}, {Li}, {Ling}, {Liu}, {Liu}, {Liu}, {Liu}, {Luo}, {Ma}, {Maggi}, {Maitra}, {Marino}, {Ng}, {Pan}, {Rukdee}, {Soria}, {Sun}, {Tam}, {Thakur}, {Tian}, {Troja}, {Wang}, {Wang}, {Wang}, {Wei}, {Wen}, {Wu}, {Wu}, {Xiao}, {Xu}, {Xu}, {Xu}, {Xu}, {Yang}, {You}, {Yu}, {Yu}, {Zhang}, {Zhang}, {Zhang}, {Zhang}, {Zhang}, {Zhang}, {Zhou}, \& {Zou}}]{Yuan2025}
{Yuan}, W., {Dai}, L., {Feng}, H., {et~al.} 2025, \bibinfo{title}{{Science objectives of the Einstein Probe mission},} Science China Physics, Mechanics, and Astronomy, 68, 239501, \dodoi{10.1007/s11433-024-2600-3}

\bibitem[{B.~A. {Zauderer} {et~al.}(2011){Zauderer}, {Berger}, {Soderberg}, {Loeb}, {Narayan}, {Frail}, {Petitpas}, {Brunthaler}, {Chornock}, {Carpenter}, {Pooley}, {Mooley}, {Kulkarni}, {Margutti}, {Fox}, {Nakar}, {Patel}, {Volgenau}, {Culverhouse}, {Bietenholz}, {Rupen}, {Max-Moerbeck}, {Readhead}, {Richards}, {Shepherd}, {Storm}, \& {Hull}}]{Zauderer2011}
{Zauderer}, B.~A., {Berger}, E., {Soderberg}, A.~M., {et~al.} 2011, \bibinfo{title}{{Birth of a relativistic outflow in the unusual {\ensuremath{\gamma}}-ray transient Swift J164449.3+573451},} \nat, 476, 425, \dodoi{10.1038/nature10366}

\bibitem[{B. {Zhang} \& P. {M{\'e}sz{\'a}ros}(2004){Zhang} \& {M{\'e}sz{\'a}ros}}]{ZhangMeszarosREVIEW}
{Zhang}, B., \& {M{\'e}sz{\'a}ros}, P. 2004, \bibinfo{title}{{Gamma-Ray Bursts: progress, problems \& prospects},} International Journal of Modern Physics A, 19, 2385, \dodoi{10.1142/S0217751X0401746X}

\bibitem[{L.-L. {Zhang} {et~al.}(2023){Zhang}, {Ren}, {Wang}, \& {Liang}}]{Zhang2023}
{Zhang}, L.-L., {Ren}, J., {Wang}, Y., \& {Liang}, E.-W. 2023, \bibinfo{title}{{Very-high-energy Gamma-Ray Afterglows of GRB 201015A and GRB 201216C},} \apj, 952, 127, \dodoi{10.3847/1538-4357/acd190}

\bibitem[{W.~J. {Zhang} {et~al.}(2024){Zhang}, {Mao}, {Zhang}, {Liu}, {Liu}, {Zhang}, {Ling}, {Jin}, {Cheng}, {Chen}, {Cui}, {Fan}, {Hu}, {Hu}, {Huang}, {Li}, {Lian}, {Liu}, {Lv}, {Pan}, {Pan}, {Sun}, {Wang}, {Wang}, {Wu}, {Xu}, {Xu}, {Yang}, {Zhang}, {Zhang}, {Zhang}, {Zhao}, {Kuulkers}, {O'Brien}, {Yuan}, \& {Einstein Probe team}}]{2024GCN35931}
{Zhang}, W.~J., {Mao}, X., {Zhang}, W.~D., {et~al.} 2024, \bibinfo{title}{{Einstein Probe detected of a fast X-ray transient EP240315a},} GRB Coordinates Network, 35931, 1

\bibitem[{Y.~J. {Zhang} {et~al.}(2025){Zhang}, {Dai}, {Chen}, {Wang}, {Liu}, \& {Einstein Probe Team}}]{2025GCN39591}
{Zhang}, Y.~J., {Dai}, C.~Y., {Chen}, W., {et~al.} 2025, \bibinfo{title}{{EP250304a: refined analysis of the EP-WXT and EP-FXT observations},} GRB Coordinates Network, 39591, 1

\bibitem[{T. {Zhao} {et~al.}(2025){Zhao}, {Chen}, {Shui}, {Chatterjee}, {Lian}, {Jin}, \& {Einstein Probe Team}}]{2025GCN38905}
{Zhao}, T., {Chen}, X.~L., {Shui}, Q.~C., {et~al.} 2025, \bibinfo{title}{{EP250111a: a new X-ray transient detected by Einstein Probe},} GRB Coordinates Network, 38905, 1

\bibitem[{T. {Zhao} {et~al.}(2024){Zhao}, {Liang}, {Liu}, {Zhang}, \& {Einstein Probe Team}}]{2024GCN38058}
{Zhao}, T., {Liang}, Y.~F., {Liu}, H.~Y., {Zhang}, W.~D., \& {Einstein Probe Team}. 2024, \bibinfo{title}{{EP241103a: EP observation update},} GRB Coordinates Network, 38058, 1

\bibitem[{W. {Zheng} {et~al.}(2024){Zheng}, {Brink}, {Filippenko}, {Yang}, \& {KAIT GRB Team}}]{Zheng2024GCN}
{Zheng}, W., {Brink}, T.~G., {Filippenko}, A.~V., {Yang}, Y., \& {KAIT GRB Team}. 2024, \bibinfo{title}{{EP241021a and EP241026b: Keck/LRIS spectroscopic observations},} GRB Coordinates Network, 38294, 1

\bibitem[{C. {Zhou} {et~al.}(2024){Zhou}, {Zhao}, {Mao}, {Liang}, {Yuan}, \& {Einstein Probe Team}}]{2024GCN38426}
{Zhou}, C., {Zhao}, G.~Y., {Mao}, X., {et~al.} 2024, \bibinfo{title}{{EP241202b: EP detection of a fast X-ray transient},} GRB Coordinates Network, 38426, 1

\bibitem[{H. {Zhou} {et~al.}(2024{\natexlab{a}}){Zhou}, {Li}, {Liu}, {Cheng}, {Ling}, \& {Einstein Probe Team}}]{2024GCN38081}
{Zhou}, H., {Li}, R.-Z., {Liu}, Q.~C., {et~al.} 2024{\natexlab{a}}, \bibinfo{title}{{EP241104a: EP detection of GRB 241104A X-ray emission},} GRB Coordinates Network, 38081, 1

\bibitem[{H. {Zhou} {et~al.}(2024{\natexlab{b}}){Zhou}, {Zhu}, {Zhang}, {Jim}, \& {Einstein Probe team}}]{2024GCN38606}
{Zhou}, H., {Zhu}, S.~F., {Zhang}, M.~H., {Jim}, C.~C., \& {Einstein Probe team}. 2024{\natexlab{b}}, \bibinfo{title}{{EP241217b/GRB 241217A: EP detection of the prompt X-ray emssion},} GRB Coordinates Network, 38606, 1

\bibitem[{H. {Zhou} {et~al.}(2024{\natexlab{c}}){Zhou}, {Zhu}, {Zhang}, {Sun}, {Jin}, \& {Einstein Probe Team}}]{2024GCN38624}
{Zhou}, H., {Zhu}, S.~F., {Zhang}, M.~H., {et~al.} 2024{\natexlab{c}}, \bibinfo{title}{{EP241217a: refined EP-WXT analysis and EP-FXT follow-up observations},} GRB Coordinates Network, 38624, 1

\bibitem[{H. {Zhou} {et~al.}(2024{\natexlab{d}}){Zhou}, {Chen}, {Sun}, {Zhang}, {Hu}, {Li}, {Ling}, {Liu}, {Zhang}, {Jin}, {Cheng}, {Cui}, {Fan}, {Hu}, {Huang}, {Liu}, {Liu}, {Lv}, {Lian}, {Mao}, {Pan}, {Pan}, {Wang}, {Wang}, {Wu}, {Xu}, {Xu}, {Yang}, {Yuan}, {Zhang}, {Zhang}, {Zhang}, {Zhang}, {Zhao}, {Yang}, {Dai}, {Liang}, {Chen}, {Jia}, {Zhang}, {Kuulkers}, {Santovincenzo}, {O'Brien}, {Nandra}, {Rau}, {Cordier}, \& {Einstein Probe Team}}]{2024GCN36691}
{Zhou}, H., {Chen}, W., {Sun}, H., {et~al.} 2024{\natexlab{d}}, \bibinfo{title}{{EP240617a: EP-WXT detection of a fast X-ray transient},} GRB Coordinates Network, 36691, 1

\bibitem[{H. {Zhou} {et~al.}(2024{\natexlab{e}}){Zhou}, {Wang}, {Hu}, {Ling}, {Jin}, {Liu}, {Zhang}, {Cheng}, {Chen}, {Cui}, {Fan}, {Hu}, {Huang}, {Li}, {Liu}, {Liu}, {Lv}, {Lian}, {Mao}, {Pan}, {Pan}, {Sun}, {Wang}, {Wu}, {Xu}, {Xu}, {Yang}, {Yuan}, {Zhang}, {Zhang}, {Zhang}, {Zhang}, {Zhao}, {Chen}, {Jia}, {Zhang}, {Kuulkers}, {Santovincenzo}, {O'Brien}, {Nandra}, {Rau}, {Cordier}, \& {Einstein Probe Team}}]{2024GCN36997}
{Zhou}, H., {Wang}, W.~X., {Hu}, J.~W., {et~al.} 2024{\natexlab{e}}, \bibinfo{title}{{EP240801a: EP-WXT detection and EP-FXT follow-up observation of a fast X-ray transient},} GRB Coordinates Network, 36997, 1

\bibitem[{X.~Y. {Zhou} {et~al.}(2025){Zhou}, {Li}, {Zhao}, {Xu}, {Wu}, {Liu}, \& {Einstein Probe Team}}]{2025GCN39266}
{Zhou}, X.~Y., {Li}, A., {Zhao}, D.~H., {et~al.} 2025, \bibinfo{title}{{EP250207b: a new X-ray transient detected by Einstein Probe},} GRB Coordinates Network, 39266, 1

\end{thebibliography}
\bibliographystyle{aasjournalv7}



\end{document}